\documentclass[twocolumn,prd,aps,superscriptaddress,preprintnumbers,tightenlines,showpacs,nofootinbib,eqsecnum,amsfonts,amsmath]{revtex4-1}
\usepackage[utf8]{inputenc}
\usepackage{graphicx}
\usepackage{multirow}
\usepackage{subfigure} 
\usepackage{yfonts}
\usepackage[usenames, dvipsnames]{color}
\usepackage{tikz}
\usepackage{mathtools}
\usepackage{amssymb}
\usepackage[normalem]{ulem}
\usepackage{booktabs}
\usepackage{dcolumn}
\usepackage{float}

\usepackage{colortbl}
\usepackage{hyperref}

\allowdisplaybreaks

\newcommand{\Ham}{\hat{H}}

\newcommand{\subreal}{_{\text{EOB}}}

\newcommand{\R}{r}
\newcommand{\Pp}{p}
\newcommand{\p}{\hat{p}}
\newcommand{\aaa}{a}

\newcommand{\zzz}{\Delta z}
\newcommand{\zzero}{\zzz^{(0)}}
\newcommand{\zone}{\zzz^{(1)}}
\newcommand{\ztwo}{\zzz^{(2)}}
\newcommand{\EEE}{\hat{E}_{\text{SMR}}}
\newcommand{\X}{f}
\newcommand{\Ebindsmr}{\hat{E}^{\text{SF}}_{\text{bind}}}
\newcommand{\Ebindeob}{\hat{E}^{\text{EOB}}_{\text{bind}}}
\newcommand{\lnE}{\ln E_\text{S}^{-2}}

\newcommand{\AEI}{\affiliation{Max Planck Institute for Gravitational Physics (Albert Einstein Institute), Am M\"uhlenberg 1, Potsdam 14476, Germany}}
\newcommand{\Maryland}{\affiliation{Department of Physics, University of Maryland, College Park, MD 20742, USA}}

\begin{document}
\title{
Quasi-circular inspirals and plunges from non-spinning effective-one-body Hamiltonians with gravitational self-force information
}

\author{Andrea Antonelli}\AEI
\author{Maarten van de Meent}\AEI
\author{Alessandra Buonanno}\AEI\Maryland
\author{Jan Steinhoff}\AEI
\author{Justin Vines}\AEI

\date{\today}

\begin{abstract}
  The self-force program aims at accurately modeling relativistic
  two-body systems with a small mass ratio (SMR). In the context of
  the effective-one-body (EOB) framework, current results from this program
  can be used to determine the effective metric
    components at linear order in the mass ratio, resumming
  post-Newtonian (PN) dynamics around the test-particle limit in the
  process. It was shown in [Akcay et al., Phys. Rev. D \textbf{86}
  (2012)] that, in the original (standard) EOB gauge,
the SMR contribution to the metric component  $g^\text{eff}_{tt}$ exhibits a coordinate singularity at the light-ring (LR) radius. In this paper, we adopt a different gauge for the
  EOB dynamics and obtain a Hamiltonian that is free of poles at the
  LR, with complete circular-orbit information at linear order in the
  mass ratio and non-circular-orbit and higher-order-in-mass-ratio
  terms up to 3PN order.  We confirm the absence of the LR-divergence in
  such an EOB Hamiltonian via plunging trajectories through the LR radius.  Moreover, we compare the
  binding energies and inspiral waveforms of EOB models with SMR, PN and mixed
  SMR-3PN information on a quasi-circular inspiral against
  numerical-relativity predictions. We find good agreement
  between NR simulations and EOB models with SMR-3PN information for both equal and unequal
  mass ratios. In particular, when compared to EOB inspiral waveforms with only 3PN information, EOB Hamiltonians with
  SMR-3PN information improves the modeling of binary systems
  with small mass ratios $q \lesssim 1/3$, with a dephasing accumulated in $\sim$30 gravitational-wave (GW) cycles being of the order of few hundredths of a radian up to 4 GW cycles before merger.
\end{abstract}
\maketitle

\section{Introduction}

Solving the two-body problem in General Relativity (GR) remains a challenge of both theoretical interest and astrophysical relevance. 
Albeit an analytical solution is lacking, advances in numerical relativity
(NR) in the past decades provided the first numerical evolutions of
merging compact objects \cite{Pretorius:2005gq,Campanelli:2005dd,Baker:2005vv}, as 
well as catalogs of waveforms \cite{Mroue:2013xna,Jani:2016wkt,Healy:2017psd,Dietrich:2018phi, Boyle:2019kee}. On the
analytical side of the problem, approximations to the binary motion and gravitational 
radiation, via expansions in one or more small parameters, have been applied to different 
domains of validity~\cite{Blanchet:2013haa,Barack:2018yvs,Tiec:2014lba}, providing us 
with a variety of waveform models.

The effective-one-body (EOB) framework is a synergistic approach that allows us to resum information from several analytical approximations.
NR-calibrated inspiral-merger-ringdown models based on EOB theory \cite{Pan:2011gk,Taracchini:2012ig,Taracchini:2013rva,Bohe:2016gbl,Cotesta:2018fcv} were employed by LIGO-Virgo experiments to detect gravitational waves (GWs) and infer astrophysical and cosmological information from them~\cite{Abbott:2016blz,Abbott:2016nmj,Abbott:2017vtc,Abbott:2017oio,Abbott:2017gyy,TheLIGOScientific:2017qsa,LIGOScientific:2018mvr,TheLIGOScientific:2016wfe,LIGOScientific:2018mvr,TheLIGOScientific:2016src}. In view of the expected increase in the signal-to-noise ratio of signals detected with upcoming LIGO-Virgo runs, and next generation detectors in space (LISA~\cite{Audley:2017drz}) and on Earth (Einstein Telescope~\cite{Punturo:2010zz} and Cosmic Explorer~\cite{Evans:2016mbw}),
it is important and timely to include more physics and build more accurate waveforms in the EOB approach. 

Historically, physical information from the two-body problem has
mostly entered EOB theory via the post-Newtonian (PN) expansion
\cite{Buonanno:1998gg, Buonanno:2000ef, Damour:2000we}, valid for
bound orbits at large distances and for velocities smaller than the
speed of light $v^2/c^2\sim\text{G}M/\R c^2 \ll 1$ (here $M=m_1+m_2$
is the total mass, with $m_1$ the mass of the primary and $m_2$ the
mass of the secondary body).  PN conservative-dynamics information has
so far been calculated up to fourth order, in the nonspinning
  case, using the Arnowitt-Deser-Misner (ADM)
\cite{Jaranowski:2015lha,Damour:2014jta,Damour:2016abl}, Fokker
\cite{Bernard:2015njp,Bernard:2016wrg,Bernard:2017bvn} and
effective-field-theory approaches \cite{Foffa:2019rdf,Foffa:2019yfl} (which were also employed to determine the 5PN gravitational interaction in the static limit~\cite{Foffa:2019hrb,Blumlein:2019zku}). In the quasi-circular-orbit limit,
4PN information has been successfully included in the EOB dynamics in
the form of an expansion in the inverse radius $u \equiv \text{G}M/\R
c^2 \ll 1$ and in the momenta $\boldsymbol{p}^2$, with exact
dependence on the symmetric mass ratio $\nu=m_{1}m_{2}/M^{2}$
\cite{Damour:2015isa}. 
Further resummations of this PN expansion form the core of the EOB waveform models \cite{Damour:2008gu,Damour:2009kr,Barausse:2009xi,Pan:2011gk,Hinderer:2016eia}.
Post-Minkowskian (PM) information, valid in the weak field $\text{G}M/\R c^2 \ll 1$, but for all velocities $v^2/c^2\leq 1$, has also provided valuable insight in the structure of EOB Hamiltonians, for both spinning and non-spinning bound systems \cite{Damour:2017zjx, Damour:2016pm,Vines:2017pm,Vines:2018gqi,Antonelli:2019ytb}.
 
The self-force (SF) program, initiated in Refs.~\cite{Mino:1996nk,Quinn:1996am}  and based on an expansion of Einstein's equations in the small mass ratio (SMR) $q=m_2/m_1$,
has been successful in the calculation of the gravitational SF of a small body around Schwarzschild \cite{Barack:2007tm,Barack:2010tm},  and recently Kerr black-holes \cite{Shah:2012gu,vandeMeent:2015lxa,vandeMeent:2016pee,vandeMeent:2017bcc}, to first order in the mass ratio and for generic bound orbits. The results, corroborated by the use of several gauges and numerical techniques (see, e.g., Ref.~\cite{Barack:2018yvs} and references therein), have been already used to evolve extreme-mass-ratio-inspirals (EMRIs) around a Schwarzschild black-hole \cite{Warburton:2011fk,Osburn:2015duj} and they represent a key input for EMRI waveform modeling schemes recently developed \cite{vandeMeent:2018rms} and under development \cite{Hinderer:2008dm}.

As the SF program employs different gauge-dependent schemes to obtain
its results \cite{Barack:2018yvs}, it is paramount to be able to check
results via gauge-invariant quantities, such as the innermost-stable--circular-orbit (ISCO)-shift
\cite{Barack:2009ey}, periastron advance 
\cite{Barack:2010ny,LeTiec:2011bk,vandeMeent:2016hel}, spin-precession
\cite{Dolan:2013roa,Kavanagh:2017wot,Bini:2014ica,Bini:2018ylh},
tidal invariants \cite{Dolan:2014pja,Bini:2014zxa} and the Detweiler redshift
\cite{Detweiler:2008ft,Barack:2011ed,Akcay:2015pza,Kavanagh:2015lva,Shah:2013uya,Johnson-McDaniel:2015vva}.
For a particle with four-velocity $\tilde{u}^\alpha$ normalized in an
effective metric $\tilde{g}_{\alpha\beta}=g^\text{(0)}_{\alpha\beta}+h^{\text{R}}_{\alpha\beta}$
[i.e., moving around a Schwarzschild background
$g^\text{(0)}_{\alpha\beta}$ perturbed by a regularized metric
$h_{\alpha\beta}^{\text{R}}$ and such that
$\tilde{g}_{\alpha\beta}\tilde{u}^\alpha\tilde{u}^\beta=-1+\mathcal{O}(\nu)$],
the Detweiler redshift is defined as the ratio between proper time
measured in an orbit around the effective metric
$\tilde{g}_{\alpha\beta}$, $d\tilde{\tau}$, and coordinate time,
$d\tau$ \footnote{As pointed out in Ref.~\cite{Barack:2018yvs}, $z$
  does not correspond to the gravitational redshift due to the use of
  the regularized perturbation $h^{\text{R}}_{\alpha\beta}$ in its
  definition. It does only in the full geometry, e.g., including a
  singular metric $h^{\text{S}}_{\alpha\beta}$ at the location of the
  particle such that the body perturbation is $h_{\alpha\beta}\equiv
  h^{\text{R}}_{\alpha\beta}+h^{\text{S}}_{\alpha\beta}$. A sounder
  physical description can be obtained if the small companion is a
  black hole, since the Detweiler redshift can then be linked to the
  surface gravity $\kappa$ of the small
  body~\cite{Zimmerman:2016ajr}. }:
$z \equiv (\tilde{u}^t)^{-1}=d\tilde{\tau}/d\tau$
\cite{Detweiler:2008ft,Barack:2018yvs}.  Recently, the
Detweiler redshift has been used for cross-cultural studies between
approximations to the two-body problem in GR
\cite{LeTiec:2011ab,Tiec:2014lba}, and it has provided an important
benchmark to check PN and SMR results in the small-mass-ratio and
weak-field domain, in which both PN and SMR frameworks are expected to
be valid. This synergistic program has been extended 
to NR simulations of equal--mass-ratio binaries with the computation of 
the Detweiler redshift in Ref.~\cite{Zimmerman:2016ajr}.

As pointed out in Ref.~\cite{Damour:2009sf}, gauge-invariant SMR
quantities such as the Detweiler redshift can be also used to inform
the conservative sector of EOB Hamiltonians
\cite{LeTiec:2011dp,Damour:2009sf,Barausse:2011dq,LeTiec:2011bk}.
There are two ways in which this valuable information
 could be incorporated into the EOB approach: it can be either used to partially determine high-order PN coefficients of EOB Hamiltonians \cite{kavanaghetal:2015,Hopper:2015icj,Kavanagh:2016idg,Bini:2013rfa,Bini:2014nfa,Bini:2015bla, Bini:2015bfb,Bini:2015xua,Bini:2016dvs,Bini:2016qtx} or it can be used to resum PN dynamics around the test-body limit \cite{LeTiec:2011dp,Barausse:2011dq,Akcay:2012ea}. Here, we focus on the latter approach. 
 
Currently available EOB Hamiltonians informed with the Detweiler redshift cannot be reliably evolved near the Schwarzschild light-ring (LR) radius, i.e., $\R=3\text{G}M/c^2$. Such an issue, hereafter called the \textit{LR-divergence} problem,  appears as a coordinate singularity of the effective Hamiltonian at the Schwarzschild LR \cite{LeTiec:2011dp,Akcay:2012ea}. In this paper we address the problem and, adopting a different EOB gauge, we obtain a Hamiltonian with SMR information that exhibits no divergence at the LR radius. This result allows us to use the precious near-LR, strong-field information from SF calculations in the evolutions of EOB Hamiltonians.

The organization of the paper is as follows.
In Sec.~\ref{sec:prelim} we review the LR-divergence arising from informing the conservative sector of standard EOB Hamiltonians with the Detweiler redshift and we discuss how a different EOB gauge (introduced in Ref.~\cite{Damour:2017zjx} in the context of PM calculations) helps to solve the issue.
In Sec.~\ref{sec:EOBSMR}, we inform the conservative sector of EOB Hamiltonians in the alternative gauge with circular-orbit information from the Detweiler redshift, and with both non-circular-orbit and higher-order-in-mass-ratio information from the PN approximation.
In Sec.~\ref{sec:evolutions}, we evolve quasi-circular inspirals from this LR-divergence-free Hamiltonian and show that the evolution of the orbital separation crosses the LR radius without encountering singularities. Moreover, we perform systematic comparisons against NR predictions of phase and binding energy for non-spinning systems with mass ratios $1/10 \leq q \leq 1$. We conclude in Sec.~\ref{sec:conclusions}. In Appendix~\ref{sec:appred} we present high-precision fits to the Detweiler redshift with improved data in the strong field. 
We use geometric units G=$c$=1 throughout the paper.

\section{On gauges and the light-ring divergence}
\label{sec:prelim}

We begin by noting some conventions to be used in the following sections. In the present paper, we do not consider spinning systems; we denote the reduced mass by $\mu=(m_1 m_2)/M$ and the total mass by $M=m_1+m_2$.  We work with generalized (polar) coordinates ${q}_a\equiv (\R,\phi)$ in the orbital plane, with canonically conjugate momenta $p_a\equiv(p_r,p_\phi)$, and we often employ the mass-reduced inverse orbital separation $u \equiv M/\R$ and the mass-reduced momenta $\p_r \equiv \Pp_r/\mu$ and $\p_\phi\equiv \Pp_\phi/(M\mu)$.

\subsection{The light-ring divergence}

In the EOB approach, the real two-body motion is mapped to the
effective motion of a test body in an effective \textit{deformed}
Schwarzschild spacetime with coordinates $(t,r,\theta,\phi)$, with the deformation parameter being the symmetric mass ratio $\nu$. 
The mapping can be obtained via a dictionary between the action integrals $I_a=(2\pi)^{-1}\oint
{\Pp}_a\, d{q}_a$ of a two-body system in the
center-of-mass frame and those of a test-body moving in the effective
metric $g_{\mu\nu}^{\text{eff}}$. Considering orbits in the equatorial plane $\theta=\pi/2$, identifying the radial and angular
action integrals of real and effective systems, i.e., setting
$I_\text{\R}^{\text{real}}=I_\text{\R}^{\text{eff}}$ and
$I_\phi^{\text{real}}=I_\phi^{\text{eff}}$, the EOB approach allows a
simple relation between the real
$H\subreal(r,p_r,p_\phi,\nu)$ and effective
$H_{\text{eff}}(r,p_r,p_\phi,\nu)$ Hamiltonians
\cite{Buonanno:1998gg}:
\begin{equation}\label{enmap}
H\subreal \equiv M \hat H\subreal =M\sqrt{1+2\nu\bigg(\frac{H_{\text{eff}}}{\mu}-1\bigg)}\,.
\end{equation}
$H_{\text{eff}}$ describes the motion of a test body with mass $\mu$ and is determined by a mass-shell constraint of the form~\cite{Damour:2000we}
\begin{equation}\label{HamJac}
g^{\mu\nu}_\text{eff}\Pp_\mu\Pp_\nu+\mu^2+Q(r,p_r,p_\phi,\nu)
=0,
\end{equation}
where the effective metric is given by
\begin{align}\label{effmetric}
ds^{2}=-A(\R,\nu)dt^{2}+[A(\R,\nu)\bar D(\R,\nu)]^{-1}d\R^{2}+\R^{2} d\Omega^{2}\,,
\end{align}
with the potentials $A(\R,\nu)$ and $\bar D(\R,\nu)$ depending on the orbital separation $r$ and the symmetric mass ratio $\nu$. In terms of the inverse radius $u=M/r$, they reduce to $A_0(u)=1-2u$ and $\bar D_0=1$ in the test particle limit ($\nu\rightarrow0$). Inserting the inverse of the metric~\eqref{effmetric} into Eq.~\eqref{HamJac}, and using $\Pp_\mu=(-H_\text{eff},\Pp_r,\Pp_\theta=0,\Pp_\phi)$, the mass-reduced effective Hamiltonian $\Ham_{\text{eff}}\equiv H_{\text{eff}}/\mu$ is found to be \cite{Damour:2000we}
\begin{align}\label{HeffDJS}
\Ham_{\text{eff}}^2&=A(u,\nu)\big[1+\p_{\phi}^{2}u^{2}+A(u,\nu)\bar D(u,\nu)\p_{r}^{2}
\nonumber\\
&\quad+\hat{Q}(u,\p_r,\p_\phi,\nu)\big],
\end{align}
with $\hat Q\equiv Q/\mu^2$.  The non-geodesic function $ Q$ in
Eq.~\eqref{HamJac} has been introduced to extend the EOB Hamiltonian 
through 3PN order without changing the mapping (\ref{enmap}) (for a geodesic 
one-body motion at 3PN order with an energy map different from 
(\ref{enmap}) see Appendix A in Ref.~\cite{Damour:2000we}). Its
mass-reduced form $\hat{Q}(u,\p_r,\p_\phi,\nu)$ in Eq.~\eqref{HeffDJS} generically depends on both the mass-recuded
radial momentum $\p_{r}$ and the mass-recuded angular momentum $\p_{\phi}$. Reference \cite{Damour:2000we} 
  showed that at 3PN order $\hat{Q}$ must be
fourth order in the momenta, and that the non-geodesic term 
is not uniquely fixed. By setting some of the free parameters to zero, it is possible 
to make the function  $\hat{Q}(u,\p_r,\p_\phi,\nu)$ depend only on the radial momentum  
[i.e., $\hat Q(u,\p_r,\p_\phi,\nu)\rightarrow\hat
Q(u,\p_r,\nu)$]. Since 2000, this choice of $\hat{Q}$ 
has been adopted in several EOB papers [although see Refs.~\cite{Damour:2002vi,Buonanno:2007pf} for alternative choices 
of $\hat{Q}$]. Henceforth, we shall 
denote the $\hat{Q}$ function that only depends on the radial momentum 
as $\hat{Q}^{\rm DJS}({u},\p_r,\nu)$, after the 
initials of the three authors of
Ref.~\cite{Damour:2000we}. We refer to the DJS EOB Hamiltonian as 
the Hamiltonian that uses the $\hat{Q}^{\rm DJS}({u},\p_r,\nu)$ function. 
Note that in this gauge, the angular momentum $\p_{\phi}$ only appears in the second term in brackets in
Eq.~\eqref{HeffDJS}. Moreover, in the circular orbit limit ($\p_{r}=0$) the
conservative dynamics information is fully described by the $A(u,\nu)$
potential in this gauge, as found at 2PN order~\cite{Buonanno:1998gg}.  
The 4PN expressions for $A(u,\nu)$, $\bar D(u,\nu)$
and $\hat Q^{\rm DJS}({u},\p_{r},\nu)$ in the DJS gauge, for quasi-circular orbits,  
are obtained mapping Eq.~\eqref{enmap} to the 4PN-expanded Hamiltonian and can be
found in Ref.~\cite{Damour:2015isa}.

The first efforts to incorporate SMR quantities in EOB Hamiltonians sought to do so 
using the gauge of Eq.~\eqref{HeffDJS} with $\hat Q(u,\hat p_{r},\hat p_{\phi},\nu)\rightarrow \hat Q^{\rm DJS}(u,\hat p_{r},\nu)$
\cite{Barausse:2011dq,Akcay:2015pjz,Barack:2010ny,Akcay:2012ea}. In this gauge, the function $A(u, \nu)$, having the complete dynamical information for circular orbits, allows a linear-in-$\nu$ expansion about the Schwarzschild limit:
\begin{align}\label{ADnuexp}
&A(u, \nu)=1-2u+\nu \aaa (u)+\mathcal{O}(\nu^2)\,.
\end{align}
The $\aaa(u)$ function resums the complete circular-orbit PN dynamics in linear order in $\nu$.
References~\cite{LeTiec:2011dp,Barausse:2011dq} obtained an expression for $\aaa(u)$ employing the linear-in-$\nu$ correction to the Detweiler redshift. Notably, the Detweiler redshift is expanded around the Schwarzschild background, $z(x)=\sqrt{1-3x}+\nu \zzz(x)+\mathcal{O}(\nu^2)$ [where $x\equiv (M\Omega)^{2/3}$ is the gauge-independent inverse radius], and the $\zzz$ correction is linked to $\aaa(u)$ via the first law of binary black-hole mechanics \cite{LeTiec:2011ab}.
The resulting expression reads: 

\begin{equation}\label{Asf}
\aaa(u)= \zzz(u)\sqrt{1-3u}-u\bigg(1+\frac{1-4u}{\sqrt{1-3u}}\bigg)\,.
\end{equation}
In Eq.~\eqref{Asf}, $\zzz$ depends on the gauge-dependent inverse radius $u$, rather than its gauge-independent counterpart $x$. This is only correct if we restrict to first order in $\nu$, since $x=u+\mathcal{O}(\nu)$.
The quantity $\zzz(x)$, has been fitted with data extending to the LR \cite{Akcay:2012ea}, allowing precious strong-field information to enter the EOB dynamics. 

The form of $\aaa(u)$ is suggestive of trouble arising at the Schwarzschild light ring, i.e., at $u_{\text{LR}}=1/3$, where the second term in Eq.~\eqref{Asf} diverges. In principle, this divergence might be tamed by the behaviour of the redshift $\zzz(u)$ appearing in the first term in brackets, but data for the redshift up to the LR show that this is not the case and that $\aaa(u)$ indeed diverges there~\cite{Akcay:2012ea}. This is worrisome, as $\aaa(u)$ directly enters the effective Hamiltonian and, via the energy map, the EOB-resummed dynamics. The EOB dynamics thus contains a divergence \textit{for generic orbits} (e.g., for any value of $\p_{\phi}$ and $\p_{r}$).	
It was pointed out in Ref.~\cite{Akcay:2012ea} that the LR-divergence is a phase-space coordinate singularity that arises due to the use of the DJS gauge, and that can be solved adopting a different gauge in which the function $\hat Q$ grows as $\hat Q\propto \p_\phi^3$ when $\p_\phi\rightarrow\infty$ and $\p_r\rightarrow0$. 

It is worth mentioning that the argument in Ref.~\cite{Akcay:2012ea} stems from a similar LR divergence that has appeared when including tidal effects 
in the EOB approach \cite{Bini:2012gu}. Tidal effects enter the potential $A(u)$ via a correction in a tidal expansion akin to Eq.~\eqref{ADnuexp}: $A(u)=A_{\text{2pp}}+\mu_\text{T} a_\text{T}(u,\nu)+\mathcal{O}(\mu_\text{T}^2)$, where $A_{\text{2pp}}$ is the two point-particle (pp) EOB potential \cite{Bini:2012gu} and $\mu_\text{T}$ the small tidal parameter. It has been found in Ref.~\cite{Bini:2012gu} that, in the extreme-mass-ratio limit and for circular orbits, the first-order correction scales as $a_\text{T}(u,\nu)\propto (1-3u)^{-1}$ when $u\rightarrow u_\text{LR}$. An alternative EOB Hamiltonian that includes dynamical tides without introducing poles at the LR has been introduced in Ref.~\cite{Steinhoff:2016rfi}; this has been achieved by abandoning the DJS gauge (see, e.g., their Appendix D).

\subsection{The post-Schwarzschild effective-one-body gauge}

Reference~\cite{Damour:2017zjx} has shown that it is possible to obtain a different EOB gauge, hereafter the \textit{post-Schwarzschild} (PS) gauge, solving Eq.~\eqref{HamJac} with the Schwarzschild limit of the metric~\eqref{effmetric}.
The mass-reduced effective Hamiltonian thus obtained has the following form:
\begin{equation}\label{HeffEG}
\Ham_{\text{eff}}^{\text{PS}}=\sqrt{\Ham_{\text{S}}^{2}+(1-2u)\hat{Q}^{\text{PS}}(u,\nu,\Ham_{\text{S}})}\,,
\end{equation}
where $\Ham_{\text{S}}$ is the Schwarzschild Hamiltonian:
\begin{equation}\label{Hs}
\Ham_{\text{S}}(u,\p_{r},\p_{\phi})=\sqrt{(1-2u)\left [1+\p_{\phi}^{2}u^{2}+(1-2u)\p_{r}^{2}\right ]}\,.
\end{equation} 

In Ref.~\cite{Damour:2017zjx}, the PS function $\hat{Q}^{\text{PS}}$ has been derived to 2PM order via a scattering-angle calculation and to 3PN order via a canonical transformation from the DJS Hamiltonian at 3PN. In Ref.~\cite{Antonelli:2019ytb}, these calculations have been extended to 3PM and 4PN orders, respectively (the latter only in the near-circular orbit limit).

It is noticed that, in PS EOB Hamiltonians, all the information on the two-body problem with $\nu \neq 0$ is contained in $\hat{Q}^{\text{PS}}(u,\nu,\Ham_{\text{S}})$. This feature and the fact that circular-orbit dynamics is contained also in the $\hat{Q}$ function, significantly differentiate PS Hamiltonians from DJS ones. The PS gauge is uniquely fixed resumming the angular and radial momenta into the Schwarzschild Hamiltonian \eqref{Hs}. The powers of such momenta are furthermore not bound in any way, due to the generic functional dependence of $\hat{Q}^{\text{PS}}(u,\nu,\Ham_{\text{S}})$ on $\Ham_{\text{S}}$. In principle, then, arbitrary powers of $\p_{\phi}$ are contained in $\hat{Q}^{\text{PS}}(u,\nu,\Ham_{\text{S}})$ via $\Ham_{\text{S}}$. In particular, differently from $\hat{Q}^{\text{DJS}}(u,\nu,\p_r)$,  powers of momentum enter at second order 
in $\hat{Q}^{\text{PS}}(u,\nu,\Ham_{\text{S}})$ instead of fourth order.

The unconstrained dependence of $\hat{Q}^{\text{PS}}$ on $\Ham_{\text{S}}$ makes the use of PS Hamiltonians very appealing in the context of our work. It was shown in Ref.~\cite{Damour:2017zjx} that, in the high energy limit for which $\p_{\phi}\rightarrow\infty$, 
the LR-divergence can be captured by the coefficient of a term proportional to $\Ham_{\text{S}}^3$. This result is in agreement with a point made in the conclusions of Ref.~\cite{Akcay:2012ea}. As it approaches the LR radius, the effective mass moving in a deformed-Schwarzschild background described by Eqs.~\eqref{ADnuexp} and~\eqref{Asf} has a divergent-energy behaviour that must be removed with an appropriate energy-corrected mass-ratio parameter $\tilde\nu=\nu \Ham_{\text{S}}$.
In the next section, building from this knowledge and making use of a simple ansatz for $\hat{Q}^{\text{PS}}(u,\nu,\Ham_{\text{S}})$, we construct a Hamiltonian in the PS gauge that contains information from $\zzz$, while remaining analytic at the LR.

\section{Conservative dynamics of post-Schwarzschild Hamiltonians}
\label{sec:EOBSMR}

\subsection{Information from circular orbits}\label{sec:preamble}
In this section, we link the conservative sector of the PS EOB Hamiltonian to the SMR contribution to $\zzz$. 
Following Ref.~\cite{Barausse:2011dq}, we do so matching, at fixed frequency, the circular orbit binding energy at linear order in $\nu$ from the EOB Hamiltonian with the binding energy in the same limit from SF results. The latter is obtained in Ref.~\cite{LeTiec:2011dp} and is a consequence of the first law of binary-black-hole mechanics. As a function of 
$\zzz$ and the gauge-invariant inverse radius $x$, it reads \cite{LeTiec:2011dp}:

\begin{equation}\label{EsEsf}
\Ebindsmr=\frac{1-2x}{\sqrt{1-3x}}-1+\nu \EEE(x,\zzz,\zzz')+\mathcal{O}(\nu^2)\,,
\end{equation}
\begin{align}\label{Ezold}
\EEE(x,\zzz,\zzz')=&-1+\sqrt{1-3 x}-\frac{x}{3}
\zzz'(x)\nonumber\\
&+\frac{\zzz(x)}{2}+\frac{(7-24 x) x}{6
	(1-3 x)^{3/2}}\,.
\end{align}
The prime denotes differentiation with respect to $x$.
We find it useful to rewrite the redshift as:

\begin{equation}\label{zsmr}
\zzz(x)=\frac{\zzero(x)}{1-3x}+\frac{\zone(x)}{\sqrt{1-3x}}+\frac{
\ztwo(x)}{1-3x}\lnE(x)
\,,
\end{equation}
 In the above expression, we have defined $E_\text{S}(x)\equiv (1-2x)/\sqrt{1-3x}$.
In Appendix~\ref{sec:appred}, $\zzero(x)$, $\zone(x)$ and $\ztwo(x)$ are fitted to high-precision SF data and such to be analytic at the LR. 
Equation~\eqref{Ezold} then reads:
\begin{widetext}
\begin{align}\label{Esf}
\EEE=&\sqrt{1-3 x}-1+\frac{(7-24 x) x}{6 (1-3 x)^{3/2}}
+\frac{1}{2(1-3x)}
\bigg[\zzero(x)+\zone(x)\sqrt{1-3 x}
+\ztwo(x)\lnE(x)\bigg]
\nonumber\\
&
-\frac{x}{3(1-3x)}\bigg\{\frac{3\zzero(x)}{1-3 x}+\frac{3 \zone(x)}{2 \sqrt{1-3 x}}+
\bigg[\frac{1-6 x}{(1-2x)(1-3x)}+\frac{3\lnE(x)}{(1-3x)}\bigg]\ztwo(x)
\nonumber\\
&
+(\zzero)'(x)+ \sqrt{1-3 x}\,(\zone)'(x)+ (\ztwo)'(x)\lnE(x)\bigg\}\,.
\end{align}
\end{widetext}
We next consider the PS EOB Hamiltonian $H\subreal$, with an ansatz for $\hat Q^{\rm PS}$ reading:

\begin{equation}\label{HSMR}
\hat{Q}^{\text{PS}}_{\text{SMR}}(u,\nu,\Ham_{\text{S}})=\nu\big[\X_{0}(u)\Ham_{\text{S}}^{5}+\X_{1}(u)\Ham_{\text{S}}^{2}+\X_{2}(u)\Ham_{\text{S}}^{3}\ln\Ham^{-2}_{\text{S}}\big]\,.
\end{equation}
In the rest of this section, when matching to the SMR results, we limit to circular orbits; thus we use $\Ham_{\text{S}}(u,\p_r=0,\p_\phi)$ in Eq.~(\ref{HSMR}). The role of the $\Ham_{\text{S}}^5$ term is to capture the global divergence $(1-3x)^{-3/2}$ of Eq.~\eqref{Esf}\footnote{In principle, a $\Ham_{\text{S}}^3$ term will suffice to capture the divergence. However, we find that this minimal choice leads to evolutions that are not well behaved for systems with comparable masses.}, while the second term $\Ham_{\text{S}}^2$ is devised to incorporate the $\sqrt{1-3x}$ terms appearing in the numerator of the same equation, which would make the Hamiltonian imaginary after the light ring. The term proportional to $\ln\Ham_{\text{S}}^{-2}$ incorporates the logs in the fit that would make the Hamiltonian non-smooth at the light ring.
Setting $\Pp_r=0$ and using:
\begin{equation}
\dot{\Pp}_r=-\frac{\partial  H\subreal}{dr}(r,\Pp_r=0,\Pp_\phi^{\text{circ}},\nu)=0\,,
\end{equation}
the (mass-reduced) circular-orbit momentum $\p_\phi^{\text{circ}}$ as a function of the inverse radius $u$ is determined at linear order in $\nu$ (with $\X_i'(u)=d\X_i/du$):
\begin{align}\label{pu}
\p_{\phi}^{\text{circ}}(u,\nu)=&\frac{1}{\sqrt{u(1-3u)}}+\nu\, 
\frac{(1-2 u)^2 }{4 (1-3
	u)^3 \sqrt{u}}\times\nonumber\\
&\bigg[2(1-2u)^3\X_{0}(u)+2(1-3u)^{3/2}\X_1 (u)
\nonumber\\
&+2(1-2u)(1-3u)\X_2(u)\lnE (u)\nonumber\\
&-(1-2u)^4\X'_{0}(u)-(1-2u)(1-3u)^{3/2}\X'_{1}(u)\nonumber\\
&-(1-2u)^2(1-3u)\lnE(u)\X'_2(u)\bigg]+\mathcal{O}(\nu^2)\,.
\end{align}
We further use the relation:
\begin{equation}\label{omega}
\Omega=\frac{\partial H\subreal}{d\Pp_\phi}(r,\Pp_r=0,\Pp_{\phi}^{\text{circ}},\nu)\,,
\end{equation}
and exploit its link to the gauge-independent inverse radius $x$ given by $x=(M\Omega)^{2/3}$.
Inserting Eq.~\eqref{pu} in Eq.~\eqref{omega} and inverting the obtained expression at linear order in $\nu$,
we establish a link between the gauge-dependent $u$ and the gauge-independent $x$ inverse radii:
\begin{widetext}
\begin{align}\label{ux}
u^{\text{circ}}(x,\nu) = &x+
\frac{x\,\nu}{6 (1-3
	x)^{3/2}}
\bigg\{
4-20x+24 x^2-(4-12x) \sqrt{1-3 x}-10 (1-2 x)^4 \X_0(x)\nonumber\\
&-4 \sqrt{1-3 x} \left(1-5x+6 x^2\right) \X_1(x)+\big[4 - 28 x + 64 x^2 - 48 x^3 -(6-42x+96x^2-72x^3)\lnE(x)\big] \X_2(x) \nonumber\\
&+\left(1-10x+40x^2-80x^3+80x^4-32x^5\right) \X_0'(x)+\sqrt{1-3 x}\left(1-7x+16x^2-12x^3\right)
\X_1'(x)\nonumber\\
&+\left(1-9x+30x^2-44x^3+24x^4\right)\lnE(x) \X_2'(x)
\bigg\}+\mathcal{O}(\nu^2)\,.
\end{align}
\end{widetext}
To calculate the (mass-reduced) gauge-invariant, circular-orbit binding energy at linear order in $\nu$ from $H\subreal$, we employ the definition :
\begin{equation}\label{Ebind}
\Ebindeob \equiv  (H\subreal-M)/\mu\,,
\end{equation}
Inserting Eqs.~\eqref{pu} and \eqref{ux} in $H\subreal$ and retaining only terms up to first order in the mass ratio, we get:
\begin{widetext}
	\begin{align}\label{Eeob}
\Ebindeob(x,\nu)=&\frac{1-2x}{\sqrt{1-3 x}}-1-\frac{\nu}{6(1-3x)^3}
\bigg\{
6-55x+170 x^2-189x^3+36 x^4-\sqrt{1-3 x} \left(6-46 x+108 x^2-72 x^3\right)	\nonumber\\
&-\left(3-7x-18 x^2\right) (1-2 x)^4 \X_0(x)-(1-3x)^{3/2} \left(3-16 x+20 x^2\right)
\X_1(x)\nonumber\\
&+(1-3x)(1-2 x)^2  \big[2x (1-6 x)-\left(3-9x-6 x^2\right) \lnE(x)\big]\X_2(x)+2 x (1-2 x)^5 (1-3 x) \X_0'(x) 	\nonumber\\
&+2x (1-3 x)^{5/2} (1-2 x)^2 \X_1'(x)+2x (1-3 x)^2 (1-2 x)^3 \lnE(x) \X_2'(x)
\bigg\}+\mathcal{O}(\nu^2)\,.
	\end{align}
\end{widetext}
Matching Eq.~\eqref{EsEsf} [with correction given by Eq.~\eqref{Esf}] and Eq.~\eqref{Eeob}, we obtain differential equations to be solved for $\X_0(x)$, $\X_1(x)$ and $\X_2(x)$.
Further splitting the $\X_i$ coefficients as follows:
\begin{align}
&\X_0(x)=\tilde{\X}_0(x)+\sum_{i=0}^{i=2}\X_0^{(i)}(x)\Delta z^{(i)}(x)\label{X0ans1}
\\
&\X_1(x)=\tilde{\X}_1(x)+\sum_{i=0}^{i=2}\X_1^{(i)}(x)\Delta z^{(i)}(x)\label{X1ans1}\,,\\
&\X_2(x)=\tilde{\X}_2(x)+\sum_{i=0}^{i=2}\X_2^{(i)}(x)\Delta z^{(i)}(x)\label{X2ans1}\,,
\end{align}
and imposing that the Hamiltonian coefficients be analytic at the LR radius (i.e., that they do not contain  $\sqrt{1-3x}$ or $\lnE(x)$ terms),
we obtain the following non-zero solutions\footnote{Similarly to what is done in Eq.~\cite{Barausse:2011dq}, we impose that the PN expansion cannot admit half-integer powers of $x$. This allows us to set all constants of integration to zero.}:
\begin{subequations}
\begin{align}
&\tilde{\X}_0(x)=-\frac{x (1-3x)\left(1-4x
\right)}{(1-2x)^5}\,,\\
&\tilde{\X}_1(x)=-\frac{x 
}{(1-2 x)^2}\,,\\
&\X_0^{(0)}(x)=\frac{1-3x}{(1-2x)^5}\,,\\
&\X_1^{(1)}(x)=\frac{1}{(1-2x)^2}\,,\\
&\X_2^{(2)}(x)=\frac{1}{(1-2x)^3}\,.
\end{align}
\end{subequations}
The $\X_{i}(x)$ coefficients are readily found via Eqs.~\eqref{X0ans1}, ~\eqref{X1ans1} and ~\eqref{X2ans1} and then inserted in the non-geodesic term in the effective Hamiltonian \eqref{HSMR} to obtain:
\begin{align}\label{QSMR}
\frac{\hat{Q}^{\text{PS}}_{\text{SMR}}}{\nu}(u,\nu,\Ham_{\text{S}})=&(1-3u)\bigg[\frac{\zzero(u)}{(1-2u)^5}-\frac{(1-4 u)\text{ }u}{(1-2 u)^5}\bigg]\Ham_{\text{S}}^5\nonumber\\
&+\bigg[\frac{\zone(u)}{(1-2 u)^2}-\frac{u}{(1-2
	u)^2}\bigg]\Ham_{\text{S}}^2 \nonumber\\
&+\frac{\ztwo(u)}{(1-2u)^3}\Ham_{\text{S}}^{3}\ln\Ham_{\text{S}}^{-2}\,.
\end{align}
We see that the resulting Hamiltonian concisely resums the complete circular-orbit PN dynamics at linear order in $\nu$. The non-geodesic function $\hat{Q}^{\text{PS}}_{\text{SMR}}$ does not contain any term divergent at the LR, as $\zzero(u)$, $\zone(u)$ and $\ztwo(u)$ are constructed to be analytic there.

\subsection{Information from non-circular orbits and from higher orders in the mass ratio}\label{sec:noncirc}

The calculation in Sec.~\ref{sec:preamble} is carried out in the circular-orbit limit at linear order in the mass ratio. However, it is  possible to include more physical information to the Hamiltonian, coming both from non-circular-orbit terms and from terms at higher orders in the mass ratio. For instance, self-force information for mildly eccentric orbits can be obtained via the SMR correction to the periastron advance $\rho_\text{SF}$ \cite{Barack:2010ny}, which can then be linked to the EOB potentials. This was the strategy used in Refs.~\cite{Damour:2009sf,Barausse:2011dq} to obtain an expression for the potential $\bar D(r)$ in terms of $\zzz(u)$ and $\rho_\text{SF}(u)$ and introduce non-circular SF data into the EOB Hamilonian up to the Schwarzschild ISCO (i.e., $u_\text{ISCO}=1/6$). Alternatively, one can exploit the generalized redshift~\cite{Barack:2011ed} and link it to $\bar D(r)$, as done in Refs.~\cite{Tiec:2015cxa,Akcay:2015pjz}. 
Here, we insert generic-orbit PN information in our Hamiltonian and leave the inclusion of non-circular SMR information in $\hat Q^{\rm PS}$  to future work.

Post-Schwarzschild EOB Hamiltonians with PN information from generic-orbits have been already considered in the literature. For example, the PS Hamiltonians at 3PN order has been investigated in Ref.~\cite{Damour:2017zjx}.
Using the PN parameters $Y \equiv (\hat H^{2}_{\text{S}}-1)\sim \mathcal{O}(1/c^2)$ and $u$, its expression is given by:
\begin{align}
\label{QPScoeff}
\hat Q^{\text{PS}}_{\text{3PN}}=&3\nu u^{2}Y+5\nu u^{3}
\nonumber\\
&+\bigg(3\nu-\frac{9}{4}\nu^{2}\bigg)u^{2}Y^{2}+\bigg(27\nu-\frac{23}{4}\nu^{2}\bigg)u^{3}Y\nonumber\\
&+\bigg(\frac{175}{3}\nu-\frac{41\pi^{2}}{32}\nu-\frac{7}{2}\nu^{2}\bigg)u^{4}\,.
\end{align}
As discussed, the above Hamiltonian contains two-body information that is not captured by the calculation leading to $\hat{Q}^{\text{PS}}_{\text{SMR}}$ and that we wish to add to it.

To this end, we consider a mixed SMR-3PN non-geodesic function of the following form:

\begin{equation}\label{Qtot}
\hat Q^{\text{PS}}_{\text{SMR-3PN}}=\hat Q^{\text{PS}}_{\text{SMR}}+\Delta\hat Q^{\text{PS}}\,,
\end{equation}
where $\hat Q^{\text{PS}}_{\text{SMR}}$ is given by Eq.~\eqref{QSMR} and contains all the circular-orbit terms at linear order in $\nu$, while $\Delta\hat Q^{\text{PS}}$ is fixed demanding that it contains all the additional PN information from Eq.~\eqref{QPScoeff}, in such a way not to contribute to the linear-in-$\nu$ binding energy in the circular-orbit limit. 

We opt to further split $\Delta\hat Q^{\text{PS}}$ into two contributions: $\Delta\hat Q^{\text{PS}}_{\text{extra}}$ collects the extra terms up to 3PN order (including both non-circular 3PN terms at linear order in $\nu$ and $\nu^2$ terms), while $\Delta\hat Q^{\text{PS}}_{\text{count.}}$ is a counterterm whose functionality is explained below. We then have:
\begin{equation}\label{Q4pnsplit}
\Delta\hat Q^{\text{PS}} \equiv \Delta\hat Q^{\text{PS}}_{\text{extra}}-\Delta\hat Q^{\text{PS}}_{\text{count.}}\,.
\end{equation}
The former contribution is readily obtained calculating the difference between Eq.~\eqref{QPScoeff} and the 3PN expansion of Eq.~\eqref{QSMR}\footnote{That is, Eq.~\eqref{QSMR} is expanded in the PN parameters $u$ and $Y=\hat H_{\text{S}}^2-1$. The redshift functions $\zzero(u)$, $\zone(u)$ and $\ztwo(u)$ also need to be PN expanded: their expressions are obtained matching the 3PN expansion of the redshift from Ref.~\cite{Kavanagh:2016idg} and Eq.~\eqref{zsmr}.}. The result reads:
\begin{align}\label{Qpn1}
\Delta\hat Q^{\text{PS}}_{\text{extra}}=& 3\nu u^2 Y+\bigg(3\nu-\frac{9}{4} \nu^2 \bigg) u^2Y^2+3 \nu u^3 \nonumber\\
&+\bigg(22\nu-\frac{23}{4} \nu^2 \bigg) u^3Y+\bigg(16 \nu-\frac{7}{2} \nu^2 \bigg)  u^4\,.
\end{align}
In the PS gauge $\hat Q^{\text{PS}}$ depends on momenta via $\hat H_\text{S}(u,\p_{\R},\p_{\phi})$, which cannot be separated into circular and non-circular orbit contributions.
Because of that, the linear-in-$\nu$ portion of Eq.~\eqref{Qpn1} contributes to the linear-in-$\nu$ binding energy for circular orbits.
Therefore, the addition of $\Delta\hat Q^{\text{PS}}_{\text{extra}}$ to $\Delta\hat Q^{\text{PS}}_{\text{SMR}}$ spoils the matching between EOB and SF binding energies for circular orbits at linear order in the mass ratio guaranteed by the sole presence of $\Delta\hat Q^{\text{PS}}_{\text{SMR}}$.

The matching between the two binding energies can be maintained with a particular choice of the second contribution to Eq.~\eqref{Q4pnsplit}, i.e., $\Delta\hat Q^{\text{PS}}_{\text{count.}}$. We choose a counterterm that starts at 4PN, in order not to spoil the agreement at 3PN for generic orbits guaranteed by Eq.~\eqref{Qpn1}:
\begin{equation}\label{Qpn2}
	\Delta\hat Q^{\text{PS}}_{\text{count.}}=\nu \big[q_{(3,2)}u^3Y^2+q_{(4,1)}u^4Y+q_{(5,0)}u^5\big]\,.
\end{equation}
We impose that the linear-in-$\nu$ binding energy from $\Delta\hat Q^{\text{PS}}$ from Eq.~\eqref{Q4pnsplit} [calculated as done for Eq.\eqref{Eeob} in Sec.~\ref{sec:preamble}] vanishes and we obtain: 
\begin{equation}\label{qcoeff}
q_{(3,2)}=9\,;\, q_{(4,1)}=96\,;\, q_{(5,0)}=112\,.
\end{equation}

The final PN correction $\Delta\hat Q^{\text{PS}}$ thus contains all the extra information from generic orbits at 3PN that is not captured by $\hat Q^{\text{PS}}_{\text{SMR}}$, without contributing to the linear in mass ratio binding energy for circular orbits.
The exercise above can be repeated at one PN order higher to obtain $\Delta\hat Q^{\text{PS}}$ at 4PN starting from the 4PN EOB Hamiltonian in the PS gauge.~\cite{Antonelli:2019ytb}. Such a computation does not present major differences from the calculation above: the only feature changing is the counterterm, which needs to start at 5PN and include logarithmic terms. We have decided not to include $\Delta\hat Q^{\text{PS}}$ at 4PN in this paper, as the 4PN Hamiltonian from which it is constructed is only valid for near-circular orbits. The $\Delta\hat Q^{\text{PS}}$ at 3PN that we obtain here is instead valid for generic orbits.

\section{Inspirals in effective-one-body theory}
\label{sec:evolutions}

\subsection{Plunging through the light ring with small mass-ratio Hamiltonians}

\begin{figure}[t]
	\includegraphics[width=\columnwidth]{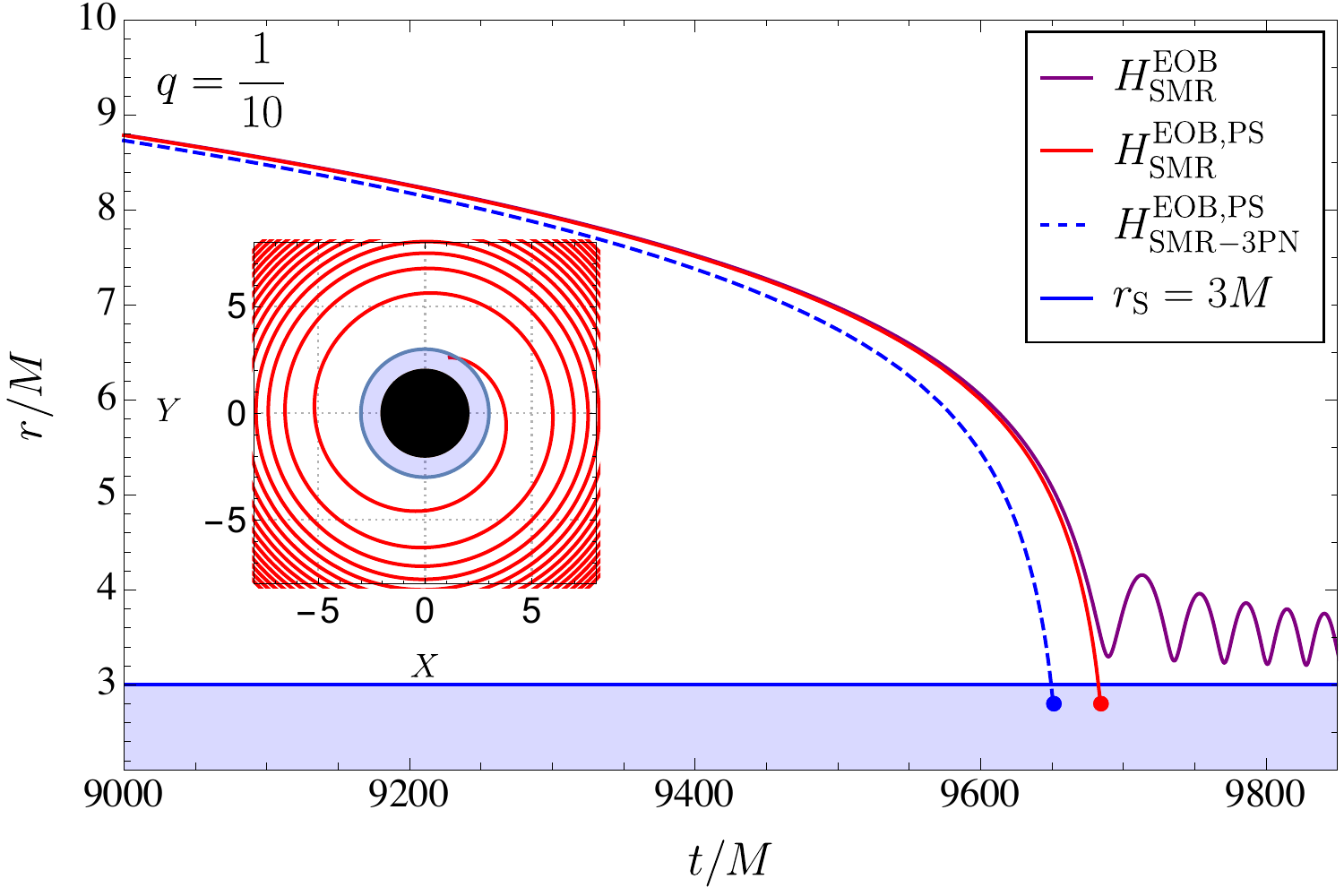}
	\caption{\textbf{Plunges through the light-ring radius:} the evolved orbital separation for the SMR Hamiltonians is presented. The effective masses of models $H_{\text{SMR}}^{\rm EOB,PS}$ and $H_{\text{SMR-3PN}}^{\rm EOB,PS}$
	plunge through the LR radius $r_S=3M$. Conversely, the plunge of the effective mass of $H_{\text{SMR}}^{\rm EOB}$ presents unphysical features associated to the LR-divergence.
	}
	\label{fig:orbitsep}
\end{figure}

In this section, we evolve the EOB Hamiltonians constructed in Secs.~\ref{sec:preamble} and ~\ref{sec:noncirc} [i.e., Eq.~\eqref{HeffEG} with non-geodesic functions~\eqref{QSMR} and~\eqref{Qtot}], and the EOB Hamiltonian with SMR information in the DJS gauge. We refer to them as $H_{\text{SMR}}^{\rm EOB,PS}$, $H_{\text{SMR-3PN}}^{\rm EOB,PS}$ and $H_{\text{SMR}}^{\rm EOB}$, see Table~\ref{table:models}. 

The EOB approach comprises of a conservative sector, discussed in detail in Sec.~\ref{sec:prelim}, and a dissipative sector, responsible for the slow GW-driven inspiral of the compact bodies towards merger. The basic set of equations for inspiraling orbits in the EOB framework are the Hamilton equations augmented with a radiation-reaction force $\mathcal{F}_{\text{RR}}$.
In terms of a generic mass-reduced EOB Hamiltonian $\Ham^{\rm EOB}(\hat r,\p_{r_{*}},\p_\phi)$, the equations read \cite{Buonanno:2000ef,Buonanno:2007pf,Damour:2007xr,Pan:2011gk}: 
\begin{subequations}
\begin{align}
& \frac{d\hat r}{d\hat t}=\frac{A(\hat r)}{\sqrt{D(\hat r)}}\frac{\partial \Ham^{\rm EOB} }{\partial \p_{r_{*}}}\,, \label{HH1}\\
&\frac{d\phi}{d\hat t}=\frac{\partial \Ham^{\rm EOB}}{\partial \p_{\phi}}\,, \label{HH2}\\
&\frac{d\p_{ r_{*}}}{d\hat t}=-\frac{A(\hat r)}{\sqrt{D(\hat r)}}\frac{\partial \Ham^{\rm EOB}}{\partial \hat r}+\mathcal{F}_{\text{RR}}\frac{\p_{r_{*}}}{\p_{\phi}}\,, \label{HH3}\\
&\frac{d\p_{\phi}}{d\hat t}=\mathcal{F}_{\text{RR}}\,, \label{HH4}
\end{align}
\end{subequations}
where we have introduced the mass-reduced radius $\hat r \equiv r/M$ and coordinate time $\hat t \equiv t/M$ and used the mass-reduced radial momentum $\p_{r_{*}}$ conjugate to the radius $r_{*}$ in tortoise coordinates,
defined for generic potentials $A(\hat r)$ and $D(\hat r)$\footnote{Here $D(\hat r)$ is the inverse of $\bar D(\hat r)$ mentioned in Sec.~\ref{sec:prelim}.} by:
\begin{equation}\label{prprstar}
\frac{d\hat r_{*}}{d\hat r} \equiv \frac{\sqrt{D(\hat r)}}{A(\hat r)}
=\frac{\p_{r}}{\p_{r_{*}}}
\,.
\end{equation}
In the evolution of the EOB Hamiltonian in the DJS gauge we use the PN-expanded expressions for $A(\hat r)$, $D(\hat r)$ and $\hat Q^{\text{DJS}}$ at the required PN order~\cite{Buonanno:1998gg,Damour:2000we,Damour:2015isa}, whereas we use their test-body limits in the evolutions of Hamiltonians in the PS gauge\footnote{The effective Hamiltonian in the PS gauge \eqref{HeffEG} is obtained solving the Hamilton-Jacobi equations with the Schwarzschild metric. The $A(\hat r)$ and $D(\hat r)$ are therefore fixed by their Schwarzschild limits.}. The Hamiltonians in both gauges depend on $\p_{ r_{*}}$, rather than $\p_{ r}$. 

The radiation reaction force $\mathcal{F}_{\text{RR}}$ drives the inspiral of the system and it contains semi-analytical two-body information \cite{Damour:2007xr,Damour:2008gu,Pan:2010hz}. In this paper, we employ its non-Keplerian form (with $\hat \Omega \equiv d\phi/d\hat t=M\Omega$):
\begin{equation}
^{\text{nK}}\mathcal{F}_{\text{RR}}=-\frac{1}{\nu\hat \Omega}\frac{dE}{dt}\,,
\end{equation}
where $dE/dt$ is the GW flux for quasi-circular orbits~\cite{Damour:2008gu}:
\begin{equation}
\frac{dE}{dt}=\frac{\hat\Omega^2}{8\pi}\sum_{l=2}^{l_\text{max}=8}\sum_{m=l-2}^{l} m^2 \big|\hat rh_{lm}\big|^2\,.
\end{equation}
The modes $h_{lm}$ are built from PN theory, but resummed multiplicatively (see e.g., Ref.~\cite{Damour:2008gu}). Here, we use the resummation of the (non-spinning) modes and flux presented in Ref.~\cite{Pan:2011gk} (which coincides with the state-of-the-art modes and flux used in the EOB waveform model for 
LIGO/Virgo data-analsyis~\cite{Bohe:2016gbl}, when spins are set to zero). We do not include 
the ``next-to-quasi-circular'' (NQC) coefficients~\cite{Bohe:2016gbl}, or any calibration parameter obtained 
imposing better agreement with numerical-relativity waveforms. Our main motivation here is to compare how well the conservative 
EOB-dynamics of SMR models compare to PN ones and with NR.

The result of the evolved orbital separations $\hat r$ of both DJS and PS  Hamiltonians  for $q=1/10$ are reported in Fig.~\ref{fig:orbitsep}.
Focusing on the evolution in the DJS case, it is seen that the pole in the conservative part of the DJS Hamiltonian affects the motion of the effective body close to the LR radius. That is, $H_{\text{SMR}}^{\rm EOB}$ diverges at $\hat r_{\text{S}}^{\text{LR}}=3$, at which point it acts as an infinite potential barrier that the effective mass cannot cross. Conversely, the effective mass plunges through the Schwarzschild LR radius in the cases of $H_{\text{SMR}}^{\rm EOB,PS}$ and $H_{\text{SMR-3PN}}^{\rm EOB,PS}$. This finding confirms that there is no unphysical behaviour at the LR radius for SMR Hamiltonians in the PS gauge. 
To conclude, we also notice that the horizons of the $H_{\text{SMR}}^{\rm EOB,PS}$ and $H_{\text{SMR-3PN}}^{\rm EOB,PS}$ models (red and blue dots) are quite close to the LR. Such a large deformation from the Schwarzschild background, while not presenting an issue by itself, could pose problems in the evolution of the EOB dynamics after the LR radius and, thus, in the modelling of EOB waveforms and frequencies during the transition between plunge and merger-ringdown phases.

\subsection{Comparisons against numerical relativity}
\label{sec:energetics}

\begin{table}
	\caption{{\bf Two-body EOB Hamiltonians evolved.} We summarize the EOB Hamiltonians evolved. 
		\label{table:models}}
	\begin{ruledtabular}
		\begin{tabular}{lp{5.2cm}p{2.5cm}}
			\addlinespace[.1cm]
			$H_{\text{SMR}}^{\rm EOB,PS}$
			& SMR Hamiltonian in PS gauge	& This paper \\
			\midrule
			\addlinespace[.1cm]
			$H_{\text{SMR-3PN}}^{\rm EOB,PS}$					
			& SMR-3PN Hamiltonian in PS gauge & This paper \\
			\midrule
			\addlinespace[.1cm]
			$H_{\text{SMR}}^{\rm EOB}$
			& SMR Hamiltonian in the DJS gauge (with LR divergence)	& \cite{Barausse:2011dq} \\
			\midrule
			\addlinespace[.1cm]
			$H_{n{\rm PN}}^{\rm EOB,PS}$
			&   $n$PN  Hamiltonian in PS gauge & \cite{Damour:2017zjx}	\\
			\midrule
			\addlinespace[.1cm]
			$H_{n\rm PN}^{\rm EOB}$	
			&	$n$PN  Hamiltonian in DJS gauge& \cite{Buonanno:1998gg,Damour:2000we} 	\\
			\addlinespace[.1cm]
		\end{tabular}
	\end{ruledtabular}
\end{table}

\begin{table}
	\caption{{\bf Set of non-spinning NR simulations and alignment time-windows.}  We list the SXS IDs, the mass ratios $q$ and the number of orbital cycles $N_{\rm orb}^{\rm merg}$ from the beginning of the simulation up to the binary black-hole merger (peak of $h^{\rm NR}_{22}$), as reported in the SXS catalog. We further include the time $t^\text{alig}_\text{in}$ at which the alignment procedure starts, the time $t^\text{alig}_\text{fin}$ at which it ends (in units of $M$) and the estimated NR error at merger $\Delta \phi^\text{merg}_\text{NR}$ (in radians).
		\label{table:setofNR}}
	\begin{ruledtabular}
		\begin{tabular}{ccc  ccc}
			\toprule
			SXS ID:						&	q$^{-1}$	& $N_{\rm orb}^{\rm merg}$	&    $t^\text{alig}_\text{in}$					&	$t^\text{alig}_\text{fin}$		&$\Delta \phi^\text{merg}_\text{NR}$ 	\\
			\midrule
			\addlinespace[.02cm]
			\hline
			\addlinespace[.1cm]
			0180	&	1	& 28.18	& 820	&	2250   & $\pm$0.25 \\
			1222	&	2	& 28.76	& 1000	&	2555  & $\pm$1.26 \\
			1221	&	3	& 27.18	& 1800	&	3000  & $\pm$0.21 \\
			1220	&	4	& 26.26	& 1800	&	3000   & $\pm$1.82 \\
			0056	&	5	& 28.81	& 1500	&	3000   & $\pm$0.39 \\
			0181	&	6	& 26.47	& 1000	&	2500   & $\pm$0.01 \\
			0298	&	7	& 19.68	& 780	&	2180   & $\pm$0.10 \\
			0063	&	8	& 25.83	& 1140	&	2540   & $\pm$0.85 \\
			0301	&	9	& 18.93	& 780	&	2180   & $\pm$0.13 \\
			0303	&	10	& 19.27	& 700	&	1900   & $\pm$0.49 \\
			
		\end{tabular}
	\end{ruledtabular}
\end{table}

\begin{table*}
	\caption{{\bf Details of the dephasing comparison.} We report the dephasing (in radians) of the SMR and 3PN models in both gauges at 8 and 4 GW cycles before NR merger, as found using the time-windows of Table~\ref{table:setofNR}. We also report the corresponding estimated NR error, which we denote by $\Delta \phi_\text{NR}$. The error for each NR simulation is estimated taking the phase differences between the highest two resolutions of the NR simulation (at fixed extrapolation order) and between two successive extrapolation orders (at fixed resolution), and adding them in quadrature.
		\label{table:simulations0}}
	\begin{ruledtabular}
		\begin{tabular}{cc | cccc c | cccc c  cccc c| ccc ccc }
			\toprule
			&\multirow{2}{*}{$q^{-1}$}
			&\multicolumn{4}{c}{8 GW cycles before merger}
			& \multirow{3}{*}{$\Delta\phi_{\text{NR}}$ }
			&\multicolumn{4}{c}{4 GW cycles before merger }  
			& \multirow{3}{*}{$\Delta\phi_{\text{NR}}$ }
		    \\
		     
		    & 
		     
		    & $\Delta \phi_{\text{SMR}}^{\rm EOB,PS}$
		    & $\Delta \phi_{\text{SMR-3PN}}^{\rm EOB,PS}$
		    & $\Delta \phi_{\text{3PN}}^{\rm EOB,PS}$
		    & $\Delta \phi_{\text{3PN}}^{\rm EOB}$
		    &
		    	
		    & $\Delta \phi_{\text{SMR}}^{\rm EOB,PS}$
		    & $\Delta \phi_{\text{SMR-3PN}}^{\rm EOB,PS}$
		    & $\Delta \phi_{\text{3PN}}^{\rm EOB,PS}$
		    & $\Delta \phi_{\text{3PN}}^{\rm EOB}$
		    &
		    
			\\
			\midrule
			\addlinespace[.02cm]
			\hline
			\addlinespace[.1cm]
			&	1	& ~0.111	& -0.033	&	-0.971   & ~0.032 & $\pm$0.032 & ~0.352 & -0.012 & -2.630 & ~0.084 & $\pm$0.056\\
			&	2	& ~0.112	& -0.061	&	-1.342   & -0.023 & $\pm$0.105 & ~0.512 & -0.021 & -5.586 & -0.043 & $\pm$0.224\\
		    & 	3	& ~0.050	& -0.021	&	-0.617   & -0.023 & $\pm$0.093 & ~0.111 & -0.026 & -1.209 & -0.048 & $\pm$0.144\\
			&	4	& ~0.046	& -0.038	&	-0.859   & -0.078 & $\pm$0.203 & ~0.187 & -0.041 & -2.540 & -0.212 & $\pm$0.372\\
			&	5	& ~0.037	& -0.034	&	-0.846   & -0.086 & $\pm$0.023 & ~0.125 & -0.044 & -2.077 & -0.211 & $\pm$0.064\\
			&	6	& -0.035	& -0.064	&	-0.433   & -0.093 & $\pm$0.006 & -0.041 & -0.082 & -0.599 & -0.126 & $\pm$0.007\\
			&	7	& ~0.024	& -0.009	&	-0.462   & -0.070 & $\pm$0.001 & ~0.092 & -0.003 & -1.403 & -0.211 & $\pm$0.009\\
			&	8	& ~0.021	& -0.021	&	-0.676   & -0.107 & $\pm$0.057 & ~0.076 & -0.025 & -1.660 & -0.260 & $\pm$0.155\\
			&	9	& ~0.017	& -0.005	&	-0.368   & -0.068 & $\pm$0.002 & ~0.063 & -0.005 & -1.185 & -0.220 & $\pm$0.012\\
			&	10	& ~0.022	& -0.001	&	-0.413   & -0.076 & $\pm$0.033 & ~0.070 & -0.004 & -1.245 & -0.233 & $\pm$0.083\\
			
		\end{tabular}
	\end{ruledtabular}
\end{table*}

\begin{table}
	\caption{{\bf Alternative alignment time-windows.}Time-windows (in units of $M$) employed for Fig.~\ref{fig:dphivsq}: here, $t^\text{alig}_\text{in}$ is the time corresponding to 34 GW cycles before merger for each NR simulation, whereas $t^\text{alig}_\text{fin}$ is chosen to encompass 10 GW cycles. The time at merger is given by $t_\text{merg}$.
	\label{table:simulations}}
	\begin{ruledtabular}
		\begin{tabular}{cccc | cccc }
			\toprule
		    q$^{-1}$			&   $t^\text{alig}_\text{in}$					&	$t^\text{alig}_\text{fin}$ &	$t_\text{merg}$			& q$^{-1}$			&   $t^\text{alig}_\text{in}$					&	$t^\text{alig}_\text{fin}$ &	$t_\text{merg}$	\\
			\midrule
			\addlinespace[.02cm]
			\hline
			\addlinespace[.1cm]
				1	& 5107	& 6911 & 9517	&	6   & 2971 & 4254 & 6000 \\
				2	& 5406	& 7078 & 9384   &   7   & 776 & 2083 & 4142\\
				3   & 3940	& 5532 & 7858   & 8 & 2652 & 3918& 5956\\
				4   & 3479	& 4975 & 7200	& 9 & 513 & 1732 & 3692\\
				5   & 4206	& 5641 & 7864   & 10  &  587 & 1771 & 3691	\\

		\end{tabular}
	\end{ruledtabular}
\end{table}

\begin{figure}
	\includegraphics[width=\columnwidth]{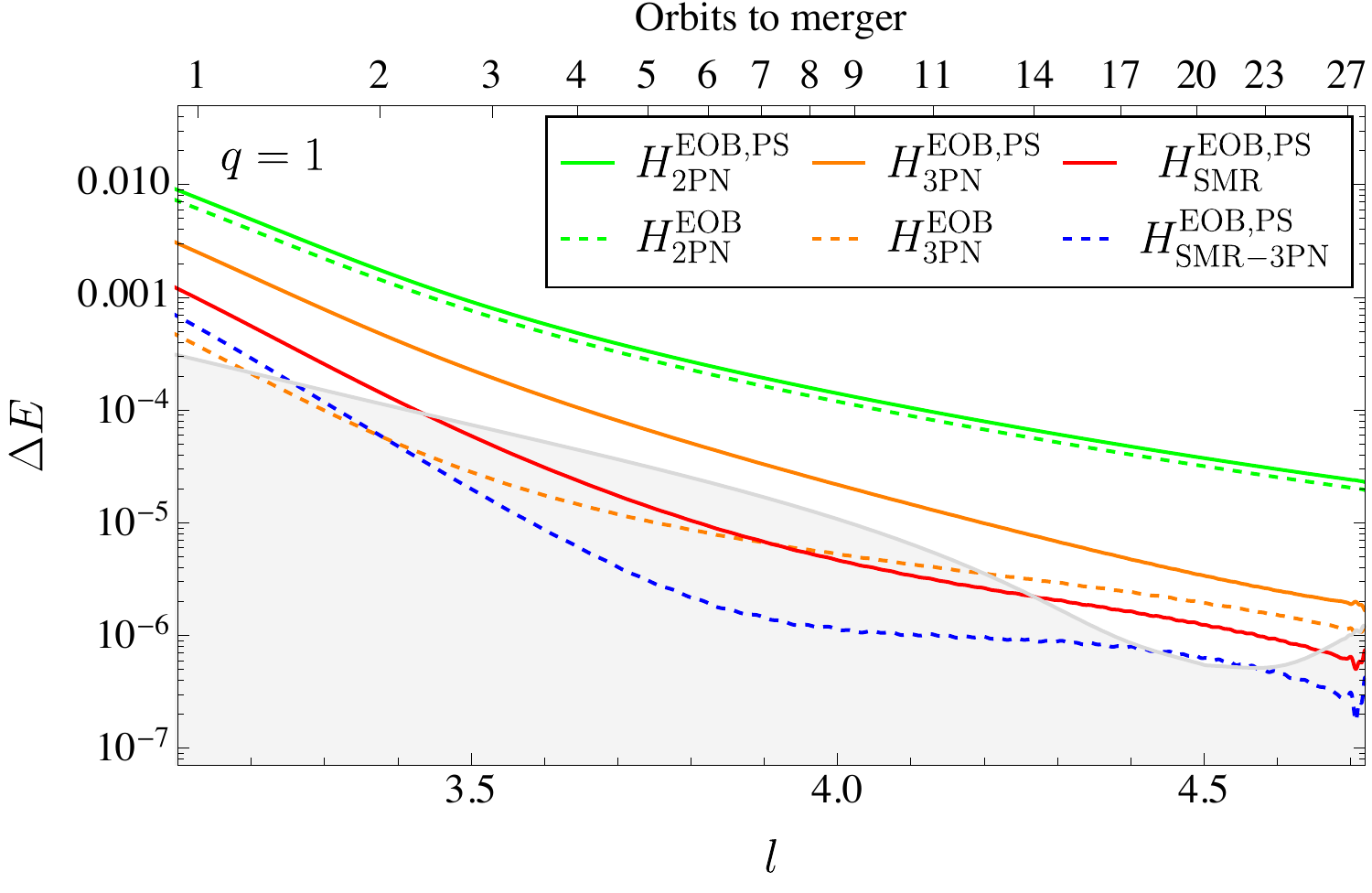}
	\includegraphics[width=\columnwidth]{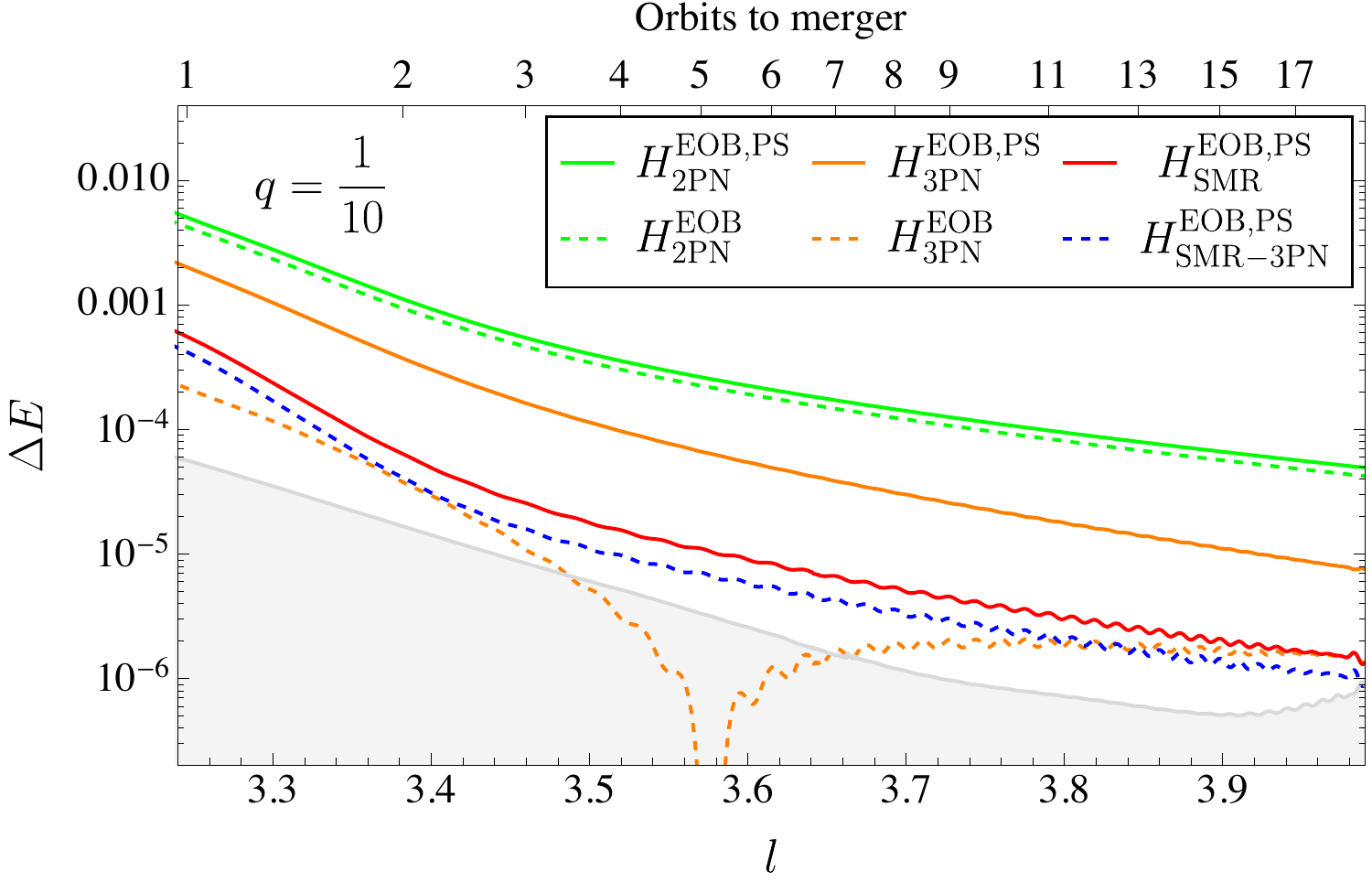}
	\caption{\textbf{SMR vs PN binding energies:} we compare the difference $\Delta E$ in binding energy from NR for our SMR Hamiltonians versus angular momentum $l$. We compare it to similar results for PN models up to third order, in both PS and DJS gauges. The \textit{estimated} NR error is shown in grey.}
	\label{fig:compEJ}
\end{figure}

\begin{figure}
	\includegraphics[width=\columnwidth]{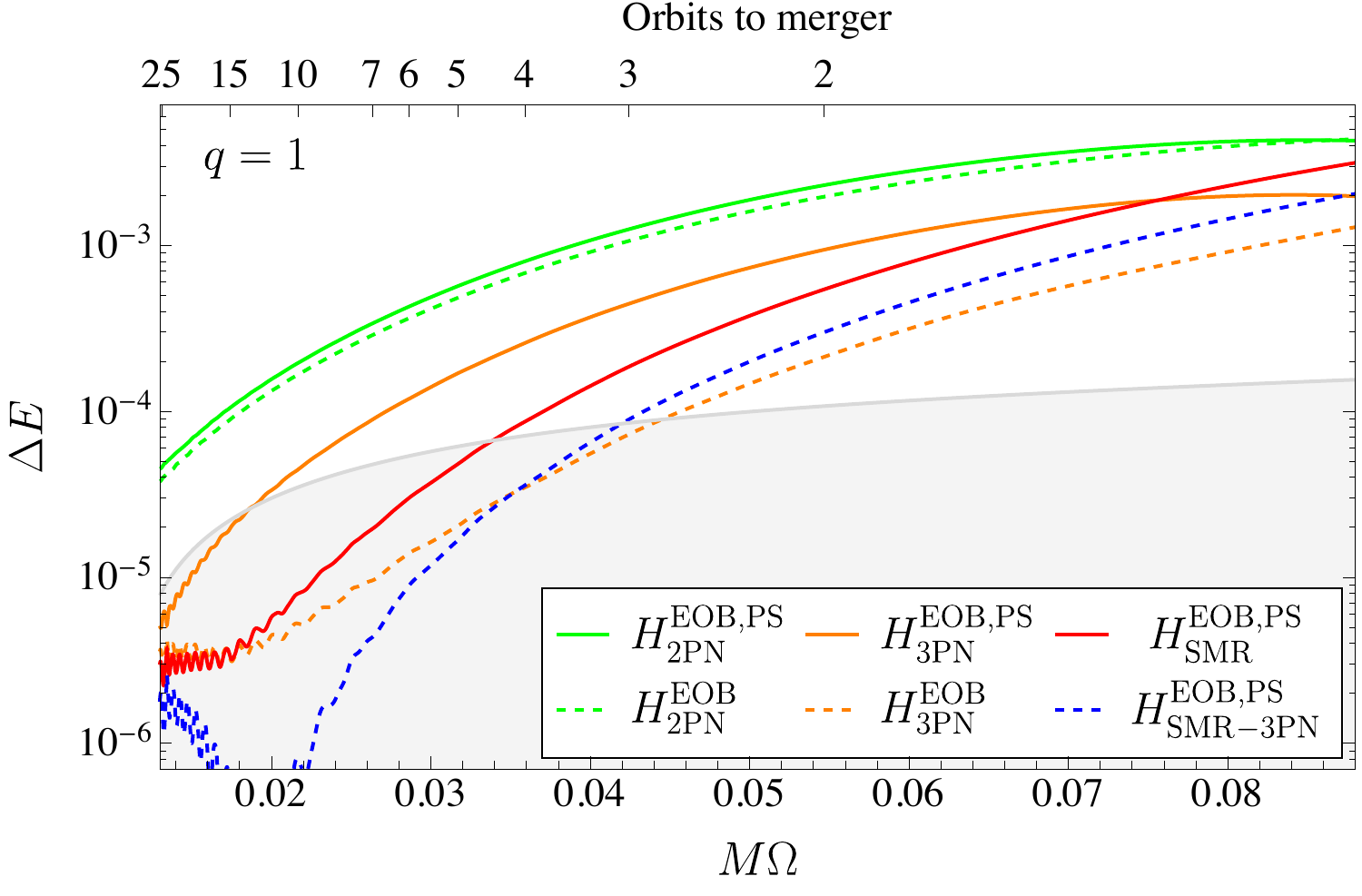}
	\includegraphics[width=\columnwidth]{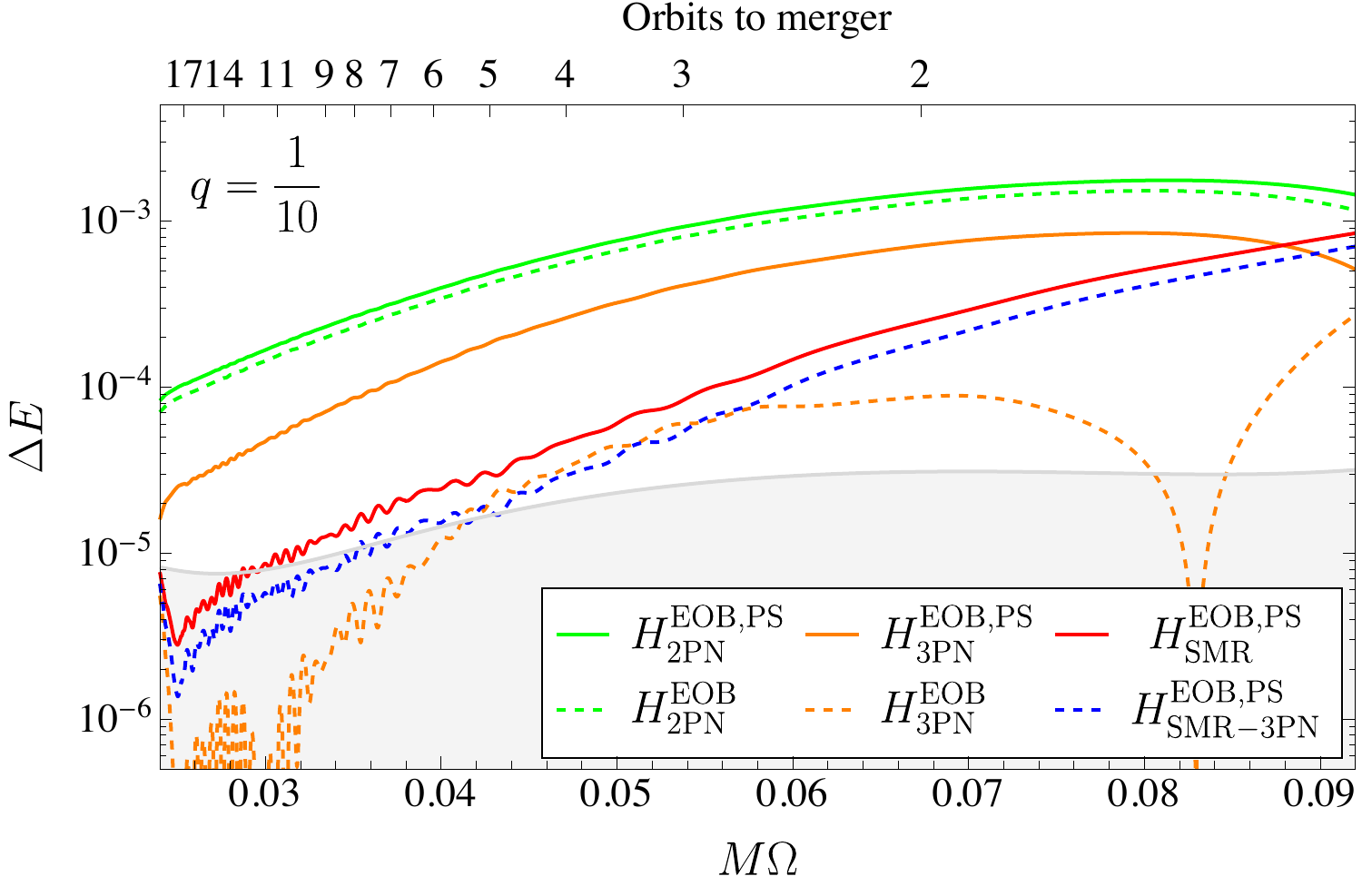}
	\caption{\textbf{SMR vs PN binding energies:} we compare the difference $\Delta E$ in binding energy from NR for our SMR Hamiltonians versus frequency $(M\Omega)$. We compare it to similar results for PN models up to third order, in both PS and DJS gauges. The \textit{estimated} NR error is shown in grey.}
	\label{fig:compEO}
\end{figure}

Here we study the energetics of the $H_{\text{SMR}}^{\rm EOB,PS}$ and $H_{\text{SMR-3PN}}^{\rm EOB,PS}$ models and the PN EOB models in both gauges via comparisons of their binding energies against NR predictions. The EOB Hamiltonians evolved and their notation are summarized in Table~\ref{table:models}.
The (quasi) gauge-invariant relations between the dimensionless circular orbit binding energy $E \equiv (H-M)/\mu$ and angular momentum $ l \equiv \p_\phi=\Pp_\phi/(M\mu)$ (and orbital frequency $\hat\Omega$) are used to draw comparisons against NR.
This type of comparisons is useful to understand how information of the real two-body motion is resummed into the conservative dynamics~\cite{Antonelli:2019ytb}.
In contrast to Ref.~\cite{Antonelli:2019ytb} and Sec.~\ref{sec:EOBSMR} of this paper, where the binding energy is calculated in the circular-orbit limit, the binding energies appearing in this section are obtained evolving the EOB Hamiltonians along quasi-circular orbits. This more closely matches the procedure used to extract the binding energy from NR simulations of quasi-circular inspirals, providing clearer comparisons \cite{Ossokine:2017dge}.
Finally, we calculate the dephasing $\Delta\phi_{22} \equiv  \phi_\text{NR}-\phi_\text{EOB}$ of the ($\ell,m$)=(2,2) modes of the $H_{\text{SMR}}^{\rm EOB,PS}$ and $H_{\text{SMR-3PN}}^{\rm EOB,PS}$ models against NR results. While more thorough comparisons aimed at using the models for LIGO inference studies would need a systematic calculation of the unfaithfulness (see e.g., Refs.~\cite{Pan:2011gk,Taracchini:2013rva,Bohe:2016gbl,Cotesta:2018fcv}), we find these comparisons illustrative to contextualize the $H_{\text{SMR}}^{\rm EOB,PS}$ and $H_{\text{SMR-3PN}}^{\rm EOB,PS}$ models in this paper.

We employ a set of ten non-spinning NR simulations from the Simulating eXtreme Spacetimes (SXS) collaboration \cite{Mroue:2013xna,Chu:2015kft}, with mass ratios $1/10\leq q\leq 1$. We summarize the details of these simulations in Table~\ref{table:setofNR}.
A description of how the $E(l)$ and $E(\hat\Omega)$ curves were calculated for a subset of these simulations can be found in Ref.~\cite{Ossokine:2017dge}.

We evolve EOB Hamiltonians with PN information up to third order, since 3PN is the order at which PS-gauge Hamiltonians can be uniquely derived for generic orbits (see the Appendix of Ref.~\cite{Antonelli:2019ytb} for more details). It is worthwhile to mention that the $H_{\rm 3PN}^{\rm EOB}$ Hamiltonian has better energetics and phases performances against NR than both $H_{\rm 4PN}^{\rm EOB}$ and the SEOBNR Hamiltonian used as a baseline for the current generation of EOB waveform models (defined, e.g., in the Appendix of Ref.~\cite{Steinhoff:2016rfi}), when calibration and NQC parameters are turned off. Restricting ourselves to comparisons with $H_{\rm 3PN}^{\rm EOB}$ only, we are therefore not running the risk to overestimate the performance of SMR models when comparing them to PN results.

Let us begin comparing the $E(l)$ and $E(\hat\Omega)$ curves. The difference $\Delta E \equiv |E_\text{NR}-E_\text{EOB}|$ is plotted for a variety of EOB models in Figs.~\ref{fig:compEJ} and \ref{fig:compEO}.
Considering the $E(l)$ relations first and focusing on the SMR models, it is seen that for $q=1/10$ both $H^{\rm EOB,PS}_{\text{SMR}}$ and $H^{\rm EOB,PS}_{\text{SMR-3PN}}$ perform better against NR than the 3PN model in the same gauge, e.g., $H^{\rm EOB,PS}_{{\rm 3PN}}$.
The $H^{\rm EOB,PS}_{\text{SMR-3PN}}$ model also performs better than both in the comparable-mass case.
A similar finding is obtained investigating the $E(\hat\Omega)$ curves, see Fig.~\ref{fig:compEO}. 
Taken together, these results highlight the importance of SMR results to improve the modeling of both equal- and unequal-mass systems within the EOB approach.
It is also seen that, for both mass ratios considered and for both $E(l)$ and $E(\hat\Omega)$ curves, $H^{\rm EOB,PS}_{\text{SMR-3PN}}$ improves the predictions of $H^{\rm EOB,PS}_{\text{SMR}}$, suggesting that generic orbit terms are important when considering quasi-circular orbit binding energies (especially in the equal-mass-ratio case).

PN Hamiltonians in the PS gauge generically perform worse in binding energy comparisons than Hamiltonians in the DJS gauge, as found out in the adiabatic approximation already in Ref.~\cite{Antonelli:2019ytb}.
This finding suggests that, notwithstanding the already good agreement between SMR models and NR simulations for both mass ratios, a better description for the EOB dynamics than the one provided by the PS gauge could be pursued in order to maximize the performance of evolutions from both PN and SMR EOB models.

\begin{figure*}[t]
	\includegraphics[width=\columnwidth]{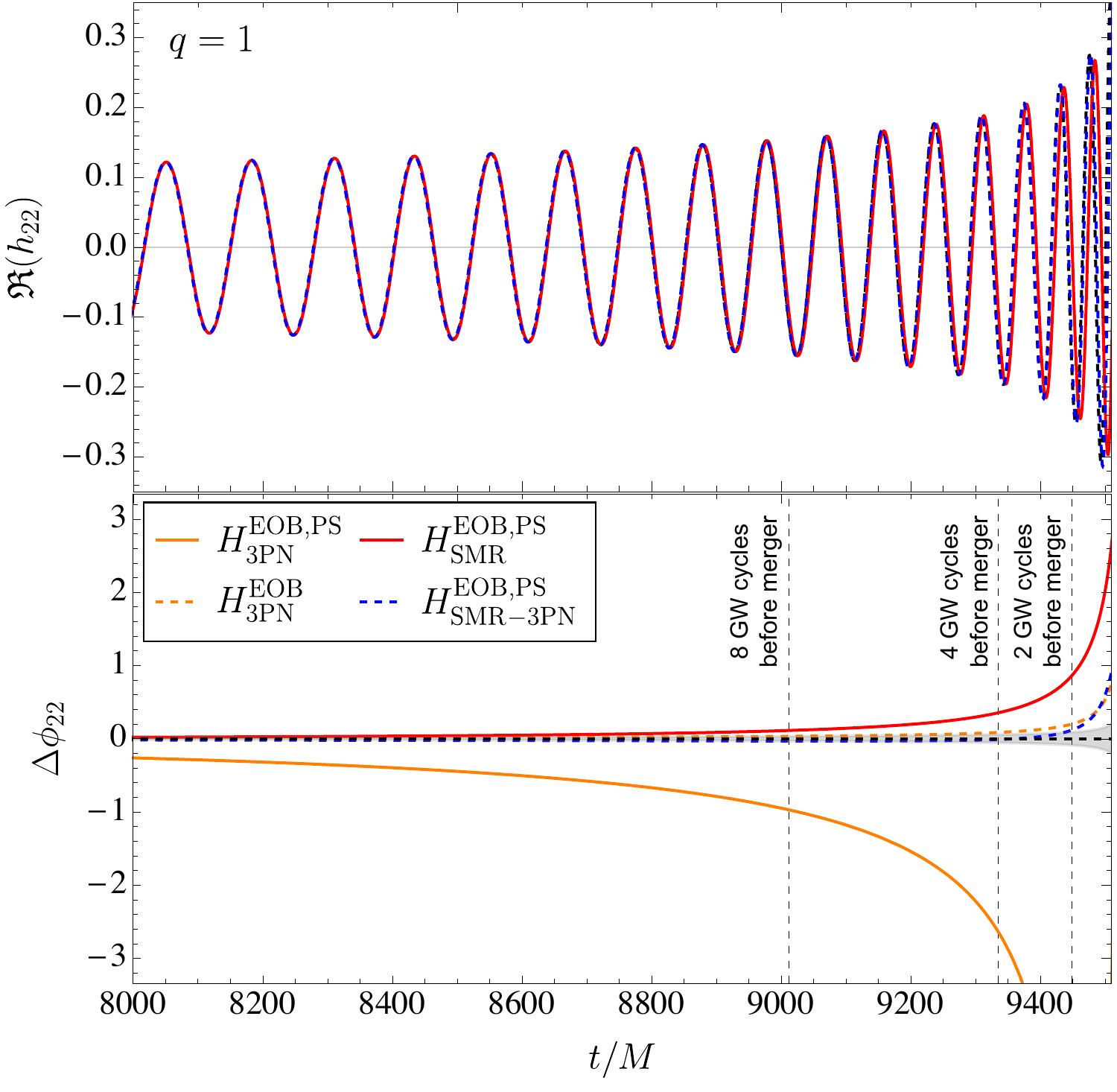}
	\hspace{12pt}
	\includegraphics[width=\columnwidth]{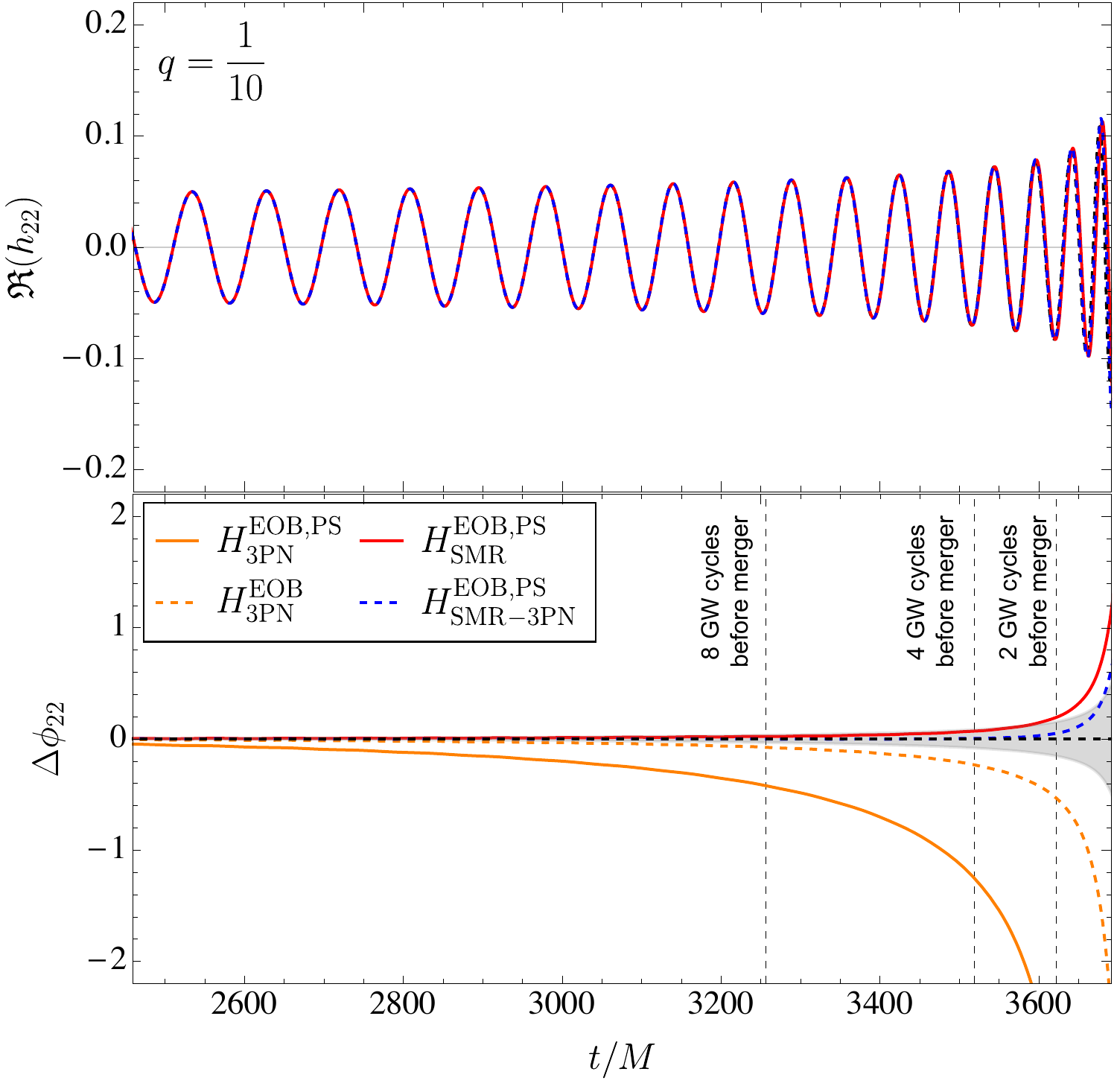}
	\caption{\textbf{Dephasing of EOB models:} in the top panels, the real parts $\mathfrak{R}(h_{22})$ of the ($\ell,m$)=(2,2) mode EOB waveform for the SMR, SMR-3PN models are shown and compared to the NR waveforms (in dashed-black, overlapping with the EOB waveforms up to few GW cycles to merger).
	In the lower panels, the dephasing of SMR and PN EOB models from the NR simulations is calculated.
	Also shown are the times corresponding to 8, 4 and 2 GW cycles before NR merger.  }
\label{fig:wavephase}
\end{figure*}
\begin{figure*}[t]
	\includegraphics[width=\columnwidth]{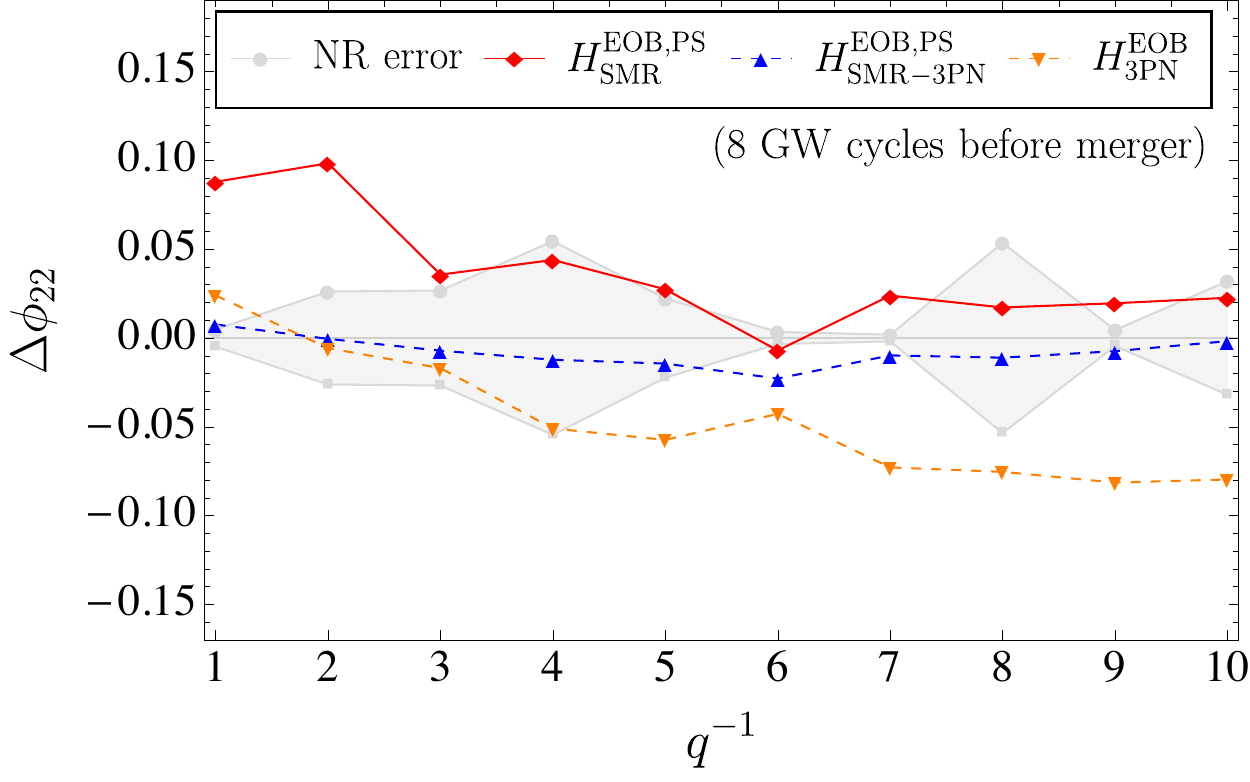}
	\hspace{12pt}
	\includegraphics[width=\columnwidth]{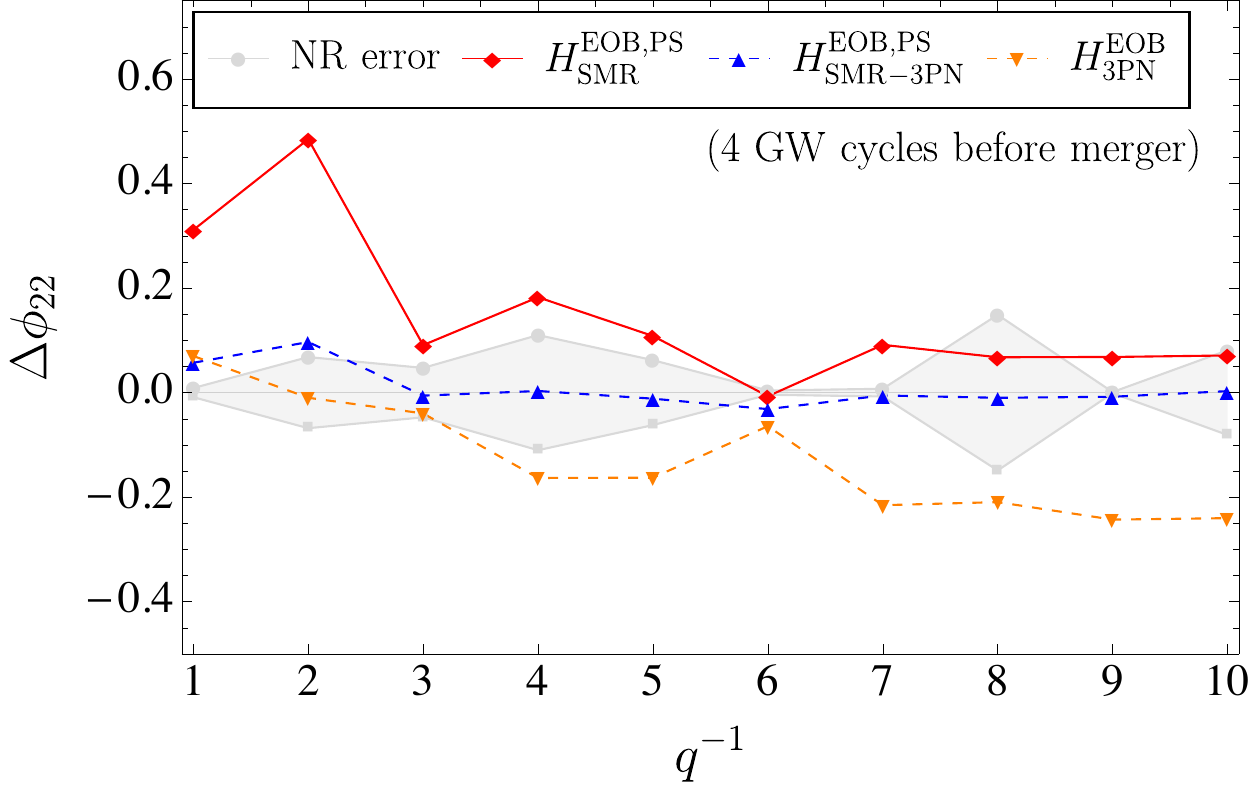}
	\caption{\textbf{Dephasing vs mass ratio:} we compare the dephasing of $H_{\text{SMR}}^{\rm EOB,PS}$, $H_{\text{SMR-PN}}^{\rm EOB,PS}$ and $H_{\text{3PN}}^{\rm EOB}$ after they have been aligned with the NR simulations from Table~\ref{table:simulations}. For each $q$, we snapshot the dephasing of the EOB models and the NR simulation at a time corresponding to 4 and 2 orbits before the merger of the binary system in the NR simulation. }
	\label{fig:dphivsq}
\end{figure*}

We complete our comparison study with the dephasing $\Delta\phi_{22}$ of the ($\ell,m$)=(2,2) modes from the EOB models and the NR simulations.
For a proper comparison, the EOB and NR waveforms must be aligned for each $q$. Here we use the alignment procedure outlined in Ref.~\cite{Pan:2011gk}, which amounts to minimizing the function:
\begin{equation}
\Xi(\Delta t,\Delta \phi)=\int_{t^\text{alig}_1}^{t^\text{alig}_2}[\phi_\text{NR}(t)-\phi_\text{EOB}(t+\Delta t)-\Delta\phi]^2 dt\\,
\end{equation} 
over the time and phase shifts, $\Delta t$ and $\Delta\phi$. 
The integrating interval [$t^\text{alig}_1,t^\text{alig}_2$] defines the time-domain window in which the alignment is performed: conservatively, it must be chosen in the inspiral of the NR simulation, large enough to average out the numerical noise and such as to avoid junk radiation at the beginning of the NR simulation \cite{Pan:2011gk}. From the alignment procedure described above, one can obtain the phase and amplitude time-shift to be applied to the EOB model to align it with the NR waveforms, i.e., the aligned waveforms are:
\begin{align}
&h^\text{NR}_{22}=A_\text{NR}(t)e^{i\phi_\text{NR}(t)}\,,\label{h22nr}\\
&h^\text{EOB}_{22}=A_\text{EOB}(t+\Delta t)e^{i[\phi_\text{EOB}(t+\Delta t)+\Delta\phi]}\label{h22eob}\,.
\end{align} 

Our choices for the time-windows are reported in Table~\ref{table:setofNR}.
In Fig.~\ref{fig:wavephase}, we show the results of our phase comparisons for $q=1$ and $q=1/10$ up to merger. For clarity, the upper panels only include the $H_{\text{SMR}}^{\rm EOB,PS}$ and $H_{\text{SMR-3PN}}^{\rm EOB,PS}$ models and the NR simulations. They show the real parts of Eqs.~\eqref{h22nr} and ~\eqref{h22eob}, from which we infer that the SMR models do not accumulate a significant amount of dephasing. Overall, they are in very good agreement with NR for both $q=1$ and $q=1/10$.
It is important to place the above results in context. In the lower panel, the dephasing of SMR models from NR is compared to that of 3PN models\footnote{In this comparison we do not include 2PN models, which we find to have much larger dephasing than the 3PN models shown.}.
Interestingly, even in the equal-mass-ratio case $H_{\text{SMR}}^{\rm EOB,PS}$ and $H_{\text{SMR-3PN}}^{\rm EOB,PS}$ compare much better than the 3PN model in the same gauge, e.g., $H_{\rm 3PN}^{\rm EOB,PS}$. Their dephasing is comparable to  $H_{\rm 3PN}^{\rm EOB}$. In the $q=1/10$ case,
they have a smaller dephasing than any other PN model considered in this study. 
In Table~\ref{table:simulations0}, we report the dephasing that the $H_{\text{SMR}}^{\rm EOB,PS}$, $H_{\text{SMR-3PN}}^{\rm EOB,PS}$, $H_{\text{3PN}}^{\rm EOB}$ and $H_{\text{3PN}}^{\rm EOB,PS}$ models accumulate up to 8 and 4 GW cycles before merger for all mass ratios (with the corresponding estimated NR error)\footnote{We have checked that shifting the time-windows by $\Delta t=\pm 100\,M$,  our $\Delta \phi$'s only change by a few hundredths of a radian.}.

Next, we want to study how the dephasing of the above models varies as a function of $q$. It would be tempting to compare the $\Delta \phi$'s reported in Table~\ref{table:simulations0} at a fixed number of cycles before merger. While this remains a valid possibility, such a comparison would neither take into account the different lengths of the NR simulations used in this set, nor the different number of GW cycles encompassed by the time-windows of Table~\ref{table:setofNR}.
To keep both parameters under control, we realign our models with alternative time-windows that are dictated by the number of GW cycles to merger $\Delta N_\text{GW}(t) \equiv N_\text{GW}(t)-N_\text{GW}^\text{merg}
$ of the NR simulations. That is, for each mass ratio we fix a different time-window [$t^\text{alig}_1,t^\text{alig}_2$], corresponding to the \textit{same} interval of cycles to merger [$\Delta N_\text{GW}(t^\text{alig}_1)$, $\Delta N_\text{GW}(t^\text{alig}_2)$]. 
The benefits of this choice are two-fold. To begin with, the alignment windows thus calculated depends on the position of the NR merger (peak of $h_{22}^{\rm NR}$), which is a quantifiable feature of every NR simulation.
Moreover, this choice allows us to assess trends across the mass ratios fairly, since the waveforms thus aligned are compared in the same range of GW cycles. A caveat for this alignment method is that the GW cycles of evolutions with smaller $q$ lie in a regime of stronger gravity.

We choose to align the EOB models to NR in an interval of $N_\text{GW}$ such that 
$[\Delta N_\text{GW}(t^\text{alig}_1), \Delta N_\text{GW}(t^\text{alig}_2)]=[-34, -24]$, corresponding to the time-windows reported in Table~\ref{table:simulations}.
This choice stems from the length of the shortest NR simulation, e.g., $q=1/9$, which counts $N_\text{GW}^\text{merg}=37.96$ GW cycles at merger (the first ~3GW cycles of this simulation are neglected in order to avoid junk radiation). In Fig.~\ref{fig:dphivsq}, we plot the dephasing for the three models that perform best in Fig.~\ref{fig:wavephase}: that is, $H_{\rm 3PN}^{\rm EOB}$, $H_{\text{SMR}}^{\rm EOB,PS}$ and $H_{\text{SMR-3PN}}^{\rm EOB,PS}$ and study the trends across $q$. For every simulation, we calculate the dephasing 8 and 4 GW cycles before merger to show the robustness of the trends\footnote{We have also checked that the trends are unaffected by variations in the number of orbital cycles in the alignment window.}.
Noticeably, the 3PN EOB waveform in the DJS gauge starts degrading in accuracy as the mass ratio is increased, while the SMR and SMR-3PN ones improve: remarkably, for most $q$'s, the SMR-3PN model only dephases by a few hundredths of a radian
up to a 4 GW cycles before merger.
Moreover, we notice that SMR models start performing better than $H_{\rm 3PN}^{\rm EOB}$ for $q \lesssim 1/3$, hinting again to the 
fact that SMR information, when reorganized in the EOB framework, could be used to model systems that are very close to the equal-mass-ratio regime \cite{LeTiec:2011bk,Barausse:2011dq}.

The picture emerging from Fig.~\ref{fig:dphivsq} is that the SMR-3PN model is the most consistent of the two models with SMR information, corroborating the findings for $q=1$ and $q=1/10$ in the binding energy comparisons. The small dephasing of the SMR-3PN model suggests that the Hamiltonian upon which it is based is a possible starting point to develop a new generation of EOB waveform models able to tackle the currently challenging intermediate-mass-ratio regime.

\section{Discussion and Conclusions}
\label{sec:conclusions}

The complete EOB Hamiltonian at linear order in SMR from Ref.~\cite{LeTiec:2011dp} suffers from a coordinate singularity at the LR radius in the deformed Schwarzschild background. Building on Refs.~\cite{Akcay:2012ea,Damour:2017zjx}, we have constructed two Hamiltonians in the post-Schwarzschild (PS) reformulation of the EOB approach \cite{Damour:2017zjx,Antonelli:2019ytb} (both with the SMR correction to the Detweiler redshift and with mixed SMR-3PN information), and checked 
that they are not affected by poles  at the LR radius (and related unphysical features) by studying plunging trajectories.

We have then explored the merits of the SMR and mixed SMR-3PN Hamiltonians via comparisons of their waveforms and binding energies, and those of PN Hamiltonians in different gauges, against NR predictions. Ultimately, we find that:
\begin{enumerate} 
	\item For both $q=1$ and $q=1/10$, the binding energies of SMR and SMR-3PN EOB models (see Figs.~\ref{fig:compEJ} and~\ref{fig:compEO}) generally compare better against NR than the binding energy of the PS Hamiltonian with 3PN information.
	\item The generic orbit 3PN information in the SMR-3PN EOB Hamiltonian improves the binding energy and phase comparisons of SMR EOB models.
	\item PN Hamiltonians in the EOB-PS gauge have binding energies that compare worse than those from PN Hamiltonians in the standard EOB gauge, confirming the findings of Ref.~\cite{Antonelli:2019ytb} and extending their validity to non-adiabatic evolutions.
	\item The SMR-3PN EOB model agrees remarkably well against NR simulations, see Fig.~\ref{fig:dphivsq}. The dephasing up to 4 GW cycles before merger is a few hundredths of a radian for $q \lesssim 1/3$ and a tenth of a radian for $q>1/3$. The only EOB PN model with comparable dephasing is the 3PN EOB Hamiltonian in the DJS gauge for $q \gtrsim 1/3$.
\end{enumerate}

The construction of the SMR EOB Hamiltonian in this paper depends on a number of choices. First of all, we chose to fix the coordinate freedom in the effective Hamiltonian using the PS gauge. This was chosen because of its relative simplicity, while allowing a natural path towards avoiding singularities at the light ring. However, there may exist different choices that are equally (or more) effective. Second, while the EOB Hamiltonian in principle applies to generic orbits, we fix the linear-in-$\nu$ part only by comparing to the circular-orbit binding energy. Consequently, there is considerable freedom in the ``non-circular-orbit'' part of the Hamiltonian. In practice, we fix this freedom by choosing the specific functional dependence of the effective Hamiltonian on $\Ham_{\text{S}}$ given by Eq.~\eqref{HSMR}. This choice is in part restricted by the requirement that the Hamiltonian be analytic, but other options are available. Third and finally, SMR data for the binding energy extends only to the light ring. The Hamiltonian in the region $u>\tfrac{1}{3}$ therefore depends only on the analytic extension of the redshift data. Given that this data is known only to finite numerical precision, there is some freedom in the choice of the exact analytical form of its fit. This choice can also affect the relative size of the different coefficient functions in Eq.~\eqref{HSMR}.

Our investigation opens up further avenues of research. To begin with, one can study whether it is possible to uniquely fix other EOB gauges that could accommodate the Detweiler redshift (without introducing a LR-divergence) and study their merits via comparisons against NR. As discussed already in Ref.~\cite{Akcay:2012ea}, to solve the LR-divergence arising in this context the non-geodesic function $\hat{Q}$ needs a term proportional to $p_{\phi}^3$, possibly resummed in another quantity (as done in the PS gauge using $\Ham_\text{S}$). It would be quite interesting to see whether other gauges that allow solving the LR-divergence also improve the comparisons against NR predictions. One concrete example of different resummation that was shown to improve the comparisons of the conservative sector of post-Minkowskian Hamiltonians in PS form has been given in the Appendix of Ref.~\cite{Antonelli:2019ytb}. It is worthwhile to study whether a similar choice  could work for the SMR and SMR-3PN models herein presented.
The hope is that using different resummations, and including information from the second order in the SMR, one could obtain a considerably improved EOB Hamiltonian that, after further calibration to NR, would be very useful for LIGO/Virgo analyses in the near-future.

Further research endeavours could be directed towards informing the EOB with different SMR quantities than the circular orbit Detweiler redshift. An example of a quantity that still needs to be fully exploited is the generalized redshift \cite{Barack:2011ed,Akcay:2015pza}, which includes information for arbitrarily eccentric orbits.
We envision using EOB Hamiltonians at linear and higher orders in the
mass ratio for inference studies in the future detectors' era, when
precise models will be needed to properly characterize high
signal-to-noise systems, possibly having rather small mass ratios. In
order for this program to be achieved, not only should the
conservative sector be optimized with both results at second order in
$q$ and (potentially) a better resummation, but information from other
crucial physical quantities should also be incorporated: notably missing
features in our analysis are the spin and eccentricity. Furthermore, a
more comprehensive study of the dissipative sector must be pursued. It
would be desirable, for instance, to include more self-force
information in the flux. Lastly, we would also need to build the full 
inspiral, merger and ringdown waveforms, and calibrate them to NR simulations.  
We leave these important investigations to future work.

\begin{acknowledgments}
It is a pleasure to thank Sergei Ossokine for providing us with the NR data used for the comparisons, and Tanja Hinderer for the original version of the EOB evolution code in Mathematica used in this paper. A.A. would like to further thank Sergei Ossokine and Roberto Cotesta for many fruitful discussions over the preparation of this paper. MvdM was supported by European Union's Horizon 2020 research and innovation programme under grant agreement No.~705229.
This work makes use of the Black-Hole Perturbation Toolkit \cite{BHPToolkit}.
\end{acknowledgments}

\appendix

\section{Detweiler-redshift data and fit}\label{sec:appred}

The  linear-in-$\nu$ Detweiler redshift $\zzz$ at a fixed $x$ is given by Refs.~\cite{Detweiler:2008ft,Akcay:2015pza}:
\begin{equation}\label{eq:zfromhuu}
\zzz = -\tfrac{1}{2}\sqrt{1-3x} h_{uu}^{\rm R}(x)+ \frac{x}{\sqrt{1-3x}},
\end{equation}
where  $h_{uu}^{\rm R}$ is the double contraction of the regular part of the metric perturbation generated by a particle on a circular orbit with its 4-velocity. We determine $h_{uu}^{\rm R}$ in a range $0<x<1/3$ to a high precision using the numerical code developed in Ref.~\cite{vandeMeent:2015lxa}. In this code the regular part of the metric perturbation is extracted using the mode-sum formalism. As noted in Ref.~\cite{Akcay:2012ea}, the convergence of the mode-sum decreases drastically as circular orbits approach the light ring. This limits the accuracy with which $h_{uu}^{\rm R}$ can be obtained. The code from Ref.~\cite{vandeMeent:2015lxa} allows calculations using arbitrary precision arithmetic, which allows us to calculate $\zzz$ much closer to the light ring and at much higher precision than previously done in Ref.~\cite{Akcay:2012ea}. For this paper, we have generated data for $\zzz$ using up to 120 $\ell$-modes, which allows us to obtain $\zzz$ up to $(1-3x)\approx 4\times 10^{-5}$, with relative accuracy $\lesssim 2.5\times10^{-5}$.

To utilize the $\zzz$ data in our SMR EOB model we need an analytic fit to the data. Two aspects of this fit are important to control for the behaviour of the model. First, the model is sensitive to the precise analytical structure of the fit near the light ring. Second, we need to control the behaviour of the fit beyond the light ring $x>1/3$, where we have no self-force data. In light of these two considerations, we want to fit the data with a model having a relatively low number of parameters. To achieve this, we leverage the analytic knowledge of the PN expansion of $\zzz$, which Ref.~\cite{Kavanagh:2015lva} calculated up to 21.5PN order.
We construct a fit of the overall form:
\begin{equation}\label{eq:fitform}
\zzz = Z_0(x) + \frac{(1-2x)^5}{1-3x} Z_{\rm PN}(x)\left[1+\alpha(x)Z_{\rm fit}(x)\right].
\end{equation}
The leading term:
\begin{equation}
Z_0(x) = x\frac{1-4x}{1-3x}+\frac{x}{\sqrt{1-3x}},
\end{equation}
is constructed such that it will exactly cancel the coefficients $\tilde{\X}_0$ and $\tilde{\X}_1$ when matched to the SMR EOB Hamiltonian.

The number of factors $(1-2x)$ in front of the second term has been chosen such that the resulting contribution to the effective Hamiltonian $\hat{Q}^{\text{PS}}_{\text{SMR}}$ vanishes at the horizon of the effective spacetime, $x=1/2$. 
The coefficient function, $Z_{\rm PN}(x)$ has the form:
\begin{equation}
 Z_{\rm PN}(x) = 2x^3 \sum_{i,j} a_{i,j} x^{i/2} \log^j x,
\end{equation}
where the coefficients $a_{i,j}$ are obtained by requiring that the series expansion of Eq.~\eqref{eq:fitform} matches the 21.5PN expression from Ref.~\cite{Kavanagh:2015lva}. Since these coefficients are numerous and lengthy, and are easily obtained using computer algebra and the expressions available for the Black Hole Peturbation Toolkit \cite{BHPToolkit}, we do not reproduce them explicitly here.

The actual fit $Z_{\rm fit}$ is multiplied by an attenuation function:
\begin{equation}
\alpha(x) = \exp\Big(\frac{4-x^{-2}}{6}\Big),
\end{equation}
that suppresses the fit exponentially in the weak field regime, ensuring that the PN behaviour of $\zzz$ is unaffected by the fit. The function $\alpha(x)$ has been chosen such that $\alpha(1/2)=1$ and is at its steepest at $x=1/3$.

The fit $Z_{\rm fit}$ itself is a polynomial in $\beta \equiv 9x(1-3x)(1-2x)$ and $\log[\frac{1-3x}{(1-2x)^2}]$ with arbitrary coefficients. We perform a large number of linear fits for varying combinations of five terms, and compare various ``goodness of fit'' indicators such as the adjusted $R^2$ value and Bayesian Information Criterion. One model that consistently outperformed the others is:
\begin{equation}
Z_{\rm fit} = c_0 + c_1 \beta + c_2 \beta^4 + (c_3 \beta+c_4\beta^4)\log\Big[\frac{1-3x}{(1-2x)^2}\Big],
\end{equation}
with:
\begin{subequations}
\begin{alignat}{2}
c_0 &=&    0.555947
&,\\
c_1 &=&   -2.589868
&,\\
c_2 &=&   31.144986
&,\\
c_3 &=&    2.440115
&,\\
c_4 &=& -179.175818
&.
\end{alignat}
\end{subequations}

With this fit the coefficient functions $\X_i$ in Eq.~\eqref{HSMR} become,
\begin{align}
\X_0(x) &= (1-3x) Z_{\rm PN}(x)\left[1+\alpha(x)(c_0 + c_1 \beta + c_2 \beta^4)\right],\\
\X_1(x) &= 0,\\
\X_2(x) &= (1-2x)^2 Z_{\rm PN}(x)\left[1+\alpha(x)(c_3 \beta+c_4\beta^4)\right].
\end{align}

\raggedright
\bibliography{refs}

\begin{thebibliography}{106}%
\makeatletter
\providecommand \@ifxundefined [1]{%
 \@ifx{#1\undefined}
}%
\providecommand \@ifnum [1]{%
 \ifnum #1\expandafter \@firstoftwo
 \else \expandafter \@secondoftwo
 \fi
}%
\providecommand \@ifx [1]{%
 \ifx #1\expandafter \@firstoftwo
 \else \expandafter \@secondoftwo
 \fi
}%
\providecommand \natexlab [1]{#1}%
\providecommand \enquote  [1]{``#1''}%
\providecommand \bibnamefont  [1]{#1}%
\providecommand \bibfnamefont [1]{#1}%
\providecommand \citenamefont [1]{#1}%
\providecommand \href@noop [0]{\@secondoftwo}%
\providecommand \href [0]{\begingroup \@sanitize@url \@href}%
\providecommand \@href[1]{\@@startlink{#1}\@@href}%
\providecommand \@@href[1]{\endgroup#1\@@endlink}%
\providecommand \@sanitize@url [0]{\catcode `\\12\catcode `\$12\catcode
  `\&12\catcode `\#12\catcode `\^12\catcode `\_12\catcode `\%12\relax}%
\providecommand \@@startlink[1]{}%
\providecommand \@@endlink[0]{}%
\providecommand \url  [0]{\begingroup\@sanitize@url \@url }%
\providecommand \@url [1]{\endgroup\@href {#1}{\urlprefix }}%
\providecommand \urlprefix  [0]{URL }%
\providecommand \Eprint [0]{\href }%
\providecommand \doibase [0]{http://dx.doi.org/}%
\providecommand \selectlanguage [0]{\@gobble}%
\providecommand \bibinfo  [0]{\@secondoftwo}%
\providecommand \bibfield  [0]{\@secondoftwo}%
\providecommand \translation [1]{[#1]}%
\providecommand \BibitemOpen [0]{}%
\providecommand \bibitemStop [0]{}%
\providecommand \bibitemNoStop [0]{.\EOS\space}%
\providecommand \EOS [0]{\spacefactor3000\relax}%
\providecommand \BibitemShut  [1]{\csname bibitem#1\endcsname}%
\let\auto@bib@innerbib\@empty
\bibitem [{\citenamefont {Pretorius}(2005)}]{Pretorius:2005gq}%
  \BibitemOpen
  \bibfield  {author} {\bibinfo {author} {\bibfnamefont {F.}~\bibnamefont
  {Pretorius}},\ }\href {\doibase 10.1103/PhysRevLett.95.121101} {\bibfield
  {journal} {\bibinfo  {journal} {Phys. Rev. Lett.}\ }\textbf {\bibinfo
  {volume} {95}},\ \bibinfo {pages} {121101} (\bibinfo {year} {2005})},\
  \Eprint {http://arxiv.org/abs/gr-qc/0507014} {arXiv:gr-qc/0507014 [gr-qc]}
  \BibitemShut {NoStop}%
\bibitem [{\citenamefont {Campanelli}\ \emph {et~al.}(2006)\citenamefont
  {Campanelli}, \citenamefont {Lousto}, \citenamefont {Marronetti},\ and\
  \citenamefont {Zlochower}}]{Campanelli:2005dd}%
  \BibitemOpen
  \bibfield  {author} {\bibinfo {author} {\bibfnamefont {M.}~\bibnamefont
  {Campanelli}}, \bibinfo {author} {\bibfnamefont {C.~O.}\ \bibnamefont
  {Lousto}}, \bibinfo {author} {\bibfnamefont {P.}~\bibnamefont {Marronetti}},
  \ and\ \bibinfo {author} {\bibfnamefont {Y.}~\bibnamefont {Zlochower}},\
  }\href {\doibase 10.1103/PhysRevLett.96.111101} {\bibfield  {journal}
  {\bibinfo  {journal} {Phys. Rev. Lett.}\ }\textbf {\bibinfo {volume} {96}},\
  \bibinfo {pages} {111101} (\bibinfo {year} {2006})},\ \Eprint
  {http://arxiv.org/abs/gr-qc/0511048} {arXiv:gr-qc/0511048 [gr-qc]}
  \BibitemShut {NoStop}%
\bibitem [{\citenamefont {Baker}\ \emph {et~al.}(2006)\citenamefont {Baker},
  \citenamefont {Centrella}, \citenamefont {Choi}, \citenamefont {Koppitz},\
  and\ \citenamefont {van Meter}}]{Baker:2005vv}%
  \BibitemOpen
  \bibfield  {author} {\bibinfo {author} {\bibfnamefont {J.~G.}\ \bibnamefont
  {Baker}}, \bibinfo {author} {\bibfnamefont {J.}~\bibnamefont {Centrella}},
  \bibinfo {author} {\bibfnamefont {D.-I.}\ \bibnamefont {Choi}}, \bibinfo
  {author} {\bibfnamefont {M.}~\bibnamefont {Koppitz}}, \ and\ \bibinfo
  {author} {\bibfnamefont {J.}~\bibnamefont {van Meter}},\ }\href {\doibase
  10.1103/PhysRevLett.96.111102} {\bibfield  {journal} {\bibinfo  {journal}
  {Phys. Rev. Lett.}\ }\textbf {\bibinfo {volume} {96}},\ \bibinfo {pages}
  {111102} (\bibinfo {year} {2006})},\ \Eprint
  {http://arxiv.org/abs/gr-qc/0511103} {arXiv:gr-qc/0511103 [gr-qc]}
  \BibitemShut {NoStop}%
\bibitem [{\citenamefont {Mroue}\ \emph {et~al.}(2013)\citenamefont {Mroue}
  \emph {et~al.}}]{Mroue:2013xna}%
  \BibitemOpen
  \bibfield  {author} {\bibinfo {author} {\bibfnamefont {A.~H.}\ \bibnamefont
  {Mroue}} \emph {et~al.},\ }\href {\doibase 10.1103/PhysRevLett.111.241104}
  {\bibfield  {journal} {\bibinfo  {journal} {Phys. Rev. Lett.}\ }\textbf
  {\bibinfo {volume} {111}},\ \bibinfo {pages} {241104} (\bibinfo {year}
  {2013})},\ \Eprint {http://arxiv.org/abs/1304.6077} {arXiv:1304.6077 [gr-qc]}
  \BibitemShut {NoStop}%
\bibitem [{\citenamefont {Jani}\ \emph {et~al.}(2016)\citenamefont {Jani},
  \citenamefont {Healy}, \citenamefont {Clark}, \citenamefont {London},
  \citenamefont {Laguna},\ and\ \citenamefont {Shoemaker}}]{Jani:2016wkt}%
  \BibitemOpen
  \bibfield  {author} {\bibinfo {author} {\bibfnamefont {K.}~\bibnamefont
  {Jani}}, \bibinfo {author} {\bibfnamefont {J.}~\bibnamefont {Healy}},
  \bibinfo {author} {\bibfnamefont {J.~A.}\ \bibnamefont {Clark}}, \bibinfo
  {author} {\bibfnamefont {L.}~\bibnamefont {London}}, \bibinfo {author}
  {\bibfnamefont {P.}~\bibnamefont {Laguna}}, \ and\ \bibinfo {author}
  {\bibfnamefont {D.}~\bibnamefont {Shoemaker}},\ }\href {\doibase
  10.1088/0264-9381/33/20/204001} {\bibfield  {journal} {\bibinfo  {journal}
  {Class. Quant. Grav.}\ }\textbf {\bibinfo {volume} {33}},\ \bibinfo {pages}
  {204001} (\bibinfo {year} {2016})},\ \Eprint
  {http://arxiv.org/abs/1605.03204} {arXiv:1605.03204 [gr-qc]} \BibitemShut
  {NoStop}%
\bibitem [{\citenamefont {Healy}\ \emph {et~al.}(2017)\citenamefont {Healy},
  \citenamefont {Lousto}, \citenamefont {Zlochower},\ and\ \citenamefont
  {Campanelli}}]{Healy:2017psd}%
  \BibitemOpen
  \bibfield  {author} {\bibinfo {author} {\bibfnamefont {J.}~\bibnamefont
  {Healy}}, \bibinfo {author} {\bibfnamefont {C.~O.}\ \bibnamefont {Lousto}},
  \bibinfo {author} {\bibfnamefont {Y.}~\bibnamefont {Zlochower}}, \ and\
  \bibinfo {author} {\bibfnamefont {M.}~\bibnamefont {Campanelli}},\ }\href
  {\doibase 10.1088/1361-6382/aa91b1} {\bibfield  {journal} {\bibinfo
  {journal} {Class. Quant. Grav.}\ }\textbf {\bibinfo {volume} {34}},\ \bibinfo
  {pages} {224001} (\bibinfo {year} {2017})},\ \Eprint
  {http://arxiv.org/abs/1703.03423} {arXiv:1703.03423 [gr-qc]} \BibitemShut
  {NoStop}%
\bibitem [{\citenamefont {Dietrich}\ \emph {et~al.}(2018)\citenamefont
  {Dietrich}, \citenamefont {Radice}, \citenamefont {Bernuzzi}, \citenamefont
  {Zappa}, \citenamefont {Perego}, \citenamefont {Brügmann}, \citenamefont
  {Chaurasia}, \citenamefont {Dudi}, \citenamefont {Tichy},\ and\ \citenamefont
  {Ujevic}}]{Dietrich:2018phi}%
  \BibitemOpen
  \bibfield  {author} {\bibinfo {author} {\bibfnamefont {T.}~\bibnamefont
  {Dietrich}}, \bibinfo {author} {\bibfnamefont {D.}~\bibnamefont {Radice}},
  \bibinfo {author} {\bibfnamefont {S.}~\bibnamefont {Bernuzzi}}, \bibinfo
  {author} {\bibfnamefont {F.}~\bibnamefont {Zappa}}, \bibinfo {author}
  {\bibfnamefont {A.}~\bibnamefont {Perego}}, \bibinfo {author} {\bibfnamefont
  {B.}~\bibnamefont {Brügmann}}, \bibinfo {author} {\bibfnamefont {S.~V.}\
  \bibnamefont {Chaurasia}}, \bibinfo {author} {\bibfnamefont {R.}~\bibnamefont
  {Dudi}}, \bibinfo {author} {\bibfnamefont {W.}~\bibnamefont {Tichy}}, \ and\
  \bibinfo {author} {\bibfnamefont {M.}~\bibnamefont {Ujevic}},\ }\href
  {\doibase 10.1088/1361-6382/aaebc0} {\bibfield  {journal} {\bibinfo
  {journal} {Class. Quant. Grav.}\ }\textbf {\bibinfo {volume} {35}},\ \bibinfo
  {pages} {24LT01} (\bibinfo {year} {2018})},\ \Eprint
  {http://arxiv.org/abs/1806.01625} {arXiv:1806.01625 [gr-qc]} \BibitemShut
  {NoStop}%
\bibitem [{\citenamefont {Boyle}\ \emph {et~al.}(2019)\citenamefont {Boyle}
  \emph {et~al.}}]{Boyle:2019kee}%
  \BibitemOpen
  \bibfield  {author} {\bibinfo {author} {\bibfnamefont {M.}~\bibnamefont
  {Boyle}} \emph {et~al.},\ }\href@noop {} {\  (\bibinfo {year} {2019})},\
  \Eprint {http://arxiv.org/abs/1904.04831} {arXiv:1904.04831 [gr-qc]}
  \BibitemShut {NoStop}%
\bibitem [{\citenamefont {Blanchet}(2014)}]{Blanchet:2013haa}%
  \BibitemOpen
  \bibfield  {author} {\bibinfo {author} {\bibfnamefont {L.}~\bibnamefont
  {Blanchet}},\ }\href {\doibase 10.12942/lrr-2014-2} {\bibfield  {journal}
  {\bibinfo  {journal} {Living Rev. Rel.}\ }\textbf {\bibinfo {volume} {17}},\
  \bibinfo {pages} {2} (\bibinfo {year} {2014})},\ \Eprint
  {http://arxiv.org/abs/1310.1528} {arXiv:1310.1528 [gr-qc]} \BibitemShut
  {NoStop}%
\bibitem [{\citenamefont {Barack}\ and\ \citenamefont
  {Pound}(2019)}]{Barack:2018yvs}%
  \BibitemOpen
  \bibfield  {author} {\bibinfo {author} {\bibfnamefont {L.}~\bibnamefont
  {Barack}}\ and\ \bibinfo {author} {\bibfnamefont {A.}~\bibnamefont {Pound}},\
  }\href {\doibase 10.1088/1361-6633/aae552} {\bibfield  {journal} {\bibinfo
  {journal} {Rept. Prog. Phys.}\ }\textbf {\bibinfo {volume} {82}},\ \bibinfo
  {pages} {016904} (\bibinfo {year} {2019})},\ \Eprint
  {http://arxiv.org/abs/1805.10385} {arXiv:1805.10385 [gr-qc]} \BibitemShut
  {NoStop}%
\bibitem [{\citenamefont {Le~Tiec}(2014)}]{Tiec:2014lba}%
  \BibitemOpen
  \bibfield  {author} {\bibinfo {author} {\bibfnamefont {A.}~\bibnamefont
  {Le~Tiec}},\ }\href {\doibase 10.1142/S0218271814300225} {\bibfield
  {journal} {\bibinfo  {journal} {Int. J. Mod. Phys.}\ }\textbf {\bibinfo
  {volume} {D23}},\ \bibinfo {pages} {1430022} (\bibinfo {year} {2014})},\
  \Eprint {http://arxiv.org/abs/1408.5505} {arXiv:1408.5505 [gr-qc]}
  \BibitemShut {NoStop}%
\bibitem [{\citenamefont {Pan}\ \emph {et~al.}(2011{\natexlab{a}})\citenamefont
  {Pan}, \citenamefont {Buonanno}, \citenamefont {Boyle}, \citenamefont
  {Buchman}, \citenamefont {Kidder}, \citenamefont {Pfeiffer},\ and\
  \citenamefont {Scheel}}]{Pan:2011gk}%
  \BibitemOpen
  \bibfield  {author} {\bibinfo {author} {\bibfnamefont {Y.}~\bibnamefont
  {Pan}}, \bibinfo {author} {\bibfnamefont {A.}~\bibnamefont {Buonanno}},
  \bibinfo {author} {\bibfnamefont {M.}~\bibnamefont {Boyle}}, \bibinfo
  {author} {\bibfnamefont {L.~T.}\ \bibnamefont {Buchman}}, \bibinfo {author}
  {\bibfnamefont {L.~E.}\ \bibnamefont {Kidder}}, \bibinfo {author}
  {\bibfnamefont {H.~P.}\ \bibnamefont {Pfeiffer}}, \ and\ \bibinfo {author}
  {\bibfnamefont {M.~A.}\ \bibnamefont {Scheel}},\ }\href {\doibase
  10.1103/PhysRevD.84.124052} {\bibfield  {journal} {\bibinfo  {journal} {Phys.
  Rev.}\ }\textbf {\bibinfo {volume} {D84}},\ \bibinfo {pages} {124052}
  (\bibinfo {year} {2011}{\natexlab{a}})},\ \Eprint
  {http://arxiv.org/abs/1106.1021} {arXiv:1106.1021 [gr-qc]} \BibitemShut
  {NoStop}%
\bibitem [{\citenamefont {Taracchini}\ \emph {et~al.}(2012)\citenamefont
  {Taracchini}, \citenamefont {Pan}, \citenamefont {Buonanno}, \citenamefont
  {Barausse}, \citenamefont {Boyle}, \citenamefont {Chu}, \citenamefont
  {Lovelace}, \citenamefont {Pfeiffer},\ and\ \citenamefont
  {Scheel}}]{Taracchini:2012ig}%
  \BibitemOpen
  \bibfield  {author} {\bibinfo {author} {\bibfnamefont {A.}~\bibnamefont
  {Taracchini}}, \bibinfo {author} {\bibfnamefont {Y.}~\bibnamefont {Pan}},
  \bibinfo {author} {\bibfnamefont {A.}~\bibnamefont {Buonanno}}, \bibinfo
  {author} {\bibfnamefont {E.}~\bibnamefont {Barausse}}, \bibinfo {author}
  {\bibfnamefont {M.}~\bibnamefont {Boyle}}, \bibinfo {author} {\bibfnamefont
  {T.}~\bibnamefont {Chu}}, \bibinfo {author} {\bibfnamefont {G.}~\bibnamefont
  {Lovelace}}, \bibinfo {author} {\bibfnamefont {H.~P.}\ \bibnamefont
  {Pfeiffer}}, \ and\ \bibinfo {author} {\bibfnamefont {M.~A.}\ \bibnamefont
  {Scheel}},\ }\href {\doibase 10.1103/PhysRevD.86.024011} {\bibfield
  {journal} {\bibinfo  {journal} {Phys. Rev.}\ }\textbf {\bibinfo {volume}
  {D86}},\ \bibinfo {pages} {024011} (\bibinfo {year} {2012})},\ \Eprint
  {http://arxiv.org/abs/1202.0790} {arXiv:1202.0790 [gr-qc]} \BibitemShut
  {NoStop}%
\bibitem [{\citenamefont {Taracchini}\ \emph {et~al.}(2014)\citenamefont
  {Taracchini} \emph {et~al.}}]{Taracchini:2013rva}%
  \BibitemOpen
  \bibfield  {author} {\bibinfo {author} {\bibfnamefont {A.}~\bibnamefont
  {Taracchini}} \emph {et~al.},\ }\href {\doibase 10.1103/PhysRevD.89.061502}
  {\bibfield  {journal} {\bibinfo  {journal} {Phys. Rev.}\ }\textbf {\bibinfo
  {volume} {D89}},\ \bibinfo {pages} {061502} (\bibinfo {year} {2014})},\
  \Eprint {http://arxiv.org/abs/1311.2544} {arXiv:1311.2544 [gr-qc]}
  \BibitemShut {NoStop}%
\bibitem [{\citenamefont {Bohé}\ \emph {et~al.}(2017)\citenamefont {Bohé}
  \emph {et~al.}}]{Bohe:2016gbl}%
  \BibitemOpen
  \bibfield  {author} {\bibinfo {author} {\bibfnamefont {A.}~\bibnamefont
  {Bohé}} \emph {et~al.},\ }\href {\doibase 10.1103/PhysRevD.95.044028}
  {\bibfield  {journal} {\bibinfo  {journal} {Phys. Rev.}\ }\textbf {\bibinfo
  {volume} {D95}},\ \bibinfo {pages} {044028} (\bibinfo {year} {2017})},\
  \Eprint {http://arxiv.org/abs/1611.03703} {arXiv:1611.03703 [gr-qc]}
  \BibitemShut {NoStop}%
\bibitem [{\citenamefont {Cotesta}\ \emph {et~al.}(2018)\citenamefont
  {Cotesta}, \citenamefont {Buonanno}, \citenamefont {Bohé}, \citenamefont
  {Taracchini}, \citenamefont {Hinder},\ and\ \citenamefont
  {Ossokine}}]{Cotesta:2018fcv}%
  \BibitemOpen
  \bibfield  {author} {\bibinfo {author} {\bibfnamefont {R.}~\bibnamefont
  {Cotesta}}, \bibinfo {author} {\bibfnamefont {A.}~\bibnamefont {Buonanno}},
  \bibinfo {author} {\bibfnamefont {A.}~\bibnamefont {Bohé}}, \bibinfo
  {author} {\bibfnamefont {A.}~\bibnamefont {Taracchini}}, \bibinfo {author}
  {\bibfnamefont {I.}~\bibnamefont {Hinder}}, \ and\ \bibinfo {author}
  {\bibfnamefont {S.}~\bibnamefont {Ossokine}},\ }\href {\doibase
  10.1103/PhysRevD.98.084028} {\bibfield  {journal} {\bibinfo  {journal} {Phys.
  Rev.}\ }\textbf {\bibinfo {volume} {D98}},\ \bibinfo {pages} {084028}
  (\bibinfo {year} {2018})},\ \Eprint {http://arxiv.org/abs/1803.10701}
  {arXiv:1803.10701 [gr-qc]} \BibitemShut {NoStop}%
\bibitem [{\citenamefont {Abbott}\ \emph
  {et~al.}(2016{\natexlab{a}})\citenamefont {Abbott} \emph
  {et~al.}}]{Abbott:2016blz}%
  \BibitemOpen
  \bibfield  {author} {\bibinfo {author} {\bibfnamefont {B.~P.}\ \bibnamefont
  {Abbott}} \emph {et~al.} (\bibinfo {collaboration} {Virgo, LIGO
  Scientific}),\ }\href {\doibase 10.1103/PhysRevLett.116.061102} {\bibfield
  {journal} {\bibinfo  {journal} {Phys. Rev. Lett.}\ }\textbf {\bibinfo
  {volume} {116}},\ \bibinfo {pages} {061102} (\bibinfo {year}
  {2016}{\natexlab{a}})},\ \Eprint {http://arxiv.org/abs/1602.03837}
  {arXiv:1602.03837 [gr-qc]} \BibitemShut {NoStop}%
\bibitem [{\citenamefont {Abbott}\ \emph
  {et~al.}(2016{\natexlab{b}})\citenamefont {Abbott} \emph
  {et~al.}}]{Abbott:2016nmj}%
  \BibitemOpen
  \bibfield  {author} {\bibinfo {author} {\bibfnamefont {B.~P.}\ \bibnamefont
  {Abbott}} \emph {et~al.} (\bibinfo {collaboration} {Virgo, LIGO
  Scientific}),\ }\href {\doibase 10.1103/PhysRevLett.116.241103} {\bibfield
  {journal} {\bibinfo  {journal} {Phys. Rev. Lett.}\ }\textbf {\bibinfo
  {volume} {116}},\ \bibinfo {pages} {241103} (\bibinfo {year}
  {2016}{\natexlab{b}})},\ \Eprint {http://arxiv.org/abs/1606.04855}
  {arXiv:1606.04855 [gr-qc]} \BibitemShut {NoStop}%
\bibitem [{\citenamefont {Abbott}\ \emph
  {et~al.}(2017{\natexlab{a}})\citenamefont {Abbott} \emph
  {et~al.}}]{Abbott:2017vtc}%
  \BibitemOpen
  \bibfield  {author} {\bibinfo {author} {\bibfnamefont {B.~P.}\ \bibnamefont
  {Abbott}} \emph {et~al.} (\bibinfo {collaboration} {VIRGO, LIGO
  Scientific}),\ }\href {\doibase 10.1103/PhysRevLett.118.221101,
  10.1103/PhysRevLett.121.129901} {\bibfield  {journal} {\bibinfo  {journal}
  {Phys. Rev. Lett.}\ }\textbf {\bibinfo {volume} {118}},\ \bibinfo {pages}
  {221101} (\bibinfo {year} {2017}{\natexlab{a}})},\ \bibinfo {note} {[Erratum:
  Phys. Rev. Lett.121,no.12,129901(2018)]},\ \Eprint
  {http://arxiv.org/abs/1706.01812} {arXiv:1706.01812 [gr-qc]} \BibitemShut
  {NoStop}%
\bibitem [{\citenamefont {Abbott}\ \emph
  {et~al.}(2017{\natexlab{b}})\citenamefont {Abbott} \emph
  {et~al.}}]{Abbott:2017oio}%
  \BibitemOpen
  \bibfield  {author} {\bibinfo {author} {\bibfnamefont {B.~P.}\ \bibnamefont
  {Abbott}} \emph {et~al.} (\bibinfo {collaboration} {Virgo, LIGO
  Scientific}),\ }\href {\doibase 10.1103/PhysRevLett.119.141101} {\bibfield
  {journal} {\bibinfo  {journal} {Phys. Rev. Lett.}\ }\textbf {\bibinfo
  {volume} {119}},\ \bibinfo {pages} {141101} (\bibinfo {year}
  {2017}{\natexlab{b}})},\ \Eprint {http://arxiv.org/abs/1709.09660}
  {arXiv:1709.09660 [gr-qc]} \BibitemShut {NoStop}%
\bibitem [{\citenamefont {Abbott}\ \emph
  {et~al.}(2017{\natexlab{c}})\citenamefont {Abbott} \emph
  {et~al.}}]{Abbott:2017gyy}%
  \BibitemOpen
  \bibfield  {author} {\bibinfo {author} {\bibfnamefont {B.~P.}\ \bibnamefont
  {Abbott}} \emph {et~al.} (\bibinfo {collaboration} {Virgo, LIGO
  Scientific}),\ }\href {\doibase 10.3847/2041-8213/aa9f0c} {\bibfield
  {journal} {\bibinfo  {journal} {Astrophys. J.}\ }\textbf {\bibinfo {volume}
  {851}},\ \bibinfo {pages} {L35} (\bibinfo {year} {2017}{\natexlab{c}})},\
  \Eprint {http://arxiv.org/abs/1711.05578} {arXiv:1711.05578 [astro-ph.HE]}
  \BibitemShut {NoStop}%
\bibitem [{\citenamefont {Abbott}\ \emph
  {et~al.}(2017{\natexlab{d}})\citenamefont {Abbott} \emph
  {et~al.}}]{TheLIGOScientific:2017qsa}%
  \BibitemOpen
  \bibfield  {author} {\bibinfo {author} {\bibfnamefont {B.}~\bibnamefont
  {Abbott}} \emph {et~al.} (\bibinfo {collaboration} {Virgo, LIGO
  Scientific}),\ }\href {\doibase 10.1103/PhysRevLett.119.161101} {\bibfield
  {journal} {\bibinfo  {journal} {Phys. Rev. Lett.}\ }\textbf {\bibinfo
  {volume} {119}},\ \bibinfo {pages} {161101} (\bibinfo {year}
  {2017}{\natexlab{d}})},\ \Eprint {http://arxiv.org/abs/1710.05832}
  {arXiv:1710.05832 [gr-qc]} \BibitemShut {NoStop}%
\bibitem [{LIG(2018)}]{LIGOScientific:2018mvr}%
  \BibitemOpen
  \href@noop {} {\  (\bibinfo {year} {2018})},\ \Eprint
  {http://arxiv.org/abs/1811.12907} {arXiv:1811.12907 [astro-ph.HE]}
  \BibitemShut {NoStop}%
\bibitem [{\citenamefont {Abbott}\ \emph
  {et~al.}(2016{\natexlab{c}})\citenamefont {Abbott} \emph
  {et~al.}}]{TheLIGOScientific:2016wfe}%
  \BibitemOpen
  \bibfield  {author} {\bibinfo {author} {\bibfnamefont {B.~P.}\ \bibnamefont
  {Abbott}} \emph {et~al.} (\bibinfo {collaboration} {Virgo, LIGO
  Scientific}),\ }\href {\doibase 10.1103/PhysRevLett.116.241102} {\bibfield
  {journal} {\bibinfo  {journal} {Phys. Rev. Lett.}\ }\textbf {\bibinfo
  {volume} {116}},\ \bibinfo {pages} {241102} (\bibinfo {year}
  {2016}{\natexlab{c}})},\ \Eprint {http://arxiv.org/abs/1602.03840}
  {arXiv:1602.03840 [gr-qc]} \BibitemShut {NoStop}%
\bibitem [{\citenamefont {Abbott}\ \emph
  {et~al.}(2016{\natexlab{d}})\citenamefont {Abbott} \emph
  {et~al.}}]{TheLIGOScientific:2016src}%
  \BibitemOpen
  \bibfield  {author} {\bibinfo {author} {\bibfnamefont {B.~P.}\ \bibnamefont
  {Abbott}} \emph {et~al.} (\bibinfo {collaboration} {Virgo, LIGO
  Scientific}),\ }\href {\doibase 10.1103/PhysRevLett.116.221101} {\bibfield
  {journal} {\bibinfo  {journal} {Phys. Rev. Lett.}\ }\textbf {\bibinfo
  {volume} {116}},\ \bibinfo {pages} {221101} (\bibinfo {year}
  {2016}{\natexlab{d}})},\ \Eprint {http://arxiv.org/abs/1602.03841}
  {arXiv:1602.03841 [gr-qc]} \BibitemShut {NoStop}%
\bibitem [{\citenamefont {Audley}\ \emph {et~al.}(2017)\citenamefont {Audley}
  \emph {et~al.}}]{Audley:2017drz}%
  \BibitemOpen
  \bibfield  {author} {\bibinfo {author} {\bibfnamefont {H.}~\bibnamefont
  {Audley}} \emph {et~al.} (\bibinfo {collaboration} {LISA}),\ }\href@noop {}
  {\  (\bibinfo {year} {2017})},\ \Eprint {http://arxiv.org/abs/1702.00786}
  {arXiv:1702.00786 [astro-ph.IM]} \BibitemShut {NoStop}%
\bibitem [{\citenamefont {Punturo}\ \emph {et~al.}(2010)\citenamefont {Punturo}
  \emph {et~al.}}]{Punturo:2010zz}%
  \BibitemOpen
  \bibfield  {author} {\bibinfo {author} {\bibfnamefont {M.}~\bibnamefont
  {Punturo}} \emph {et~al.},\ }\bibfield  {booktitle} {\emph {\bibinfo
  {booktitle} {{Proceedings, 14th Workshop on Gravitational wave data analysis
  (GWDAW-14): Rome, Italy, January 26-29, 2010}}},\ }\href {\doibase
  10.1088/0264-9381/27/19/194002} {\bibfield  {journal} {\bibinfo  {journal}
  {Class. Quant. Grav.}\ }\textbf {\bibinfo {volume} {27}},\ \bibinfo {pages}
  {194002} (\bibinfo {year} {2010})}\BibitemShut {NoStop}%
\bibitem [{\citenamefont {Abbott}\ \emph
  {et~al.}(2017{\natexlab{e}})\citenamefont {Abbott} \emph
  {et~al.}}]{Evans:2016mbw}%
  \BibitemOpen
  \bibfield  {author} {\bibinfo {author} {\bibfnamefont {B.~P.}\ \bibnamefont
  {Abbott}} \emph {et~al.} (\bibinfo {collaboration} {LIGO Scientific}),\
  }\href {\doibase 10.1088/1361-6382/aa51f4} {\bibfield  {journal} {\bibinfo
  {journal} {Class. Quant. Grav.}\ }\textbf {\bibinfo {volume} {34}},\ \bibinfo
  {pages} {044001} (\bibinfo {year} {2017}{\natexlab{e}})},\ \Eprint
  {http://arxiv.org/abs/1607.08697} {arXiv:1607.08697 [astro-ph.IM]}
  \BibitemShut {NoStop}%
\bibitem [{\citenamefont {Buonanno}\ and\ \citenamefont
  {Damour}(1999)}]{Buonanno:1998gg}%
  \BibitemOpen
  \bibfield  {author} {\bibinfo {author} {\bibfnamefont {A.}~\bibnamefont
  {Buonanno}}\ and\ \bibinfo {author} {\bibfnamefont {T.}~\bibnamefont
  {Damour}},\ }\href {\doibase 10.1103/PhysRevD.59.084006} {\bibfield
  {journal} {\bibinfo  {journal} {Phys. Rev.}\ }\textbf {\bibinfo {volume}
  {D59}},\ \bibinfo {pages} {084006} (\bibinfo {year} {1999})},\ \Eprint
  {http://arxiv.org/abs/gr-qc/9811091} {arXiv:gr-qc/9811091 [gr-qc]}
  \BibitemShut {NoStop}%
\bibitem [{\citenamefont {Buonanno}\ and\ \citenamefont
  {Damour}(2000)}]{Buonanno:2000ef}%
  \BibitemOpen
  \bibfield  {author} {\bibinfo {author} {\bibfnamefont {A.}~\bibnamefont
  {Buonanno}}\ and\ \bibinfo {author} {\bibfnamefont {T.}~\bibnamefont
  {Damour}},\ }\href {\doibase 10.1103/PhysRevD.62.064015} {\bibfield
  {journal} {\bibinfo  {journal} {Phys. Rev.}\ }\textbf {\bibinfo {volume}
  {D62}},\ \bibinfo {pages} {064015} (\bibinfo {year} {2000})},\ \Eprint
  {http://arxiv.org/abs/gr-qc/0001013} {arXiv:gr-qc/0001013 [gr-qc]}
  \BibitemShut {NoStop}%
\bibitem [{\citenamefont {Damour}\ \emph {et~al.}(2000)\citenamefont {Damour},
  \citenamefont {Jaranowski},\ and\ \citenamefont {Schäfer}}]{Damour:2000we}%
  \BibitemOpen
  \bibfield  {author} {\bibinfo {author} {\bibfnamefont {T.}~\bibnamefont
  {Damour}}, \bibinfo {author} {\bibfnamefont {P.}~\bibnamefont {Jaranowski}},
  \ and\ \bibinfo {author} {\bibfnamefont {G.}~\bibnamefont {Schäfer}},\
  }\href {\doibase 10.1103/PhysRevD.62.084011} {\bibfield  {journal} {\bibinfo
  {journal} {Phys. Rev.}\ }\textbf {\bibinfo {volume} {D62}},\ \bibinfo {pages}
  {084011} (\bibinfo {year} {2000})},\ \Eprint
  {http://arxiv.org/abs/gr-qc/0005034} {arXiv:gr-qc/0005034 [gr-qc]}
  \BibitemShut {NoStop}%
\bibitem [{\citenamefont {Jaranowski}\ and\ \citenamefont
  {Schäfer}(2015)}]{Jaranowski:2015lha}%
  \BibitemOpen
  \bibfield  {author} {\bibinfo {author} {\bibfnamefont {P.}~\bibnamefont
  {Jaranowski}}\ and\ \bibinfo {author} {\bibfnamefont {G.}~\bibnamefont
  {Schäfer}},\ }\href {\doibase 10.1103/PhysRevD.92.124043} {\bibfield
  {journal} {\bibinfo  {journal} {Phys. Rev.}\ }\textbf {\bibinfo {volume}
  {D92}},\ \bibinfo {pages} {124043} (\bibinfo {year} {2015})},\ \Eprint
  {http://arxiv.org/abs/1508.01016} {arXiv:1508.01016 [gr-qc]} \BibitemShut
  {NoStop}%
\bibitem [{\citenamefont {Damour}\ \emph {et~al.}(2014)\citenamefont {Damour},
  \citenamefont {Jaranowski},\ and\ \citenamefont {Schäfer}}]{Damour:2014jta}%
  \BibitemOpen
  \bibfield  {author} {\bibinfo {author} {\bibfnamefont {T.}~\bibnamefont
  {Damour}}, \bibinfo {author} {\bibfnamefont {P.}~\bibnamefont {Jaranowski}},
  \ and\ \bibinfo {author} {\bibfnamefont {G.}~\bibnamefont {Schäfer}},\
  }\href {\doibase 10.1103/PhysRevD.89.064058} {\bibfield  {journal} {\bibinfo
  {journal} {Phys. Rev.}\ }\textbf {\bibinfo {volume} {D89}},\ \bibinfo {pages}
  {064058} (\bibinfo {year} {2014})},\ \Eprint {http://arxiv.org/abs/1401.4548}
  {arXiv:1401.4548 [gr-qc]} \BibitemShut {NoStop}%
\bibitem [{\citenamefont {Damour}\ \emph {et~al.}(2016)\citenamefont {Damour},
  \citenamefont {Jaranowski},\ and\ \citenamefont {Schäfer}}]{Damour:2016abl}%
  \BibitemOpen
  \bibfield  {author} {\bibinfo {author} {\bibfnamefont {T.}~\bibnamefont
  {Damour}}, \bibinfo {author} {\bibfnamefont {P.}~\bibnamefont {Jaranowski}},
  \ and\ \bibinfo {author} {\bibfnamefont {G.}~\bibnamefont {Schäfer}},\
  }\href {\doibase 10.1103/PhysRevD.93.084014} {\bibfield  {journal} {\bibinfo
  {journal} {Phys. Rev.}\ }\textbf {\bibinfo {volume} {D93}},\ \bibinfo {pages}
  {084014} (\bibinfo {year} {2016})},\ \Eprint
  {http://arxiv.org/abs/1601.01283} {arXiv:1601.01283 [gr-qc]} \BibitemShut
  {NoStop}%
\bibitem [{\citenamefont {Bernard}\ \emph {et~al.}(2016)\citenamefont
  {Bernard}, \citenamefont {Blanchet}, \citenamefont {Bohé}, \citenamefont
  {Faye},\ and\ \citenamefont {Marsat}}]{Bernard:2015njp}%
  \BibitemOpen
  \bibfield  {author} {\bibinfo {author} {\bibfnamefont {L.}~\bibnamefont
  {Bernard}}, \bibinfo {author} {\bibfnamefont {L.}~\bibnamefont {Blanchet}},
  \bibinfo {author} {\bibfnamefont {A.}~\bibnamefont {Bohé}}, \bibinfo
  {author} {\bibfnamefont {G.}~\bibnamefont {Faye}}, \ and\ \bibinfo {author}
  {\bibfnamefont {S.}~\bibnamefont {Marsat}},\ }\href {\doibase
  10.1103/PhysRevD.93.084037} {\bibfield  {journal} {\bibinfo  {journal} {Phys.
  Rev.}\ }\textbf {\bibinfo {volume} {D93}},\ \bibinfo {pages} {084037}
  (\bibinfo {year} {2016})},\ \Eprint {http://arxiv.org/abs/1512.02876}
  {arXiv:1512.02876 [gr-qc]} \BibitemShut {NoStop}%
\bibitem [{\citenamefont {Bernard}\ \emph
  {et~al.}(2017{\natexlab{a}})\citenamefont {Bernard}, \citenamefont
  {Blanchet}, \citenamefont {Bohé}, \citenamefont {Faye},\ and\ \citenamefont
  {Marsat}}]{Bernard:2016wrg}%
  \BibitemOpen
  \bibfield  {author} {\bibinfo {author} {\bibfnamefont {L.}~\bibnamefont
  {Bernard}}, \bibinfo {author} {\bibfnamefont {L.}~\bibnamefont {Blanchet}},
  \bibinfo {author} {\bibfnamefont {A.}~\bibnamefont {Bohé}}, \bibinfo
  {author} {\bibfnamefont {G.}~\bibnamefont {Faye}}, \ and\ \bibinfo {author}
  {\bibfnamefont {S.}~\bibnamefont {Marsat}},\ }\href {\doibase
  10.1103/PhysRevD.95.044026} {\bibfield  {journal} {\bibinfo  {journal} {Phys.
  Rev.}\ }\textbf {\bibinfo {volume} {D95}},\ \bibinfo {pages} {044026}
  (\bibinfo {year} {2017}{\natexlab{a}})},\ \Eprint
  {http://arxiv.org/abs/1610.07934} {arXiv:1610.07934 [gr-qc]} \BibitemShut
  {NoStop}%
\bibitem [{\citenamefont {Bernard}\ \emph
  {et~al.}(2017{\natexlab{b}})\citenamefont {Bernard}, \citenamefont
  {Blanchet}, \citenamefont {Bohé}, \citenamefont {Faye},\ and\ \citenamefont
  {Marsat}}]{Bernard:2017bvn}%
  \BibitemOpen
  \bibfield  {author} {\bibinfo {author} {\bibfnamefont {L.}~\bibnamefont
  {Bernard}}, \bibinfo {author} {\bibfnamefont {L.}~\bibnamefont {Blanchet}},
  \bibinfo {author} {\bibfnamefont {A.}~\bibnamefont {Bohé}}, \bibinfo
  {author} {\bibfnamefont {G.}~\bibnamefont {Faye}}, \ and\ \bibinfo {author}
  {\bibfnamefont {S.}~\bibnamefont {Marsat}},\ }\href {\doibase
  10.1103/PhysRevD.96.104043} {\bibfield  {journal} {\bibinfo  {journal} {Phys.
  Rev.}\ }\textbf {\bibinfo {volume} {D96}},\ \bibinfo {pages} {104043}
  (\bibinfo {year} {2017}{\natexlab{b}})},\ \Eprint
  {http://arxiv.org/abs/1706.08480} {arXiv:1706.08480 [gr-qc]} \BibitemShut
  {NoStop}%
\bibitem [{\citenamefont {Foffa}\ and\ \citenamefont
  {Sturani}(2019)}]{Foffa:2019rdf}%
  \BibitemOpen
  \bibfield  {author} {\bibinfo {author} {\bibfnamefont {S.}~\bibnamefont
  {Foffa}}\ and\ \bibinfo {author} {\bibfnamefont {R.}~\bibnamefont
  {Sturani}},\ }\href@noop {} {\  (\bibinfo {year} {2019})},\ \Eprint
  {http://arxiv.org/abs/1903.05113} {arXiv:1903.05113 [gr-qc]} \BibitemShut
  {NoStop}%
\bibitem [{\citenamefont {Foffa}\ \emph
  {et~al.}(2019{\natexlab{a}})\citenamefont {Foffa}, \citenamefont {Porto},
  \citenamefont {Rothstein},\ and\ \citenamefont {Sturani}}]{Foffa:2019yfl}%
  \BibitemOpen
  \bibfield  {author} {\bibinfo {author} {\bibfnamefont {S.}~\bibnamefont
  {Foffa}}, \bibinfo {author} {\bibfnamefont {R.~A.}\ \bibnamefont {Porto}},
  \bibinfo {author} {\bibfnamefont {I.}~\bibnamefont {Rothstein}}, \ and\
  \bibinfo {author} {\bibfnamefont {R.}~\bibnamefont {Sturani}},\ }\href@noop
  {} {\  (\bibinfo {year} {2019}{\natexlab{a}})},\ \Eprint
  {http://arxiv.org/abs/1903.05118} {arXiv:1903.05118 [gr-qc]} \BibitemShut
  {NoStop}%
\bibitem [{\citenamefont {Foffa}\ \emph
  {et~al.}(2019{\natexlab{b}})\citenamefont {Foffa}, \citenamefont {Mastrolia},
  \citenamefont {Sturani}, \citenamefont {Sturm},\ and\ \citenamefont
  {Torres~Bobadilla}}]{Foffa:2019hrb}%
  \BibitemOpen
  \bibfield  {author} {\bibinfo {author} {\bibfnamefont {S.}~\bibnamefont
  {Foffa}}, \bibinfo {author} {\bibfnamefont {P.}~\bibnamefont {Mastrolia}},
  \bibinfo {author} {\bibfnamefont {R.}~\bibnamefont {Sturani}}, \bibinfo
  {author} {\bibfnamefont {C.}~\bibnamefont {Sturm}}, \ and\ \bibinfo {author}
  {\bibfnamefont {W.~J.}\ \bibnamefont {Torres~Bobadilla}},\ }\href {\doibase
  10.1103/PhysRevLett.122.241605} {\bibfield  {journal} {\bibinfo  {journal}
  {Phys. Rev. Lett.}\ }\textbf {\bibinfo {volume} {122}},\ \bibinfo {pages}
  {241605} (\bibinfo {year} {2019}{\natexlab{b}})},\ \Eprint
  {http://arxiv.org/abs/1902.10571} {arXiv:1902.10571 [gr-qc]} \BibitemShut
  {NoStop}%
\bibitem [{\citenamefont {Blümlein}\ \emph {et~al.}(2019)\citenamefont
  {Blümlein}, \citenamefont {Maier},\ and\ \citenamefont
  {Marquard}}]{Blumlein:2019zku}%
  \BibitemOpen
  \bibfield  {author} {\bibinfo {author} {\bibfnamefont {J.}~\bibnamefont
  {Blümlein}}, \bibinfo {author} {\bibfnamefont {A.}~\bibnamefont {Maier}}, \
  and\ \bibinfo {author} {\bibfnamefont {P.}~\bibnamefont {Marquard}},\
  }\href@noop {} {\  (\bibinfo {year} {2019})},\ \Eprint
  {http://arxiv.org/abs/1902.11180} {arXiv:1902.11180 [gr-qc]} \BibitemShut
  {NoStop}%
\bibitem [{\citenamefont {Damour}\ \emph {et~al.}(2015)\citenamefont {Damour},
  \citenamefont {Jaranowski},\ and\ \citenamefont {Schäfer}}]{Damour:2015isa}%
  \BibitemOpen
  \bibfield  {author} {\bibinfo {author} {\bibfnamefont {T.}~\bibnamefont
  {Damour}}, \bibinfo {author} {\bibfnamefont {P.}~\bibnamefont {Jaranowski}},
  \ and\ \bibinfo {author} {\bibfnamefont {G.}~\bibnamefont {Schäfer}},\
  }\href {\doibase 10.1103/PhysRevD.91.084024} {\bibfield  {journal} {\bibinfo
  {journal} {Phys. Rev.}\ }\textbf {\bibinfo {volume} {D91}},\ \bibinfo {pages}
  {084024} (\bibinfo {year} {2015})},\ \Eprint
  {http://arxiv.org/abs/1502.07245} {arXiv:1502.07245 [gr-qc]} \BibitemShut
  {NoStop}%
\bibitem [{\citenamefont {Damour}\ \emph {et~al.}(2009)\citenamefont {Damour},
  \citenamefont {Iyer},\ and\ \citenamefont {Nagar}}]{Damour:2008gu}%
  \BibitemOpen
  \bibfield  {author} {\bibinfo {author} {\bibfnamefont {T.}~\bibnamefont
  {Damour}}, \bibinfo {author} {\bibfnamefont {B.~R.}\ \bibnamefont {Iyer}}, \
  and\ \bibinfo {author} {\bibfnamefont {A.}~\bibnamefont {Nagar}},\ }\href
  {\doibase 10.1103/PhysRevD.79.064004} {\bibfield  {journal} {\bibinfo
  {journal} {Phys. Rev.}\ }\textbf {\bibinfo {volume} {D79}},\ \bibinfo {pages}
  {064004} (\bibinfo {year} {2009})},\ \Eprint {http://arxiv.org/abs/0811.2069}
  {arXiv:0811.2069 [gr-qc]} \BibitemShut {NoStop}%
\bibitem [{\citenamefont {Damour}\ and\ \citenamefont
  {Nagar}(2009)}]{Damour:2009kr}%
  \BibitemOpen
  \bibfield  {author} {\bibinfo {author} {\bibfnamefont {T.}~\bibnamefont
  {Damour}}\ and\ \bibinfo {author} {\bibfnamefont {A.}~\bibnamefont {Nagar}},\
  }\href {\doibase 10.1103/PhysRevD.79.081503} {\bibfield  {journal} {\bibinfo
  {journal} {Phys. Rev.}\ }\textbf {\bibinfo {volume} {D79}},\ \bibinfo {pages}
  {081503} (\bibinfo {year} {2009})},\ \Eprint {http://arxiv.org/abs/0902.0136}
  {arXiv:0902.0136 [gr-qc]} \BibitemShut {NoStop}%
\bibitem [{\citenamefont {Barausse}\ and\ \citenamefont
  {Buonanno}(2010)}]{Barausse:2009xi}%
  \BibitemOpen
  \bibfield  {author} {\bibinfo {author} {\bibfnamefont {E.}~\bibnamefont
  {Barausse}}\ and\ \bibinfo {author} {\bibfnamefont {A.}~\bibnamefont
  {Buonanno}},\ }\href {\doibase 10.1103/PhysRevD.81.084024} {\bibfield
  {journal} {\bibinfo  {journal} {Phys. Rev.}\ }\textbf {\bibinfo {volume}
  {D81}},\ \bibinfo {pages} {084024} (\bibinfo {year} {2010})},\ \Eprint
  {http://arxiv.org/abs/0912.3517} {arXiv:0912.3517 [gr-qc]} \BibitemShut
  {NoStop}%
\bibitem [{\citenamefont {Hinderer}\ \emph {et~al.}(2016)\citenamefont
  {Hinderer} \emph {et~al.}}]{Hinderer:2016eia}%
  \BibitemOpen
  \bibfield  {author} {\bibinfo {author} {\bibfnamefont {T.}~\bibnamefont
  {Hinderer}} \emph {et~al.},\ }\href {\doibase 10.1103/PhysRevLett.116.181101}
  {\bibfield  {journal} {\bibinfo  {journal} {Phys. Rev. Lett.}\ }\textbf
  {\bibinfo {volume} {116}},\ \bibinfo {pages} {181101} (\bibinfo {year}
  {2016})},\ \Eprint {http://arxiv.org/abs/1602.00599} {arXiv:1602.00599
  [gr-qc]} \BibitemShut {NoStop}%
\bibitem [{\citenamefont {Damour}(2018)}]{Damour:2017zjx}%
  \BibitemOpen
  \bibfield  {author} {\bibinfo {author} {\bibfnamefont {T.}~\bibnamefont
  {Damour}},\ }\href {\doibase 10.1103/PhysRevD.97.044038} {\bibfield
  {journal} {\bibinfo  {journal} {Phys. Rev.}\ }\textbf {\bibinfo {volume}
  {D97}},\ \bibinfo {pages} {044038} (\bibinfo {year} {2018})},\ \Eprint
  {http://arxiv.org/abs/1710.10599} {arXiv:1710.10599 [gr-qc]} \BibitemShut
  {NoStop}%
\bibitem [{\citenamefont {Damour}(2016)}]{Damour:2016pm}%
  \BibitemOpen
  \bibfield  {author} {\bibinfo {author} {\bibfnamefont {T.}~\bibnamefont
  {Damour}},\ }\href {\doibase 10.1103/PhysRevD.94.104015} {\bibfield
  {journal} {\bibinfo  {journal} {Phys.Rev.}\ }\textbf {\bibinfo {volume}
  {D94}},\ \bibinfo {pages} {104015} (\bibinfo {year} {2016})},\ \Eprint
  {http://arxiv.org/abs/1609.00354} {arXiv:1609.00354 [gr-qc]} \BibitemShut
  {NoStop}%
\bibitem [{\citenamefont {Vines}(2018)}]{Vines:2017pm}%
  \BibitemOpen
  \bibfield  {author} {\bibinfo {author} {\bibfnamefont {J.}~\bibnamefont
  {Vines}},\ }\href {\doibase 10.1088/1361-6382/aaa3a8} {\bibfield  {journal}
  {\bibinfo  {journal} {Class.Quant.Grav.}\ }\textbf {\bibinfo {volume} {35}},\
  \bibinfo {pages} {084002} (\bibinfo {year} {2018})},\ \Eprint
  {http://arxiv.org/abs/1709.06016} {arXiv:1709.06016 [gr-qc]} \BibitemShut
  {NoStop}%
\bibitem [{\citenamefont {Vines}\ \emph {et~al.}(2018)\citenamefont {Vines},
  \citenamefont {Steinhoff},\ and\ \citenamefont {Buonanno}}]{Vines:2018gqi}%
  \BibitemOpen
  \bibfield  {author} {\bibinfo {author} {\bibfnamefont {J.}~\bibnamefont
  {Vines}}, \bibinfo {author} {\bibfnamefont {J.}~\bibnamefont {Steinhoff}}, \
  and\ \bibinfo {author} {\bibfnamefont {A.}~\bibnamefont {Buonanno}},\
  }\href@noop {} {\  (\bibinfo {year} {2018})},\ \Eprint
  {http://arxiv.org/abs/1812.00956} {arXiv:1812.00956 [gr-qc]} \BibitemShut
  {NoStop}%
\bibitem [{\citenamefont {Antonelli}\ \emph {et~al.}(2019)\citenamefont
  {Antonelli}, \citenamefont {Buonanno}, \citenamefont {Steinhoff},
  \citenamefont {van~de Meent},\ and\ \citenamefont
  {Vines}}]{Antonelli:2019ytb}%
  \BibitemOpen
  \bibfield  {author} {\bibinfo {author} {\bibfnamefont {A.}~\bibnamefont
  {Antonelli}}, \bibinfo {author} {\bibfnamefont {A.}~\bibnamefont {Buonanno}},
  \bibinfo {author} {\bibfnamefont {J.}~\bibnamefont {Steinhoff}}, \bibinfo
  {author} {\bibfnamefont {M.}~\bibnamefont {van~de Meent}}, \ and\ \bibinfo
  {author} {\bibfnamefont {J.}~\bibnamefont {Vines}},\ }\href {\doibase
  10.1103/PhysRevD.99.104004} {\bibfield  {journal} {\bibinfo  {journal} {Phys.
  Rev.}\ }\textbf {\bibinfo {volume} {D99}},\ \bibinfo {pages} {104004}
  (\bibinfo {year} {2019})},\ \Eprint {http://arxiv.org/abs/1901.07102}
  {arXiv:1901.07102 [gr-qc]} \BibitemShut {NoStop}%
\bibitem [{\citenamefont {Mino}\ \emph {et~al.}(1997)\citenamefont {Mino},
  \citenamefont {Sasaki},\ and\ \citenamefont {Tanaka}}]{Mino:1996nk}%
  \BibitemOpen
  \bibfield  {author} {\bibinfo {author} {\bibfnamefont {Y.}~\bibnamefont
  {Mino}}, \bibinfo {author} {\bibfnamefont {M.}~\bibnamefont {Sasaki}}, \ and\
  \bibinfo {author} {\bibfnamefont {T.}~\bibnamefont {Tanaka}},\ }\href
  {\doibase 10.1103/PhysRevD.55.3457} {\bibfield  {journal} {\bibinfo
  {journal} {Phys. Rev.}\ }\textbf {\bibinfo {volume} {D55}},\ \bibinfo {pages}
  {3457} (\bibinfo {year} {1997})},\ \Eprint
  {http://arxiv.org/abs/gr-qc/9606018} {arXiv:gr-qc/9606018 [gr-qc]}
  \BibitemShut {NoStop}%
\bibitem [{\citenamefont {Quinn}\ and\ \citenamefont
  {Wald}(1997)}]{Quinn:1996am}%
  \BibitemOpen
  \bibfield  {author} {\bibinfo {author} {\bibfnamefont {T.~C.}\ \bibnamefont
  {Quinn}}\ and\ \bibinfo {author} {\bibfnamefont {R.~M.}\ \bibnamefont
  {Wald}},\ }\href {\doibase 10.1103/PhysRevD.56.3381} {\bibfield  {journal}
  {\bibinfo  {journal} {Phys. Rev.}\ }\textbf {\bibinfo {volume} {D56}},\
  \bibinfo {pages} {3381} (\bibinfo {year} {1997})},\ \Eprint
  {http://arxiv.org/abs/gr-qc/9610053} {arXiv:gr-qc/9610053 [gr-qc]}
  \BibitemShut {NoStop}%
\bibitem [{\citenamefont {Barack}\ and\ \citenamefont
  {Sago}(2007)}]{Barack:2007tm}%
  \BibitemOpen
  \bibfield  {author} {\bibinfo {author} {\bibfnamefont {L.}~\bibnamefont
  {Barack}}\ and\ \bibinfo {author} {\bibfnamefont {N.}~\bibnamefont {Sago}},\
  }\href {\doibase 10.1103/PhysRevD.75.064021} {\bibfield  {journal} {\bibinfo
  {journal} {Phys. Rev.}\ }\textbf {\bibinfo {volume} {D75}},\ \bibinfo {pages}
  {064021} (\bibinfo {year} {2007})},\ \Eprint
  {http://arxiv.org/abs/gr-qc/0701069} {arXiv:gr-qc/0701069 [gr-qc]}
  \BibitemShut {NoStop}%
\bibitem [{\citenamefont {Barack}\ and\ \citenamefont
  {Sago}(2010)}]{Barack:2010tm}%
  \BibitemOpen
  \bibfield  {author} {\bibinfo {author} {\bibfnamefont {L.}~\bibnamefont
  {Barack}}\ and\ \bibinfo {author} {\bibfnamefont {N.}~\bibnamefont {Sago}},\
  }\href {\doibase 10.1103/PhysRevD.81.084021} {\bibfield  {journal} {\bibinfo
  {journal} {Phys. Rev.}\ }\textbf {\bibinfo {volume} {D81}},\ \bibinfo {pages}
  {084021} (\bibinfo {year} {2010})},\ \Eprint {http://arxiv.org/abs/1002.2386}
  {arXiv:1002.2386 [gr-qc]} \BibitemShut {NoStop}%
\bibitem [{\citenamefont {Shah}\ \emph {et~al.}(2012)\citenamefont {Shah},
  \citenamefont {Friedman},\ and\ \citenamefont {Keidl}}]{Shah:2012gu}%
  \BibitemOpen
  \bibfield  {author} {\bibinfo {author} {\bibfnamefont {A.~G.}\ \bibnamefont
  {Shah}}, \bibinfo {author} {\bibfnamefont {J.~L.}\ \bibnamefont {Friedman}},
  \ and\ \bibinfo {author} {\bibfnamefont {T.~S.}\ \bibnamefont {Keidl}},\
  }\href {\doibase 10.1103/PhysRevD.86.084059} {\bibfield  {journal} {\bibinfo
  {journal} {Phys. Rev.}\ }\textbf {\bibinfo {volume} {D86}},\ \bibinfo {pages}
  {084059} (\bibinfo {year} {2012})},\ \Eprint {http://arxiv.org/abs/1207.5595}
  {arXiv:1207.5595 [gr-qc]} \BibitemShut {NoStop}%
\bibitem [{\citenamefont {van~de Meent}\ and\ \citenamefont
  {Shah}(2015)}]{vandeMeent:2015lxa}%
  \BibitemOpen
  \bibfield  {author} {\bibinfo {author} {\bibfnamefont {M.}~\bibnamefont
  {van~de Meent}}\ and\ \bibinfo {author} {\bibfnamefont {A.~G.}\ \bibnamefont
  {Shah}},\ }\href {\doibase 10.1103/PhysRevD.92.064025} {\bibfield  {journal}
  {\bibinfo  {journal} {Phys. Rev.}\ }\textbf {\bibinfo {volume} {D92}},\
  \bibinfo {pages} {064025} (\bibinfo {year} {2015})},\ \Eprint
  {http://arxiv.org/abs/1506.04755} {arXiv:1506.04755 [gr-qc]} \BibitemShut
  {NoStop}%
\bibitem [{\citenamefont {van~de Meent}(2016)}]{vandeMeent:2016pee}%
  \BibitemOpen
  \bibfield  {author} {\bibinfo {author} {\bibfnamefont {M.}~\bibnamefont
  {van~de Meent}},\ }\href {\doibase 10.1103/PhysRevD.94.044034} {\bibfield
  {journal} {\bibinfo  {journal} {Phys. Rev.}\ }\textbf {\bibinfo {volume}
  {D94}},\ \bibinfo {pages} {044034} (\bibinfo {year} {2016})},\ \Eprint
  {http://arxiv.org/abs/1606.06297} {arXiv:1606.06297 [gr-qc]} \BibitemShut
  {NoStop}%
\bibitem [{\citenamefont {van~de Meent}(2018)}]{vandeMeent:2017bcc}%
  \BibitemOpen
  \bibfield  {author} {\bibinfo {author} {\bibfnamefont {M.}~\bibnamefont
  {van~de Meent}},\ }\href {\doibase 10.1103/PhysRevD.97.104033} {\bibfield
  {journal} {\bibinfo  {journal} {Phys. Rev.}\ }\textbf {\bibinfo {volume}
  {D97}},\ \bibinfo {pages} {104033} (\bibinfo {year} {2018})},\ \Eprint
  {http://arxiv.org/abs/1711.09607} {arXiv:1711.09607 [gr-qc]} \BibitemShut
  {NoStop}%
\bibitem [{\citenamefont {Warburton}\ \emph {et~al.}(2012)\citenamefont
  {Warburton}, \citenamefont {Akcay}, \citenamefont {Barack}, \citenamefont
  {Gair},\ and\ \citenamefont {Sago}}]{Warburton:2011fk}%
  \BibitemOpen
  \bibfield  {author} {\bibinfo {author} {\bibfnamefont {N.}~\bibnamefont
  {Warburton}}, \bibinfo {author} {\bibfnamefont {S.}~\bibnamefont {Akcay}},
  \bibinfo {author} {\bibfnamefont {L.}~\bibnamefont {Barack}}, \bibinfo
  {author} {\bibfnamefont {J.~R.}\ \bibnamefont {Gair}}, \ and\ \bibinfo
  {author} {\bibfnamefont {N.}~\bibnamefont {Sago}},\ }\href {\doibase
  10.1103/PhysRevD.85.061501} {\bibfield  {journal} {\bibinfo  {journal} {Phys.
  Rev.}\ }\textbf {\bibinfo {volume} {D85}},\ \bibinfo {pages} {061501}
  (\bibinfo {year} {2012})},\ \Eprint {http://arxiv.org/abs/1111.6908}
  {arXiv:1111.6908 [gr-qc]} \BibitemShut {NoStop}%
\bibitem [{\citenamefont {Osburn}\ \emph {et~al.}(2016)\citenamefont {Osburn},
  \citenamefont {Warburton},\ and\ \citenamefont {Evans}}]{Osburn:2015duj}%
  \BibitemOpen
  \bibfield  {author} {\bibinfo {author} {\bibfnamefont {T.}~\bibnamefont
  {Osburn}}, \bibinfo {author} {\bibfnamefont {N.}~\bibnamefont {Warburton}}, \
  and\ \bibinfo {author} {\bibfnamefont {C.~R.}\ \bibnamefont {Evans}},\ }\href
  {\doibase 10.1103/PhysRevD.93.064024} {\bibfield  {journal} {\bibinfo
  {journal} {Phys. Rev.}\ }\textbf {\bibinfo {volume} {D93}},\ \bibinfo {pages}
  {064024} (\bibinfo {year} {2016})},\ \Eprint
  {http://arxiv.org/abs/1511.01498} {arXiv:1511.01498 [gr-qc]} \BibitemShut
  {NoStop}%
\bibitem [{\citenamefont {van~de Meent}\ and\ \citenamefont
  {Warburton}(2018)}]{vandeMeent:2018rms}%
  \BibitemOpen
  \bibfield  {author} {\bibinfo {author} {\bibfnamefont {M.}~\bibnamefont
  {van~de Meent}}\ and\ \bibinfo {author} {\bibfnamefont {N.}~\bibnamefont
  {Warburton}},\ }\href {\doibase 10.1088/1361-6382/aac8ce} {\bibfield
  {journal} {\bibinfo  {journal} {Class. Quant. Grav.}\ }\textbf {\bibinfo
  {volume} {35}},\ \bibinfo {pages} {144003} (\bibinfo {year} {2018})},\
  \Eprint {http://arxiv.org/abs/1802.05281} {arXiv:1802.05281 [gr-qc]}
  \BibitemShut {NoStop}%
\bibitem [{\citenamefont {Hinderer}\ and\ \citenamefont
  {Flanagan}(2008)}]{Hinderer:2008dm}%
  \BibitemOpen
  \bibfield  {author} {\bibinfo {author} {\bibfnamefont {T.}~\bibnamefont
  {Hinderer}}\ and\ \bibinfo {author} {\bibfnamefont {E.~E.}\ \bibnamefont
  {Flanagan}},\ }\href {\doibase 10.1103/PhysRevD.78.064028} {\bibfield
  {journal} {\bibinfo  {journal} {Phys. Rev.}\ }\textbf {\bibinfo {volume}
  {D78}},\ \bibinfo {pages} {064028} (\bibinfo {year} {2008})},\ \Eprint
  {http://arxiv.org/abs/0805.3337} {arXiv:0805.3337 [gr-qc]} \BibitemShut
  {NoStop}%
\bibitem [{\citenamefont {Barack}\ and\ \citenamefont
  {Sago}(2009)}]{Barack:2009ey}%
  \BibitemOpen
  \bibfield  {author} {\bibinfo {author} {\bibfnamefont {L.}~\bibnamefont
  {Barack}}\ and\ \bibinfo {author} {\bibfnamefont {N.}~\bibnamefont {Sago}},\
  }\href {\doibase 10.1103/PhysRevLett.102.191101} {\bibfield  {journal}
  {\bibinfo  {journal} {Phys. Rev. Lett.}\ }\textbf {\bibinfo {volume} {102}},\
  \bibinfo {pages} {191101} (\bibinfo {year} {2009})},\ \Eprint
  {http://arxiv.org/abs/0902.0573} {arXiv:0902.0573 [gr-qc]} \BibitemShut
  {NoStop}%
\bibitem [{\citenamefont {Barack}\ \emph {et~al.}(2010)\citenamefont {Barack},
  \citenamefont {Damour},\ and\ \citenamefont {Sago}}]{Barack:2010ny}%
  \BibitemOpen
  \bibfield  {author} {\bibinfo {author} {\bibfnamefont {L.}~\bibnamefont
  {Barack}}, \bibinfo {author} {\bibfnamefont {T.}~\bibnamefont {Damour}}, \
  and\ \bibinfo {author} {\bibfnamefont {N.}~\bibnamefont {Sago}},\ }\href
  {\doibase 10.1103/PhysRevD.82.084036} {\bibfield  {journal} {\bibinfo
  {journal} {Phys. Rev.}\ }\textbf {\bibinfo {volume} {D82}},\ \bibinfo {pages}
  {084036} (\bibinfo {year} {2010})},\ \Eprint {http://arxiv.org/abs/1008.0935}
  {arXiv:1008.0935 [gr-qc]} \BibitemShut {NoStop}%
\bibitem [{\citenamefont {Le~Tiec}\ \emph {et~al.}(2011)\citenamefont
  {Le~Tiec}, \citenamefont {Mroue}, \citenamefont {Barack}, \citenamefont
  {Buonanno}, \citenamefont {Pfeiffer}, \citenamefont {Sago},\ and\
  \citenamefont {Taracchini}}]{LeTiec:2011bk}%
  \BibitemOpen
  \bibfield  {author} {\bibinfo {author} {\bibfnamefont {A.}~\bibnamefont
  {Le~Tiec}}, \bibinfo {author} {\bibfnamefont {A.~H.}\ \bibnamefont {Mroue}},
  \bibinfo {author} {\bibfnamefont {L.}~\bibnamefont {Barack}}, \bibinfo
  {author} {\bibfnamefont {A.}~\bibnamefont {Buonanno}}, \bibinfo {author}
  {\bibfnamefont {H.~P.}\ \bibnamefont {Pfeiffer}}, \bibinfo {author}
  {\bibfnamefont {N.}~\bibnamefont {Sago}}, \ and\ \bibinfo {author}
  {\bibfnamefont {A.}~\bibnamefont {Taracchini}},\ }\href {\doibase
  10.1103/PhysRevLett.107.141101} {\bibfield  {journal} {\bibinfo  {journal}
  {Phys. Rev. Lett.}\ }\textbf {\bibinfo {volume} {107}},\ \bibinfo {pages}
  {141101} (\bibinfo {year} {2011})},\ \Eprint {http://arxiv.org/abs/1106.3278}
  {arXiv:1106.3278 [gr-qc]} \BibitemShut {NoStop}%
\bibitem [{\citenamefont {van~de Meent}(2017)}]{vandeMeent:2016hel}%
  \BibitemOpen
  \bibfield  {author} {\bibinfo {author} {\bibfnamefont {M.}~\bibnamefont
  {van~de Meent}},\ }\href {\doibase 10.1103/PhysRevLett.118.011101} {\bibfield
   {journal} {\bibinfo  {journal} {Phys. Rev. Lett.}\ }\textbf {\bibinfo
  {volume} {118}},\ \bibinfo {pages} {011101} (\bibinfo {year} {2017})},\
  \Eprint {http://arxiv.org/abs/1610.03497} {arXiv:1610.03497 [gr-qc]}
  \BibitemShut {NoStop}%
\bibitem [{\citenamefont {Dolan}\ \emph {et~al.}(2014)\citenamefont {Dolan},
  \citenamefont {Warburton}, \citenamefont {Harte}, \citenamefont {Le~Tiec},
  \citenamefont {Wardell},\ and\ \citenamefont {Barack}}]{Dolan:2013roa}%
  \BibitemOpen
  \bibfield  {author} {\bibinfo {author} {\bibfnamefont {S.~R.}\ \bibnamefont
  {Dolan}}, \bibinfo {author} {\bibfnamefont {N.}~\bibnamefont {Warburton}},
  \bibinfo {author} {\bibfnamefont {A.~I.}\ \bibnamefont {Harte}}, \bibinfo
  {author} {\bibfnamefont {A.}~\bibnamefont {Le~Tiec}}, \bibinfo {author}
  {\bibfnamefont {B.}~\bibnamefont {Wardell}}, \ and\ \bibinfo {author}
  {\bibfnamefont {L.}~\bibnamefont {Barack}},\ }\href {\doibase
  10.1103/PhysRevD.89.064011} {\bibfield  {journal} {\bibinfo  {journal} {Phys.
  Rev.}\ }\textbf {\bibinfo {volume} {D89}},\ \bibinfo {pages} {064011}
  (\bibinfo {year} {2014})},\ \Eprint {http://arxiv.org/abs/1312.0775}
  {arXiv:1312.0775 [gr-qc]} \BibitemShut {NoStop}%
\bibitem [{\citenamefont {Kavanagh}\ \emph {et~al.}(2017)\citenamefont
  {Kavanagh}, \citenamefont {Bini}, \citenamefont {Damour}, \citenamefont
  {Hopper}, \citenamefont {Ottewill},\ and\ \citenamefont
  {Wardell}}]{Kavanagh:2017wot}%
  \BibitemOpen
  \bibfield  {author} {\bibinfo {author} {\bibfnamefont {C.}~\bibnamefont
  {Kavanagh}}, \bibinfo {author} {\bibfnamefont {D.}~\bibnamefont {Bini}},
  \bibinfo {author} {\bibfnamefont {T.}~\bibnamefont {Damour}}, \bibinfo
  {author} {\bibfnamefont {S.}~\bibnamefont {Hopper}}, \bibinfo {author}
  {\bibfnamefont {A.~C.}\ \bibnamefont {Ottewill}}, \ and\ \bibinfo {author}
  {\bibfnamefont {B.}~\bibnamefont {Wardell}},\ }\href {\doibase
  10.1103/PhysRevD.96.064012} {\bibfield  {journal} {\bibinfo  {journal} {Phys.
  Rev.}\ }\textbf {\bibinfo {volume} {D96}},\ \bibinfo {pages} {064012}
  (\bibinfo {year} {2017})},\ \Eprint {http://arxiv.org/abs/1706.00459}
  {arXiv:1706.00459 [gr-qc]} \BibitemShut {NoStop}%
\bibitem [{\citenamefont {Bini}\ and\ \citenamefont
  {Damour}(2014{\natexlab{a}})}]{Bini:2014ica}%
  \BibitemOpen
  \bibfield  {author} {\bibinfo {author} {\bibfnamefont {D.}~\bibnamefont
  {Bini}}\ and\ \bibinfo {author} {\bibfnamefont {T.}~\bibnamefont {Damour}},\
  }\href {\doibase 10.1103/PhysRevD.90.024039} {\bibfield  {journal} {\bibinfo
  {journal} {Phys. Rev.}\ }\textbf {\bibinfo {volume} {D90}},\ \bibinfo {pages}
  {024039} (\bibinfo {year} {2014}{\natexlab{a}})},\ \Eprint
  {http://arxiv.org/abs/1404.2747} {arXiv:1404.2747 [gr-qc]} \BibitemShut
  {NoStop}%
\bibitem [{\citenamefont {Bini}\ \emph {et~al.}(2018)\citenamefont {Bini},
  \citenamefont {Damour}, \citenamefont {Geralico}, \citenamefont {Kavanagh},\
  and\ \citenamefont {van~de Meent}}]{Bini:2018ylh}%
  \BibitemOpen
  \bibfield  {author} {\bibinfo {author} {\bibfnamefont {D.}~\bibnamefont
  {Bini}}, \bibinfo {author} {\bibfnamefont {T.}~\bibnamefont {Damour}},
  \bibinfo {author} {\bibfnamefont {A.}~\bibnamefont {Geralico}}, \bibinfo
  {author} {\bibfnamefont {C.}~\bibnamefont {Kavanagh}}, \ and\ \bibinfo
  {author} {\bibfnamefont {M.}~\bibnamefont {van~de Meent}},\ }\href {\doibase
  10.1103/PhysRevD.98.104062} {\bibfield  {journal} {\bibinfo  {journal} {Phys.
  Rev.}\ }\textbf {\bibinfo {volume} {D98}},\ \bibinfo {pages} {104062}
  (\bibinfo {year} {2018})},\ \Eprint {http://arxiv.org/abs/1809.02516}
  {arXiv:1809.02516 [gr-qc]} \BibitemShut {NoStop}%
\bibitem [{\citenamefont {Dolan}\ \emph {et~al.}(2015)\citenamefont {Dolan},
  \citenamefont {Nolan}, \citenamefont {Ottewill}, \citenamefont {Warburton},\
  and\ \citenamefont {Wardell}}]{Dolan:2014pja}%
  \BibitemOpen
  \bibfield  {author} {\bibinfo {author} {\bibfnamefont {S.~R.}\ \bibnamefont
  {Dolan}}, \bibinfo {author} {\bibfnamefont {P.}~\bibnamefont {Nolan}},
  \bibinfo {author} {\bibfnamefont {A.~C.}\ \bibnamefont {Ottewill}}, \bibinfo
  {author} {\bibfnamefont {N.}~\bibnamefont {Warburton}}, \ and\ \bibinfo
  {author} {\bibfnamefont {B.}~\bibnamefont {Wardell}},\ }\href {\doibase
  10.1103/PhysRevD.91.023009} {\bibfield  {journal} {\bibinfo  {journal} {Phys.
  Rev.}\ }\textbf {\bibinfo {volume} {D91}},\ \bibinfo {pages} {023009}
  (\bibinfo {year} {2015})},\ \Eprint {http://arxiv.org/abs/1406.4890}
  {arXiv:1406.4890 [gr-qc]} \BibitemShut {NoStop}%
\bibitem [{\citenamefont {Bini}\ and\ \citenamefont
  {Damour}(2014{\natexlab{b}})}]{Bini:2014zxa}%
  \BibitemOpen
  \bibfield  {author} {\bibinfo {author} {\bibfnamefont {D.}~\bibnamefont
  {Bini}}\ and\ \bibinfo {author} {\bibfnamefont {T.}~\bibnamefont {Damour}},\
  }\href {\doibase 10.1103/PhysRevD.90.124037} {\bibfield  {journal} {\bibinfo
  {journal} {Phys. Rev.}\ }\textbf {\bibinfo {volume} {D90}},\ \bibinfo {pages}
  {124037} (\bibinfo {year} {2014}{\natexlab{b}})},\ \Eprint
  {http://arxiv.org/abs/1409.6933} {arXiv:1409.6933 [gr-qc]} \BibitemShut
  {NoStop}%
\bibitem [{\citenamefont {Detweiler}(2008)}]{Detweiler:2008ft}%
  \BibitemOpen
  \bibfield  {author} {\bibinfo {author} {\bibfnamefont {S.~L.}\ \bibnamefont
  {Detweiler}},\ }\href {\doibase 10.1103/PhysRevD.77.124026} {\bibfield
  {journal} {\bibinfo  {journal} {Phys. Rev.}\ }\textbf {\bibinfo {volume}
  {D77}},\ \bibinfo {pages} {124026} (\bibinfo {year} {2008})},\ \Eprint
  {http://arxiv.org/abs/0804.3529} {arXiv:0804.3529 [gr-qc]} \BibitemShut
  {NoStop}%
\bibitem [{\citenamefont {Barack}\ and\ \citenamefont
  {Sago}(2011)}]{Barack:2011ed}%
  \BibitemOpen
  \bibfield  {author} {\bibinfo {author} {\bibfnamefont {L.}~\bibnamefont
  {Barack}}\ and\ \bibinfo {author} {\bibfnamefont {N.}~\bibnamefont {Sago}},\
  }\href {\doibase 10.1103/PhysRevD.83.084023} {\bibfield  {journal} {\bibinfo
  {journal} {Phys. Rev.}\ }\textbf {\bibinfo {volume} {D83}},\ \bibinfo {pages}
  {084023} (\bibinfo {year} {2011})},\ \Eprint {http://arxiv.org/abs/1101.3331}
  {arXiv:1101.3331 [gr-qc]} \BibitemShut {NoStop}%
\bibitem [{\citenamefont {Akcay}\ \emph {et~al.}(2015)\citenamefont {Akcay},
  \citenamefont {Le~Tiec}, \citenamefont {Barack}, \citenamefont {Sago},\ and\
  \citenamefont {Warburton}}]{Akcay:2015pza}%
  \BibitemOpen
  \bibfield  {author} {\bibinfo {author} {\bibfnamefont {S.}~\bibnamefont
  {Akcay}}, \bibinfo {author} {\bibfnamefont {A.}~\bibnamefont {Le~Tiec}},
  \bibinfo {author} {\bibfnamefont {L.}~\bibnamefont {Barack}}, \bibinfo
  {author} {\bibfnamefont {N.}~\bibnamefont {Sago}}, \ and\ \bibinfo {author}
  {\bibfnamefont {N.}~\bibnamefont {Warburton}},\ }\href {\doibase
  10.1103/PhysRevD.91.124014} {\bibfield  {journal} {\bibinfo  {journal} {Phys.
  Rev.}\ }\textbf {\bibinfo {volume} {D91}},\ \bibinfo {pages} {124014}
  (\bibinfo {year} {2015})},\ \Eprint {http://arxiv.org/abs/1503.01374}
  {arXiv:1503.01374 [gr-qc]} \BibitemShut {NoStop}%
\bibitem [{\citenamefont {Kavanagh}\ \emph
  {et~al.}(2015{\natexlab{a}})\citenamefont {Kavanagh}, \citenamefont
  {Ottewill},\ and\ \citenamefont {Wardell}}]{Kavanagh:2015lva}%
  \BibitemOpen
  \bibfield  {author} {\bibinfo {author} {\bibfnamefont {C.}~\bibnamefont
  {Kavanagh}}, \bibinfo {author} {\bibfnamefont {A.~C.}\ \bibnamefont
  {Ottewill}}, \ and\ \bibinfo {author} {\bibfnamefont {B.}~\bibnamefont
  {Wardell}},\ }\href {\doibase 10.1103/PhysRevD.92.084025} {\bibfield
  {journal} {\bibinfo  {journal} {Phys. Rev.}\ }\textbf {\bibinfo {volume}
  {D92}},\ \bibinfo {pages} {084025} (\bibinfo {year} {2015}{\natexlab{a}})},\
  \Eprint {http://arxiv.org/abs/1503.02334} {arXiv:1503.02334 [gr-qc]}
  \BibitemShut {NoStop}%
\bibitem [{\citenamefont {Shah}\ \emph {et~al.}(2014)\citenamefont {Shah},
  \citenamefont {Friedman},\ and\ \citenamefont {Whiting}}]{Shah:2013uya}%
  \BibitemOpen
  \bibfield  {author} {\bibinfo {author} {\bibfnamefont {A.~G.}\ \bibnamefont
  {Shah}}, \bibinfo {author} {\bibfnamefont {J.~L.}\ \bibnamefont {Friedman}},
  \ and\ \bibinfo {author} {\bibfnamefont {B.~F.}\ \bibnamefont {Whiting}},\
  }\href {\doibase 10.1103/PhysRevD.89.064042} {\bibfield  {journal} {\bibinfo
  {journal} {Phys. Rev.}\ }\textbf {\bibinfo {volume} {D89}},\ \bibinfo {pages}
  {064042} (\bibinfo {year} {2014})},\ \Eprint {http://arxiv.org/abs/1312.1952}
  {arXiv:1312.1952 [gr-qc]} \BibitemShut {NoStop}%
\bibitem [{\citenamefont {Johnson-McDaniel}\ \emph {et~al.}(2015)\citenamefont
  {Johnson-McDaniel}, \citenamefont {Shah},\ and\ \citenamefont
  {Whiting}}]{Johnson-McDaniel:2015vva}%
  \BibitemOpen
  \bibfield  {author} {\bibinfo {author} {\bibfnamefont {N.~K.}\ \bibnamefont
  {Johnson-McDaniel}}, \bibinfo {author} {\bibfnamefont {A.~G.}\ \bibnamefont
  {Shah}}, \ and\ \bibinfo {author} {\bibfnamefont {B.~F.}\ \bibnamefont
  {Whiting}},\ }\href {\doibase 10.1103/PhysRevD.92.044007} {\bibfield
  {journal} {\bibinfo  {journal} {Phys. Rev.}\ }\textbf {\bibinfo {volume}
  {D92}},\ \bibinfo {pages} {044007} (\bibinfo {year} {2015})},\ \Eprint
  {http://arxiv.org/abs/1503.02638} {arXiv:1503.02638 [gr-qc]} \BibitemShut
  {NoStop}%
\bibitem [{\citenamefont {Zimmerman}\ \emph {et~al.}(2016)\citenamefont
  {Zimmerman}, \citenamefont {Lewis},\ and\ \citenamefont
  {Pfeiffer}}]{Zimmerman:2016ajr}%
  \BibitemOpen
  \bibfield  {author} {\bibinfo {author} {\bibfnamefont {A.}~\bibnamefont
  {Zimmerman}}, \bibinfo {author} {\bibfnamefont {A.~G.~M.}\ \bibnamefont
  {Lewis}}, \ and\ \bibinfo {author} {\bibfnamefont {H.~P.}\ \bibnamefont
  {Pfeiffer}},\ }\href {\doibase 10.1103/PhysRevLett.117.191101} {\bibfield
  {journal} {\bibinfo  {journal} {Phys. Rev. Lett.}\ }\textbf {\bibinfo
  {volume} {117}},\ \bibinfo {pages} {191101} (\bibinfo {year} {2016})},\
  \Eprint {http://arxiv.org/abs/1606.08056} {arXiv:1606.08056 [gr-qc]}
  \BibitemShut {NoStop}%
\bibitem [{\citenamefont {Le~Tiec}\ \emph
  {et~al.}(2012{\natexlab{a}})\citenamefont {Le~Tiec}, \citenamefont
  {Blanchet},\ and\ \citenamefont {Whiting}}]{LeTiec:2011ab}%
  \BibitemOpen
  \bibfield  {author} {\bibinfo {author} {\bibfnamefont {A.}~\bibnamefont
  {Le~Tiec}}, \bibinfo {author} {\bibfnamefont {L.}~\bibnamefont {Blanchet}}, \
  and\ \bibinfo {author} {\bibfnamefont {B.~F.}\ \bibnamefont {Whiting}},\
  }\href {\doibase 10.1103/PhysRevD.85.064039} {\bibfield  {journal} {\bibinfo
  {journal} {Phys. Rev.}\ }\textbf {\bibinfo {volume} {D85}},\ \bibinfo {pages}
  {064039} (\bibinfo {year} {2012}{\natexlab{a}})},\ \Eprint
  {http://arxiv.org/abs/1111.5378} {arXiv:1111.5378 [gr-qc]} \BibitemShut
  {NoStop}%
\bibitem [{\citenamefont {Damour}(2010)}]{Damour:2009sf}%
  \BibitemOpen
  \bibfield  {author} {\bibinfo {author} {\bibfnamefont {T.}~\bibnamefont
  {Damour}},\ }\href {\doibase 10.1103/PhysRevD.81.024017} {\bibfield
  {journal} {\bibinfo  {journal} {Phys.Rev.}\ }\textbf {\bibinfo {volume}
  {D81}},\ \bibinfo {pages} {024017} (\bibinfo {year} {2010})},\ \Eprint
  {http://arxiv.org/abs/0910.5533} {arXiv:0910.5533 [gr-qc]} \BibitemShut
  {NoStop}%
\bibitem [{\citenamefont {Le~Tiec}\ \emph
  {et~al.}(2012{\natexlab{b}})\citenamefont {Le~Tiec}, \citenamefont
  {Barausse},\ and\ \citenamefont {Buonanno}}]{LeTiec:2011dp}%
  \BibitemOpen
  \bibfield  {author} {\bibinfo {author} {\bibfnamefont {A.}~\bibnamefont
  {Le~Tiec}}, \bibinfo {author} {\bibfnamefont {E.}~\bibnamefont {Barausse}}, \
  and\ \bibinfo {author} {\bibfnamefont {A.}~\bibnamefont {Buonanno}},\ }\href
  {\doibase 10.1103/PhysRevLett.108.131103} {\bibfield  {journal} {\bibinfo
  {journal} {Phys. Rev. Lett.}\ }\textbf {\bibinfo {volume} {108}},\ \bibinfo
  {pages} {131103} (\bibinfo {year} {2012}{\natexlab{b}})},\ \Eprint
  {http://arxiv.org/abs/1111.5609} {arXiv:1111.5609 [gr-qc]} \BibitemShut
  {NoStop}%
\bibitem [{\citenamefont {Barausse}\ \emph {et~al.}(2012)\citenamefont
  {Barausse}, \citenamefont {Buonanno},\ and\ \citenamefont
  {Le~Tiec}}]{Barausse:2011dq}%
  \BibitemOpen
  \bibfield  {author} {\bibinfo {author} {\bibfnamefont {E.}~\bibnamefont
  {Barausse}}, \bibinfo {author} {\bibfnamefont {A.}~\bibnamefont {Buonanno}},
  \ and\ \bibinfo {author} {\bibfnamefont {A.}~\bibnamefont {Le~Tiec}},\ }\href
  {\doibase 10.1103/PhysRevD.85.064010} {\bibfield  {journal} {\bibinfo
  {journal} {Phys. Rev.}\ }\textbf {\bibinfo {volume} {D85}},\ \bibinfo {pages}
  {064010} (\bibinfo {year} {2012})},\ \Eprint {http://arxiv.org/abs/1111.5610}
  {arXiv:1111.5610 [gr-qc]} \BibitemShut {NoStop}%
\bibitem [{\citenamefont {Kavanagh}\ \emph
  {et~al.}(2015{\natexlab{b}})\citenamefont {Kavanagh}, \citenamefont
  {Ottewill},\ and\ \citenamefont {Wardell}}]{kavanaghetal:2015}%
  \BibitemOpen
  \bibfield  {author} {\bibinfo {author} {\bibfnamefont {C.}~\bibnamefont
  {Kavanagh}}, \bibinfo {author} {\bibfnamefont {A.~C.}\ \bibnamefont
  {Ottewill}}, \ and\ \bibinfo {author} {\bibfnamefont {B.}~\bibnamefont
  {Wardell}},\ }\href {\doibase 10.1103/PhysRevD.93.124038} {\bibfield
  {journal} {\bibinfo  {journal} {Phys.Rev.}\ }\textbf {\bibinfo {volume}
  {D92}},\ \bibinfo {pages} {024017} (\bibinfo {year} {2015}{\natexlab{b}})},\
  \Eprint {http://arxiv.org/abs/1601.03394} {arXiv:1601.03394 [gr-qc]}
  \BibitemShut {NoStop}%
\bibitem [{\citenamefont {Hopper}\ \emph {et~al.}(2016)\citenamefont {Hopper},
  \citenamefont {Kavanagh},\ and\ \citenamefont {Ottewill}}]{Hopper:2015icj}%
  \BibitemOpen
  \bibfield  {author} {\bibinfo {author} {\bibfnamefont {S.}~\bibnamefont
  {Hopper}}, \bibinfo {author} {\bibfnamefont {C.}~\bibnamefont {Kavanagh}}, \
  and\ \bibinfo {author} {\bibfnamefont {A.~C.}\ \bibnamefont {Ottewill}},\
  }\href {\doibase 10.1103/PhysRevD.93.044010} {\bibfield  {journal} {\bibinfo
  {journal} {Phys. Rev.}\ }\textbf {\bibinfo {volume} {D93}},\ \bibinfo {pages}
  {044010} (\bibinfo {year} {2016})},\ \Eprint
  {http://arxiv.org/abs/1512.01556} {arXiv:1512.01556 [gr-qc]} \BibitemShut
  {NoStop}%
\bibitem [{\citenamefont {Kavanagh}\ \emph {et~al.}(2016)\citenamefont
  {Kavanagh}, \citenamefont {Ottewill},\ and\ \citenamefont
  {Wardell}}]{Kavanagh:2016idg}%
  \BibitemOpen
  \bibfield  {author} {\bibinfo {author} {\bibfnamefont {C.}~\bibnamefont
  {Kavanagh}}, \bibinfo {author} {\bibfnamefont {A.~C.}\ \bibnamefont
  {Ottewill}}, \ and\ \bibinfo {author} {\bibfnamefont {B.}~\bibnamefont
  {Wardell}},\ }\href {\doibase 10.1103/PhysRevD.93.124038} {\bibfield
  {journal} {\bibinfo  {journal} {Phys. Rev.}\ }\textbf {\bibinfo {volume}
  {D93}},\ \bibinfo {pages} {124038} (\bibinfo {year} {2016})},\ \Eprint
  {http://arxiv.org/abs/1601.03394} {arXiv:1601.03394 [gr-qc]} \BibitemShut
  {NoStop}%
\bibitem [{\citenamefont {Bini}\ and\ \citenamefont
  {Damour}(2014{\natexlab{c}})}]{Bini:2013rfa}%
  \BibitemOpen
  \bibfield  {author} {\bibinfo {author} {\bibfnamefont {D.}~\bibnamefont
  {Bini}}\ and\ \bibinfo {author} {\bibfnamefont {T.}~\bibnamefont {Damour}},\
  }\href {\doibase 10.1103/PhysRevD.89.064063} {\bibfield  {journal} {\bibinfo
  {journal} {Phys. Rev.}\ }\textbf {\bibinfo {volume} {D89}},\ \bibinfo {pages}
  {064063} (\bibinfo {year} {2014}{\natexlab{c}})},\ \Eprint
  {http://arxiv.org/abs/1312.2503} {arXiv:1312.2503 [gr-qc]} \BibitemShut
  {NoStop}%
\bibitem [{\citenamefont {Bini}\ and\ \citenamefont
  {Damour}(2014{\natexlab{d}})}]{Bini:2014nfa}%
  \BibitemOpen
  \bibfield  {author} {\bibinfo {author} {\bibfnamefont {D.}~\bibnamefont
  {Bini}}\ and\ \bibinfo {author} {\bibfnamefont {T.}~\bibnamefont {Damour}},\
  }\href {\doibase 10.1103/PhysRevD.89.104047} {\bibfield  {journal} {\bibinfo
  {journal} {Phys. Rev.}\ }\textbf {\bibinfo {volume} {D89}},\ \bibinfo {pages}
  {104047} (\bibinfo {year} {2014}{\natexlab{d}})},\ \Eprint
  {http://arxiv.org/abs/1403.2366} {arXiv:1403.2366 [gr-qc]} \BibitemShut
  {NoStop}%
\bibitem [{\citenamefont {Bini}\ and\ \citenamefont
  {Damour}(2015)}]{Bini:2015bla}%
  \BibitemOpen
  \bibfield  {author} {\bibinfo {author} {\bibfnamefont {D.}~\bibnamefont
  {Bini}}\ and\ \bibinfo {author} {\bibfnamefont {T.}~\bibnamefont {Damour}},\
  }\href {\doibase 10.1103/PhysRevD.91.064050} {\bibfield  {journal} {\bibinfo
  {journal} {Phys. Rev.}\ }\textbf {\bibinfo {volume} {D91}},\ \bibinfo {pages}
  {064050} (\bibinfo {year} {2015})},\ \Eprint
  {http://arxiv.org/abs/1502.02450} {arXiv:1502.02450 [gr-qc]} \BibitemShut
  {NoStop}%
\bibitem [{\citenamefont {Bini}\ \emph
  {et~al.}(2016{\natexlab{a}})\citenamefont {Bini}, \citenamefont {Damour},\
  and\ \citenamefont {Geralico}}]{Bini:2015bfb}%
  \BibitemOpen
  \bibfield  {author} {\bibinfo {author} {\bibfnamefont {D.}~\bibnamefont
  {Bini}}, \bibinfo {author} {\bibfnamefont {T.}~\bibnamefont {Damour}}, \ and\
  \bibinfo {author} {\bibfnamefont {A.}~\bibnamefont {Geralico}},\ }\href
  {\doibase 10.1103/PhysRevD.93.064023} {\bibfield  {journal} {\bibinfo
  {journal} {Phys. Rev.}\ }\textbf {\bibinfo {volume} {D93}},\ \bibinfo {pages}
  {064023} (\bibinfo {year} {2016}{\natexlab{a}})},\ \Eprint
  {http://arxiv.org/abs/1511.04533} {arXiv:1511.04533 [gr-qc]} \BibitemShut
  {NoStop}%
\bibitem [{\citenamefont {Bini}\ \emph {et~al.}(2015)\citenamefont {Bini},
  \citenamefont {Damour},\ and\ \citenamefont {Geralico}}]{Bini:2015xua}%
  \BibitemOpen
  \bibfield  {author} {\bibinfo {author} {\bibfnamefont {D.}~\bibnamefont
  {Bini}}, \bibinfo {author} {\bibfnamefont {T.}~\bibnamefont {Damour}}, \ and\
  \bibinfo {author} {\bibfnamefont {A.}~\bibnamefont {Geralico}},\ }\href
  {\doibase 10.1103/PhysRevD.93.109902, 10.1103/PhysRevD.92.124058} {\bibfield
  {journal} {\bibinfo  {journal} {Phys. Rev.}\ }\textbf {\bibinfo {volume}
  {D92}},\ \bibinfo {pages} {124058} (\bibinfo {year} {2015})},\ \bibinfo
  {note} {[Erratum: Phys. Rev.D93,no.10,109902(2016)]},\ \Eprint
  {http://arxiv.org/abs/1510.06230} {arXiv:1510.06230 [gr-qc]} \BibitemShut
  {NoStop}%
\bibitem [{\citenamefont {Bini}\ \emph
  {et~al.}(2016{\natexlab{b}})\citenamefont {Bini}, \citenamefont {Damour},\
  and\ \citenamefont {Geralico}}]{Bini:2016dvs}%
  \BibitemOpen
  \bibfield  {author} {\bibinfo {author} {\bibfnamefont {D.}~\bibnamefont
  {Bini}}, \bibinfo {author} {\bibfnamefont {T.}~\bibnamefont {Damour}}, \ and\
  \bibinfo {author} {\bibfnamefont {A.}~\bibnamefont {Geralico}},\ }\href
  {\doibase 10.1103/PhysRevD.93.124058} {\bibfield  {journal} {\bibinfo
  {journal} {Phys. Rev.}\ }\textbf {\bibinfo {volume} {D93}},\ \bibinfo {pages}
  {124058} (\bibinfo {year} {2016}{\natexlab{b}})},\ \Eprint
  {http://arxiv.org/abs/1602.08282} {arXiv:1602.08282 [gr-qc]} \BibitemShut
  {NoStop}%
\bibitem [{\citenamefont {Bini}\ \emph
  {et~al.}(2016{\natexlab{c}})\citenamefont {Bini}, \citenamefont {Damour},\
  and\ \citenamefont {Geralico}}]{Bini:2016qtx}%
  \BibitemOpen
  \bibfield  {author} {\bibinfo {author} {\bibfnamefont {D.}~\bibnamefont
  {Bini}}, \bibinfo {author} {\bibfnamefont {T.}~\bibnamefont {Damour}}, \ and\
  \bibinfo {author} {\bibfnamefont {a.}~\bibnamefont {Geralico}},\ }\href
  {\doibase 10.1103/PhysRevD.93.104017} {\bibfield  {journal} {\bibinfo
  {journal} {Phys. Rev.}\ }\textbf {\bibinfo {volume} {D93}},\ \bibinfo {pages}
  {104017} (\bibinfo {year} {2016}{\natexlab{c}})},\ \Eprint
  {http://arxiv.org/abs/1601.02988} {arXiv:1601.02988 [gr-qc]} \BibitemShut
  {NoStop}%
\bibitem [{\citenamefont {Akcay}\ \emph {et~al.}(2012)\citenamefont {Akcay},
  \citenamefont {Barack}, \citenamefont {Damour},\ and\ \citenamefont
  {Sago}}]{Akcay:2012ea}%
  \BibitemOpen
  \bibfield  {author} {\bibinfo {author} {\bibfnamefont {S.}~\bibnamefont
  {Akcay}}, \bibinfo {author} {\bibfnamefont {L.}~\bibnamefont {Barack}},
  \bibinfo {author} {\bibfnamefont {T.}~\bibnamefont {Damour}}, \ and\ \bibinfo
  {author} {\bibfnamefont {N.}~\bibnamefont {Sago}},\ }\href {\doibase
  10.1103/PhysRevD.86.104041} {\bibfield  {journal} {\bibinfo  {journal} {Phys.
  Rev.}\ }\textbf {\bibinfo {volume} {D86}},\ \bibinfo {pages} {104041}
  (\bibinfo {year} {2012})},\ \Eprint {http://arxiv.org/abs/1209.0964}
  {arXiv:1209.0964 [gr-qc]} \BibitemShut {NoStop}%
\bibitem [{\citenamefont {Damour}\ \emph {et~al.}(2003)\citenamefont {Damour},
  \citenamefont {Iyer}, \citenamefont {Jaranowski},\ and\ \citenamefont
  {Sathyaprakash}}]{Damour:2002vi}%
  \BibitemOpen
  \bibfield  {author} {\bibinfo {author} {\bibfnamefont {T.}~\bibnamefont
  {Damour}}, \bibinfo {author} {\bibfnamefont {B.~R.}\ \bibnamefont {Iyer}},
  \bibinfo {author} {\bibfnamefont {P.}~\bibnamefont {Jaranowski}}, \ and\
  \bibinfo {author} {\bibfnamefont {B.~S.}\ \bibnamefont {Sathyaprakash}},\
  }\href {\doibase 10.1103/PhysRevD.67.064028} {\bibfield  {journal} {\bibinfo
  {journal} {Phys. Rev.}\ }\textbf {\bibinfo {volume} {D67}},\ \bibinfo {pages}
  {064028} (\bibinfo {year} {2003})},\ \Eprint
  {http://arxiv.org/abs/gr-qc/0211041} {arXiv:gr-qc/0211041 [gr-qc]}
  \BibitemShut {NoStop}%
\bibitem [{\citenamefont {Buonanno}\ \emph {et~al.}(2007)\citenamefont
  {Buonanno}, \citenamefont {Pan}, \citenamefont {Baker}, \citenamefont
  {Centrella}, \citenamefont {Kelly}, \citenamefont {McWilliams},\ and\
  \citenamefont {van Meter}}]{Buonanno:2007pf}%
  \BibitemOpen
  \bibfield  {author} {\bibinfo {author} {\bibfnamefont {A.}~\bibnamefont
  {Buonanno}}, \bibinfo {author} {\bibfnamefont {Y.}~\bibnamefont {Pan}},
  \bibinfo {author} {\bibfnamefont {J.~G.}\ \bibnamefont {Baker}}, \bibinfo
  {author} {\bibfnamefont {J.}~\bibnamefont {Centrella}}, \bibinfo {author}
  {\bibfnamefont {B.~J.}\ \bibnamefont {Kelly}}, \bibinfo {author}
  {\bibfnamefont {S.~T.}\ \bibnamefont {McWilliams}}, \ and\ \bibinfo {author}
  {\bibfnamefont {J.~R.}\ \bibnamefont {van Meter}},\ }\href {\doibase
  10.1103/PhysRevD.76.104049} {\bibfield  {journal} {\bibinfo  {journal} {Phys.
  Rev.}\ }\textbf {\bibinfo {volume} {D76}},\ \bibinfo {pages} {104049}
  (\bibinfo {year} {2007})},\ \Eprint {http://arxiv.org/abs/0706.3732}
  {arXiv:0706.3732 [gr-qc]} \BibitemShut {NoStop}%
\bibitem [{\citenamefont {Akcay}\ and\ \citenamefont {van~de
  Meent}(2016)}]{Akcay:2015pjz}%
  \BibitemOpen
  \bibfield  {author} {\bibinfo {author} {\bibfnamefont {S.}~\bibnamefont
  {Akcay}}\ and\ \bibinfo {author} {\bibfnamefont {M.}~\bibnamefont {van~de
  Meent}},\ }\href {\doibase 10.1103/PhysRevD.93.064063} {\bibfield  {journal}
  {\bibinfo  {journal} {Phys. Rev.}\ }\textbf {\bibinfo {volume} {D93}},\
  \bibinfo {pages} {064063} (\bibinfo {year} {2016})},\ \Eprint
  {http://arxiv.org/abs/1512.03392} {arXiv:1512.03392 [gr-qc]} \BibitemShut
  {NoStop}%
\bibitem [{\citenamefont {Bini}\ \emph {et~al.}(2012)\citenamefont {Bini},
  \citenamefont {Damour},\ and\ \citenamefont {Faye}}]{Bini:2012gu}%
  \BibitemOpen
  \bibfield  {author} {\bibinfo {author} {\bibfnamefont {D.}~\bibnamefont
  {Bini}}, \bibinfo {author} {\bibfnamefont {T.}~\bibnamefont {Damour}}, \ and\
  \bibinfo {author} {\bibfnamefont {G.}~\bibnamefont {Faye}},\ }\href {\doibase
  10.1103/PhysRevD.85.124034} {\bibfield  {journal} {\bibinfo  {journal} {Phys.
  Rev.}\ }\textbf {\bibinfo {volume} {D85}},\ \bibinfo {pages} {124034}
  (\bibinfo {year} {2012})},\ \Eprint {http://arxiv.org/abs/1202.3565}
  {arXiv:1202.3565 [gr-qc]} \BibitemShut {NoStop}%
\bibitem [{\citenamefont {Steinhoff}\ \emph {et~al.}(2016)\citenamefont
  {Steinhoff}, \citenamefont {Hinderer}, \citenamefont {Buonanno},\ and\
  \citenamefont {Taracchini}}]{Steinhoff:2016rfi}%
  \BibitemOpen
  \bibfield  {author} {\bibinfo {author} {\bibfnamefont {J.}~\bibnamefont
  {Steinhoff}}, \bibinfo {author} {\bibfnamefont {T.}~\bibnamefont {Hinderer}},
  \bibinfo {author} {\bibfnamefont {A.}~\bibnamefont {Buonanno}}, \ and\
  \bibinfo {author} {\bibfnamefont {A.}~\bibnamefont {Taracchini}},\ }\href
  {\doibase 10.1103/PhysRevD.94.104028} {\bibfield  {journal} {\bibinfo
  {journal} {Phys. Rev.}\ }\textbf {\bibinfo {volume} {D94}},\ \bibinfo {pages}
  {104028} (\bibinfo {year} {2016})},\ \Eprint
  {http://arxiv.org/abs/1608.01907} {arXiv:1608.01907 [gr-qc]} \BibitemShut
  {NoStop}%
\bibitem [{\citenamefont {Le~Tiec}(2015)}]{Tiec:2015cxa}%
  \BibitemOpen
  \bibfield  {author} {\bibinfo {author} {\bibfnamefont {A.}~\bibnamefont
  {Le~Tiec}},\ }\href {\doibase 10.1103/PhysRevD.92.084021} {\bibfield
  {journal} {\bibinfo  {journal} {Phys. Rev.}\ }\textbf {\bibinfo {volume}
  {D92}},\ \bibinfo {pages} {084021} (\bibinfo {year} {2015})},\ \Eprint
  {http://arxiv.org/abs/1506.05648} {arXiv:1506.05648 [gr-qc]} \BibitemShut
  {NoStop}%
\bibitem [{\citenamefont {Damour}\ and\ \citenamefont
  {Nagar}(2007)}]{Damour:2007xr}%
  \BibitemOpen
  \bibfield  {author} {\bibinfo {author} {\bibfnamefont {T.}~\bibnamefont
  {Damour}}\ and\ \bibinfo {author} {\bibfnamefont {A.}~\bibnamefont {Nagar}},\
  }\href {\doibase 10.1103/PhysRevD.76.064028} {\bibfield  {journal} {\bibinfo
  {journal} {Phys. Rev.}\ }\textbf {\bibinfo {volume} {D76}},\ \bibinfo {pages}
  {064028} (\bibinfo {year} {2007})},\ \Eprint {http://arxiv.org/abs/0705.2519}
  {arXiv:0705.2519 [gr-qc]} \BibitemShut {NoStop}%
\bibitem [{\citenamefont {Pan}\ \emph {et~al.}(2011{\natexlab{b}})\citenamefont
  {Pan}, \citenamefont {Buonanno}, \citenamefont {Fujita}, \citenamefont
  {Racine},\ and\ \citenamefont {Tagoshi}}]{Pan:2010hz}%
  \BibitemOpen
  \bibfield  {author} {\bibinfo {author} {\bibfnamefont {Y.}~\bibnamefont
  {Pan}}, \bibinfo {author} {\bibfnamefont {A.}~\bibnamefont {Buonanno}},
  \bibinfo {author} {\bibfnamefont {R.}~\bibnamefont {Fujita}}, \bibinfo
  {author} {\bibfnamefont {E.}~\bibnamefont {Racine}}, \ and\ \bibinfo {author}
  {\bibfnamefont {H.}~\bibnamefont {Tagoshi}},\ }\href {\doibase
  10.1103/PhysRevD.83.064003, 10.1103/PhysRevD.87.109901} {\bibfield  {journal}
  {\bibinfo  {journal} {Phys. Rev.}\ }\textbf {\bibinfo {volume} {D83}},\
  \bibinfo {pages} {064003} (\bibinfo {year} {2011}{\natexlab{b}})},\ \bibinfo
  {note} {[Erratum: Phys. Rev.D87,no.10,109901(2013)]},\ \Eprint
  {http://arxiv.org/abs/1006.0431} {arXiv:1006.0431 [gr-qc]} \BibitemShut
  {NoStop}%
\bibitem [{\citenamefont {Chu}\ \emph {et~al.}(2016)\citenamefont {Chu},
  \citenamefont {Fong}, \citenamefont {Kumar}, \citenamefont {Pfeiffer},
  \citenamefont {Boyle}, \citenamefont {Hemberger}, \citenamefont {Kidder},
  \citenamefont {Scheel},\ and\ \citenamefont {Szilagyi}}]{Chu:2015kft}%
  \BibitemOpen
  \bibfield  {author} {\bibinfo {author} {\bibfnamefont {T.}~\bibnamefont
  {Chu}}, \bibinfo {author} {\bibfnamefont {H.}~\bibnamefont {Fong}}, \bibinfo
  {author} {\bibfnamefont {P.}~\bibnamefont {Kumar}}, \bibinfo {author}
  {\bibfnamefont {H.~P.}\ \bibnamefont {Pfeiffer}}, \bibinfo {author}
  {\bibfnamefont {M.}~\bibnamefont {Boyle}}, \bibinfo {author} {\bibfnamefont
  {D.~A.}\ \bibnamefont {Hemberger}}, \bibinfo {author} {\bibfnamefont {L.~E.}\
  \bibnamefont {Kidder}}, \bibinfo {author} {\bibfnamefont {M.~A.}\
  \bibnamefont {Scheel}}, \ and\ \bibinfo {author} {\bibfnamefont
  {B.}~\bibnamefont {Szilagyi}},\ }\href {\doibase
  10.1088/0264-9381/33/16/165001} {\bibfield  {journal} {\bibinfo  {journal}
  {Class. Quant. Grav.}\ }\textbf {\bibinfo {volume} {33}},\ \bibinfo {pages}
  {165001} (\bibinfo {year} {2016})},\ \Eprint
  {http://arxiv.org/abs/1512.06800} {arXiv:1512.06800 [gr-qc]} \BibitemShut
  {NoStop}%
\bibitem [{\citenamefont {Ossokine}\ \emph {et~al.}(2018)\citenamefont
  {Ossokine}, \citenamefont {Dietrich}, \citenamefont {Foley}, \citenamefont
  {Katebi},\ and\ \citenamefont {Lovelace}}]{Ossokine:2017dge}%
  \BibitemOpen
  \bibfield  {author} {\bibinfo {author} {\bibfnamefont {S.}~\bibnamefont
  {Ossokine}}, \bibinfo {author} {\bibfnamefont {T.}~\bibnamefont {Dietrich}},
  \bibinfo {author} {\bibfnamefont {E.}~\bibnamefont {Foley}}, \bibinfo
  {author} {\bibfnamefont {R.}~\bibnamefont {Katebi}}, \ and\ \bibinfo {author}
  {\bibfnamefont {G.}~\bibnamefont {Lovelace}},\ }\href {\doibase
  10.1103/PhysRevD.98.104057} {\bibfield  {journal} {\bibinfo  {journal} {Phys.
  Rev.}\ }\textbf {\bibinfo {volume} {D98}},\ \bibinfo {pages} {104057}
  (\bibinfo {year} {2018})},\ \Eprint {http://arxiv.org/abs/1712.06533}
  {arXiv:1712.06533 [gr-qc]} \BibitemShut {NoStop}%
\bibitem [{BHP()}]{BHPToolkit}%
  \BibitemOpen
  \href@noop {} {\enquote {\bibinfo {title} {{Black Hole Perturbation
  Toolkit}},}\ }\bibinfo {howpublished}
  {(\href{http://bhptoolkit.org/}{bhptoolkit.org})}\BibitemShut {NoStop}%
\end{thebibliography}%

\end{document}